\documentclass[12pt]{book}
\usepackage{graphicx}
\usepackage{amsmath,amssymb,bm}
\usepackage{hyperref}
\usepackage{subfigure}
\usepackage{multirow}
\usepackage{float}
\newcommand{\dis}{\displaystyle}
\newcommand{\rv}{{\bm r}}
\newcommand{\xv}{\bm{x}}
\newcommand{\q}{{\bm q}}
\newcommand{\dd}{\text{d}}
\newcommand{\U}{\text{U}}
\newcommand{\kv}{\bm{k}}
\newcommand{\pv}{\bm{p}}
\newcommand{\qv}{\bm{q}}
\newcommand{\Bv}{\bm{B}}

\newcommand{\Ev}{\bm{E}}
\newcommand{\zev}{\bm{0}}
\newcommand{\jv}{\bm{j}}

\newcommand{\e}{{\rm e}}
\renewcommand\d{\partial}
\renewcommand\i{{\rm i}}
\newcommand{\Z}{\mathbb{Z}}
\newcommand{\SU}{\rm SU}
\newcommand{\na}{\bm{\nabla}}

\newcommand{\mG}{\mathcal{G}}
\newcommand{\ra}{\rightarrow}
\newcommand{\eps}{\varepsilon}
\newcommand{\ex}[1]{\left<#1\right>}

\title{Novel dynamic critical phenomena induced by superfluidity and the chiral magnetic effect in Quantum Chromodynamics}
\author{Noriyuki Sogabe}

\begin{document}
\maketitle

\begin{figure}[t]
\includegraphics[bb=0 0 1024 768,height=0.01cm]{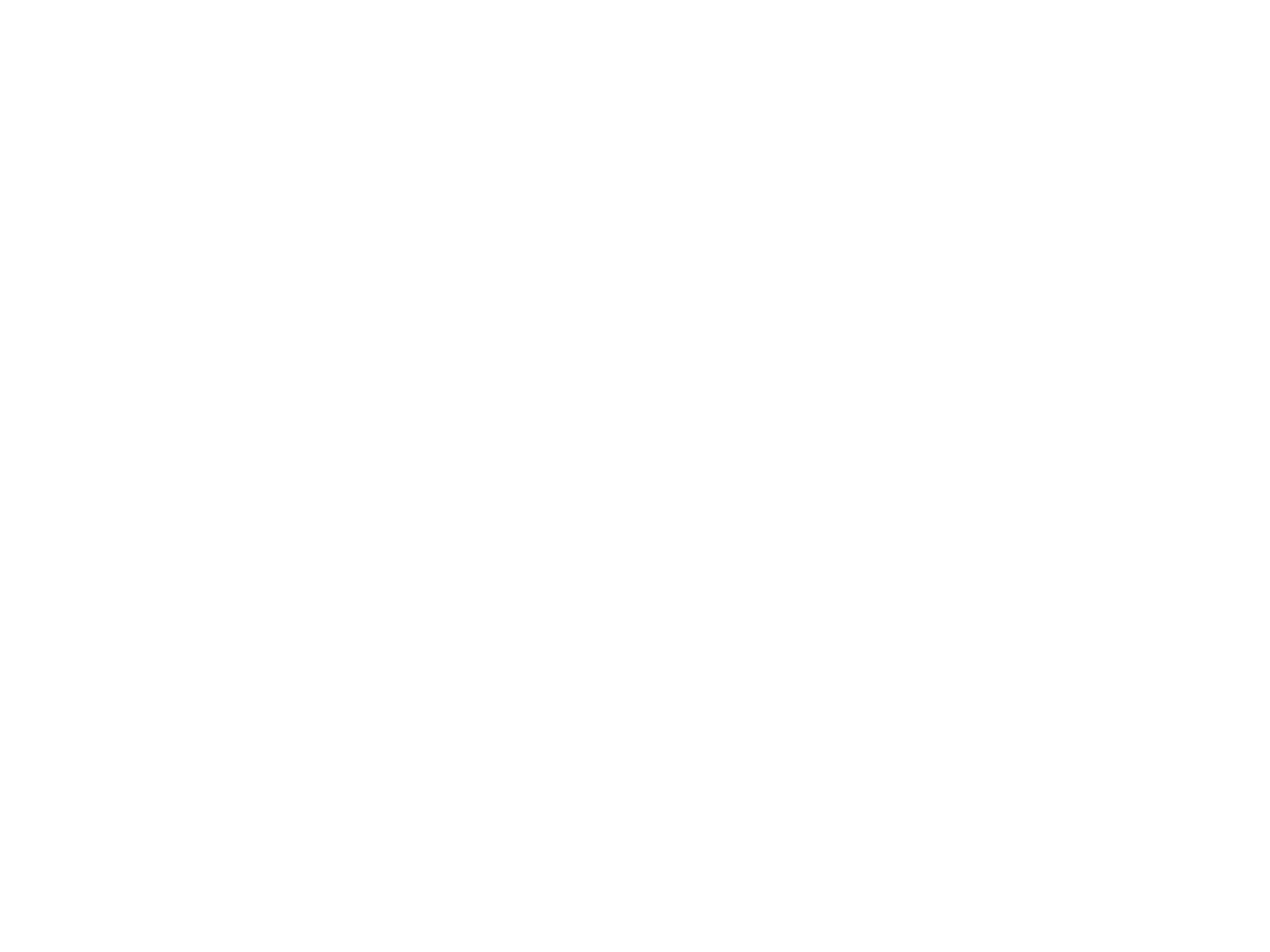}
\end{figure}

\thispagestyle{empty}
\begin{center}
\begin{minipage}{1\linewidth}
\centering{\Large A Thesis for the Degree of Ph.D. in Science\par }
\vspace{3cm}
{\Large  Novel dynamic critical phenomena induced by superfluidity and the chiral magnetic effect in Quantum Chromodynamics \par}
\vspace{3cm}
\vspace{3cm}
{\Large February 2020\par}
\vspace{2cm}
{\Large Graduate School of Science and Technology\par}
{\Large Keio University\par}
\vspace{2cm}
{\Large Noriyuki Sogabe}
\end{minipage}
\end{center}
\clearpage

\chapter*{Acknowledgements}
\addcontentsline{toc}{chapter}{\numberline{}Acknowledgements}%

First and foremost, I would like to express the most profound appreciation to my supervisor Prof. Naoki Yamamoto. He has taught me the joy of physics and led my life in an exciting direction. His broad knowledge and intuition have strongly influenced my understanding of physics. I am fortunate to have Naoki as my adviser and should thank him for his patience. It has been an honor to be his first Ph.D. student. 

I am deeply grateful to my collaborator, Masaru Hongo, for my works on the interplay between the CME and the dynamic critical phenomena. I would also like to thank many people in the community, whom I met at conferences, workshops, seminars, and summer schools. I am particularly grateful to Profs. Kenji Fukushima and Yoshimasa Hidaka. I would also like to express my gratitude to the referees of my Ph.D. defense, Profs. Youhei Fujitani, Yasuhiro Nishimura, and Kohei Soga. 

I want to thank Aron Beekman, Gergely Fej\H{o}s, Tomoyuki Ishikawa, Kentaro Nishimura, Di-Lun Yang, who I was sharing the office with for numerous discussions and relaxing chat. I am also very thankful to the other (including previous) members of the theoretical physics group at Keio University, particularly Daichi Kagamihara and Hiroaki Wakamura, for their friendship and encouragement.

I appreciate the financial support of research from the Japan Society for the Promotion of Science. When I started writing this thesis during a journey to visit Germany, I was stuck at the Frankfurt airport affected by the typhoon Hagibis. I want to thank my friend, Alexander Max Eller, and his mother, Galvi, to allow me to stay at their house. 

Last but not least, I would like to thank my family for all their love, support, encouragement, and patience. Without them, I could not have completed my Ph.D. thesis. 
\begin{flushright}
{\it Noriyuki Sogabe}
\end{flushright}

\newpage \thispagestyle{empty}

\chapter*{Summary}
\addcontentsline{toc}{chapter}{\numberline{}Summary}%

Understanding the phase structure of Quantum Chromodynamics (QCD) at finite temperature and finite baryon chemical potential is a long-standing problem in the standard model of particle physics. So far, in addition to the nuclear liquid-gas critical point, the possible existence of two critical points is theoretically suggested in the QCD phase diagram: one is the high-temperature critical point between the hadron phase and the quark-gluon plasma phase and the other is the high-density critical point between the nuclear and quark superfluid phases. Since these critical points can be potentially tested in relativistic heavy-ion collision experiments, theoretical predictions for critical phenomena near these critical points are important. On the other hand, heavy-ion collision experiments have another goal to search for the chiral transport phenomena related to the quantum anomaly. One typical example is the chiral magnetic effect, which is the electric current along the magnetic field. In particular, it is known that the chiral magnetic effect leads to the generation of a novel density wave called the chiral magnetic wave.

In this thesis, we first construct the low-energy effective field theory near the high-density QCD
critical point and study its static and dynamic critical phenomena. We find that the critical slowing
down of the speed of the superfluid phonon near the critical point. Furthermore, we show that the
dynamic universality class of the high-density critical point is not only different from that of the
high-temperature critical point, but also a new dynamic universality class beyond the conventional
classification by Hohenberg and Halperin. Since this new universality class stems from the interplay
specific to QCD between the chiral order parameter and the superfluid photon, the observation of
the dynamic critical phenomena in the vicinity of the high-density critical point would provide an
indirect evidence of the superfluidity in high-density QCD matter.

We next consider the second-order chiral phase transition in massless QCD under an external magnetic field and study the interplay between the dynamic critical phenomena and the chiral magnetic effect. For this purpose, we construct the nonlinear Langevin equations including the effects of the quantum anomaly and perform the dynamic renormalization group analysis. As a result, we show that the presence of the chiral magnetic effect and the resulting chiral magnetic wave change the dynamic universality class of the system from the so-called model E into the model A within the conventional classification. We also find that the speed of the chiral magnetic wave tends to vanish when the phase transition is approached. This phenomenon is characterized by the same critical exponents as those for the critical attenuation of the sound wave near the critical points in liquid-gas phase transitions.

\tableofcontents
\newpage \thispagestyle{empty}
\chapter{Introduction}

%Inside a nucleus which consists matter makes up everything, the force bounding the proton and the neutron is called the strong interaction. The strong interaction is a force between the 

%QCD can predicts the states of matter under extreme conditions at extremely hot and/or dense conditions at the orders of a trillion kelvin and/or a trillion kg$/{\rm cm}^3$ 
%The states of matter under extreme conditions at the orders of a trillion kelvin and/or a trillion kg$/{\rm cm}^3$ can be studied by the Quantum Chromodynamics (QCD), which is the fundamental theory of the strong interaction between quarks and gluons. 

Quantum Chromodynamics (QCD) is the fundamental theory of the strong interaction between quarks and gluons. One of the remarkable features of QCD is the asymptotic freedom: the coupling constant of the strong interaction becomes small when the typical energy scale of the system is getting large. When the energy scale is lower than the intrinsic scale of QCD, $\Lambda_{\rm QCD} \sim 200$ MeV, the coupling constant becomes large, and the perturbation theory breaks down. In this low-energy regime, quarks and gluons are confined in color-neutral hadrons, such as baryons made of three quarks and mesons consisting of a quark and an antiquark. Another property of low-energy QCD is the spontaneous breaking of chiral symmetry. The chiral symmetry of QCD in the massless quark limit is broken in the vacuum. In the real world with finite quark masses, the chiral symmetry is an approximate symmetry for the light quarks, especially up and down quarks whose masses are sufficiently smaller than $\Lambda_{\rm QCD}$.

\begin{figure}[t]
\centering\includegraphics[bb=0 0 850 460,height=7cm]{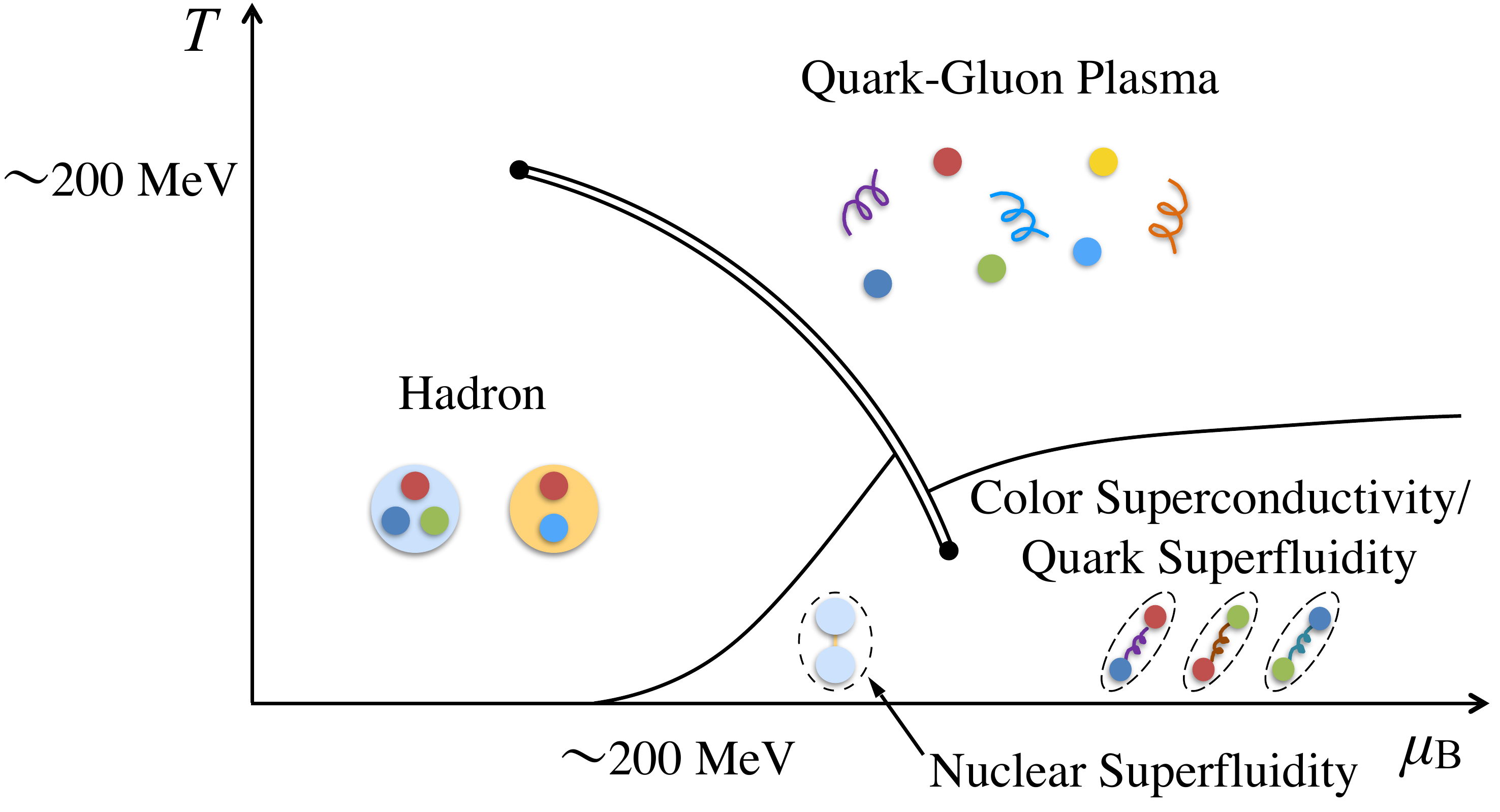}
\caption{Schematic phase diagram of QCD}
\label{fig:phase}
\end{figure}

What happens to matter when it is heated and/or squeezed up to the orders of $\Lambda_{\rm QCD}$? Such matter under extreme conditions at high temperature $T\sim10^{12}$ kelvin and/or baryon chemical potential $\mu_{\rm B}$ (corresponding to mass density $\sim 10^{12}$ ${\rm kg}/{\rm cm}^3$), can also be described by QCD. However, it has been a long-standing problem to understand the phase structure of QCD at finite $T$ and $\mu_{\rm B}$ \cite{Fukushima:2010bq}. Figure~\ref{fig:phase} shows a conjectured phase digram. At low $T$ and low $\mu_{\rm B}$, the chiral symmetry is spontaneously broken in the hadron phase, whereas it is restored at sufficiently high $T$, and the quark-gluon plasma is realized. On the other hand, at high $\mu_{\rm B}$, quarks form Cooper pairs induced by the attractive one-gluon exchange interaction. It follows that the gluons acquire finite masses in an analogy of the gapped photons in ordinary superconductors. This state is called the ``color superconductivity.'' Besides, quark matter and nuclear matter can exhibit the superfluidity as a consequence of the spontaneous breaking of the symmetry associated with the baryon number conservation.

As is also illustrated in Fig.~\ref{fig:phase}, there are two possible QCD critical points where the first-order chiral phase transition line (the doubled line in Fig.~\ref{fig:phase}) terminates. One is the high-temperature critical point between the hadron phase and the quark-gluon plasma phase \cite{Asakawa:1989bq}; the other is the high-density critical point between the nuclear and quark superfluid phases \cite{Hatsuda:2006ps,Yamamoto:2007ah,Abuki:2010jq}. An important task is to understand the dynamic critical phenomena in QCD. In general, dynamic critical phenomena can be classified based on the symmetries and the low-energy gapless modes of systems near a second-order phase transition or a critical point. Such a classification of dynamic critical phenomena is called the dynamic universality class \cite{Hohenberg:1977ym}. In QCD, the high-temperature critical point belongs to the same dynamic universality class as that of the nuclear liquid-gas critical point \cite{Fujii:2003bz,Fujii:2004jt,Son:2004iv,Minami:2011un}, the so-called model H within the conventional classification by Hohenberg and Halperin \cite{Hohenberg:1977ym}. On the other hand, the second-order chiral phase transition in massless two-flavor QCD at finite $T$ and zero $\mu_{\rm B}$ belongs to the same class as that of O(4) antiferromagnets \cite{Rajagopal:1992qz}.

Relativistic heavy-ion collision experiments can potentially test these QCD critical points \cite{Stephanov:2004wx}. In particular, one of the big goals of the Beam Energy Scan (BES) program at the Relativistic Heavy Ion Collider (RHIC) is to search for the high-temperature QCD critical point \cite{BNL}. Moreover, theoretical predictions specific to dense QCD matter would be crucial for the future low-energy heavy-ion collisions at Facility of Antiproton and Ion Research (FAIR), Nuclotron-based Ion Collider Facility (NICA), Japan Proton Accelerator Research Complex (J-PARC), and Heavy Ion Research Facility (HIRF).

On the other hand, the BES program at RHIC has another goal \cite{BNL} to search for the anomalous chiral transport phenomena related to the quantum anomaly \cite
{Adler:1969gk,Bell:1969ts}. One typical example is the chiral magnetic effect (CME), which is the generation of the electric current along the magnetic field \cite{Kharzeev:2007jp,Fukushima:2008xe,Nielsen:1983rb,Vilenkin:1980fu}. In particular, a remarkable consequence of the CME is the creation of the density-wave called the chiral magnetic wave (CMW) \cite{Kharzeev:2010gd,Newman:2005hd}. Although possible signals consistent with the presence of the CME may have been observed in RHIC \cite
{Abelev:2009ac,Abelev:2009ad} and also in Large Hadron Collider (LHC) \cite{Abelev:2012pa}, it is pointed out that there are ambiguities in this interpretation due to possible background effects not related to the CME \cite{Voloshin:2010ut}. %
\footnote{
On the other hand, the CME has already been observed in the table-top experiments of the Weyl/Dirac semimetals (see, e.g., Ref.~\cite{Li:2014bha}). In these systems, relativistic fermions appear as quasiparticles close to the band touching points.
}These ambiguities are currently tested by comparing collisions of the isobaric nuclei such as $ ^{96}_{44}$Ru and $ ^{96}_{40}$Zr, with the same background contributions but different CME signals due to the nuclear-charge difference \cite{Deng:2016knn}. 

In this thesis, we study the novel dynamic critical phenomena induced by the superfluidity and the CME in QCD, {\it respectively}. In each case, we use the low-energy effective theory based on the symmetries and the hydrodynamic variables of QCD.

For the first subject, we construct the low-energy effective theory of the system near the high-density QCD critical point and study its static and dynamic critical phenomena \cite{Sogabe:2016ywr}. In particular, we find the critical slowing down of the speed of the superfluid phonon. Moreover, we find that the dynamic universality class of the high-density critical point is not only different from that of the high-temperature critical but also is a {\it new class  beyond the Hohenberg and Halperin's conventional classification} \cite{Hohenberg:1977ym}. This new universality class stems from the interplay specific to QCD between the chiral order parameter and the superfluid photon. Therefore, observation of the dynamic critical phenomenon in the vicinity of the high-density critical point would provide indirect evidence of superfluidity in the dense-QCD matter.

For the second subject, we study the dynamic critical phenomena of the second-order chiral phase transition in massless QCD under an external magnetic field \cite{Hongo:2018cle}, and clarify the interplay between the dynamic critical phenomena and the CME in QCD. For this purpose, we first construct the nonlinear Langevin equations incorporating the quantum-anomaly effects \cite
{Jackiw,Faddeev:1984jp}, and study it by using the dynamic renormalization group \cite{Martin:1973zz,Janssen:1976,DeDominicis:1978}. As a result, we show that the inclusion of the CME changes the dynamic universality class from the model E into model A within the conventional classification. We also find that the speed of the CMW tends to zero near the phase transition. We here observe the same critical exponents as those of the critical attenuation of the sound wave known in the system near the liquid-gas critical point \cite{Onuki:1997}.

This thesis is organized as follows. In Chap.~\ref{chap:QCD}, we give a brief overview of QCD. In Chap.~\ref{chap:TDCP}, we review the theory of dynamic critical phenomena. Based on the reviews above, in Chap.~\ref{chap:DCPHD}, we study the static and dynamic critical phenomena of the high-density QCD critical point. In Chap.~\ref{chap:DRGCME}, we clarify the interplay between the dynamic critical phenomena and the CME in QCD. Finally, we conclude with Chap.~\ref{chap:Summary}. Among others, Chaps.~\ref{chap:DCPHD} and \ref{chap:DRGCME} are based on our original papers \cite{Sogabe:2016ywr} and \cite{Hongo:2018cle}, respectively. The work \cite{Sogabe:2016ywr} is a collaboration with N. Yamamoto, and the work \cite{Hongo:2018cle} is a collaboration with M. Hongo and N. Yamamoto.

%In Appendix~\ref{chap:staticRG}, we review the static renormalization group of the scalar field theory. In Appendix~\ref{chap:EM}, we show that the effects of the energy and momentum densities do not affect the findings in Chap.~\ref{chap:DCPHD} where we omit them for simplicity. In appendix~\ref{sec:calculation}, we carry out the detailed calculations on the dynamic renormalization group of Chap.~\ref{chap:DRGCME}.

In this thesis, we will work in natural units, where the speed of light, the reduced Plank constant, the elementary charge, and the Boltzmann constant are set equal to unity, $c=\hbar =e =k_{\rm B}=1$.

\chapter{Overview of QCD}
\label{chap:QCD}
We here give an overview of QCD. We start with its basic properties in Sec.~\ref{sec:QCD}. Next, in Sec.~\ref{sec:PS}, we review the  phase structure of QCD. In Sec.~\ref{sec:CME}, we provide a brief explanation of the CME.

\section{Quantum Chromodynamics}
\label{sec:QCD}
In Sec.~\ref{sec:Lagrangian}, we first introduce the Lagrangian of QCD. In section \ref{sec:Asymptotic}, we next explain the asymptotic freedom. Then, in Sec.~\ref{sec:sym}, we summarize the internal symmetries of QCD.

\subsection{Lagrangian}
\label{sec:Lagrangian}
The Lagrangian (density) of QCD is given by
\begin{align}
\label{eq:LQCD}
\mathcal L_{\rm QCD} &= \mathcal L_{\rm quark} + \mathcal L_{\rm gluon},
\end{align}
where
\begin{align}
\label{eq:Lquark}
\mathcal L_{\rm quark} &= \bar q_i (\i \gamma^\mu D_\mu-m_i) q_i,\\
\label{eq:Lglue}
\mathcal L_{\rm gluon} &=-\frac{1}{2} {\rm Tr} G_{\mu\nu} G^{\mu\nu}.
\end{align}
Here, $\mathcal L_{\rm quark}$ is the kinetic term for the quark field $q$. The quark field $q(x)$ has the internal degrees of freedom in addition to the coordinates $x^\mu=(t,\xv)$: the first one is the flavor labeled by $i=1,...,N_{\rm f}$. There are six flavors of quarks in QCD: up (u), down (d), charm (c), strange (s), top (t), bottom (b). Quarks of different flavors have different masses $ m_i $. 
U, c, and t quarks have the electric charge + 2/3, while d, s, and b quarks have the electric charge $-1/3$, in a unit of the elementary charge. The second one is the spinor component running $1,...,4$ with $\gamma^\mu$ being a $4\times4$ matrix in the Dirac space. The antiquark field is defined by $\bar q_i \equiv q_i^\dagger \gamma^0$.  
%Here, the total number of Dirac index $4$ corresponds to 2 degrees of freedom for spin $1/2$, times 2 degrees of freedom for the particle and the anti-particle. 
The last one is the color labeled by $1,...,N_{\rm c}$. The quark field is coupled to the gluon field $A^a_\mu$ $(a=1,...,N_{\rm c}^2-1)$  through the covariant derivative, 
\begin{align}
\label{eq:CD}
D_\mu &\equiv   \d_\mu -\i  g A^\mu,
\end{align}
where $g$ is the interaction strength; $t^a$ is the generators of the color ${\rm SU}(N_{\rm c})$ space with
\begin{align}
\label{eq:talgebra}
[t^a,t^b]= \i f_{abc} t^c,\quad {\rm Tr}(t^a t^b)=\frac{\delta^{ab}}{2}.
\end{align}
Here, $f_{abc}$ is called the structure constants.
% determined by using the symmetry generators, e.g., when $N_{\rm c}=3$, $f_{abc}$ can be obtained by calculating the commutation relations among the so-called Gell-Mann matrices.

Let us see that $\mathcal L_{\rm gluon}$ includes the kinetic term for the gluon fields and their self-interactions. In Eq.~(\ref{eq:Lglue}), $G^a_{\mu\nu}$ is the field strength defined in the following form:
\begin{align}
\label{eq:Gmunu}
G_{\mu\nu} &\equiv \frac{\i}{g}[D_\mu,D_\nu].
\end{align}
By using Eq.~(\ref{eq:CD}), one can write Eq.~(\ref{eq:Gmunu}) into
\begin{align}
G_{\mu\nu}
%&= \frac{\i}{g} \left[ (\d_\mu -\i  g t^a A^a_\mu)( \d_\nu -\i  g t^b A^b_\nu  )-( \d_\nu -\i  g t^b A^b_\nu  ) (\d_\mu -\i  g t^a A^a_\mu)\right]\notag\\
%&= \frac{\i}{g} \left[ \d_\mu\d_\nu -\i  g t^b (\d_\mu A^b_\nu)  -\i  g t^b A^b_\nu \d_\mu    -\i  g t^a A^a_\mu \d_\nu  -g^2 t^a t^b A^a_\mu A^b_\nu\right.\notag\\ 
%&\qquad\quad  \left.  - \d_\nu\d_\mu  +\i g t^a (\d_\nu A^a_\mu) +\i g t^a  A^a_\mu\d_\nu  +\i  g t^b A^b_\nu  \d_\mu  +  g^2  t^b t^a A^b_\nu A^a_\mu)\right]\notag\\
%&= (\d_\mu A^a_\nu - \d_\nu A^a_\mu) t^a - \i g  [t^a,t^b] A^a_\mu A^b_\nu \notag\\
&=\left( \d_\mu A^a_\nu - \d_\nu A^a_\mu  + g f_{abc}  A^b_\mu A^c_\nu\right)t^a.
\end{align}
From this expression, one can also rewrite $\mathcal L_{\rm gluon}$ into
\begin{align}
\mathcal L_{\rm gluon} 
%&=-\left( \d_\mu A^a_\nu - \d_\nu A^a_\mu  + g f_{abc}  A^b_\mu A^c_\nu\right) \left( \d^\mu A_d^\nu - \d^\nu A_d ^ \mu  + g f_{def}  A_e^\mu A_f^\nu\right)\frac{1}{2} {\rm Tr} ( t^a t^d)\notag\\
% &=-\frac{1}{4}\left( \d_\mu A^a_\nu - \d_\nu A^a_\mu  + g f_{abc}  A^b_\mu A^c_\nu\right) \left( \d^\mu A_a^\nu - \d^\nu A_a ^ \mu  + g f_{aef}  A_e^\mu A_f^\nu\right)\notag\\
%&=-\frac{1}{4}\left( \d_\mu A^a_\nu - \d_\nu A^a_\mu \right) \left( \d^\mu A_a^\nu - \d^\nu A_a ^ \mu  \right)  -\frac{g}{4} f_{abc}  \left( \d_\mu A^a_\nu - \d_\nu A^a_\mu \right)  A_b^\mu A_c^\nu\notag\\
%&\quad-\frac{g}{4}  f_{abc}   \left( \d^\mu A_a^\nu - \d^\nu A_a ^ \mu  \right) A^b_\mu A^c_\nu -\frac{g^2 }{4} f_{abc}   f_{ade}  A^b_\mu A^c_\nu A_d^\mu A_e^\nu\notag\\
&=-\frac{1}{4}\left( \d_\mu A^a_\nu - \d_\nu A^a_\mu \right)^2 -g  f_{abc}   (\d^\mu A_a^\nu)  A^b_\mu A^c_\nu -\frac{g^2 }{4} f_{abc}   f_{ade}  A^b_\mu A^c_\nu A_d^\mu A_e^\nu.
\end{align}
Here, the first term is the kinetic term for the gluon fields; the second and the third terms are 3- and 4-points interaction terms among the gluons.

\subsection{Asymptotic freedom}
\label{sec:Asymptotic}
One of the remarkable features of QCD is the asymptotic freedom \cite{Politzer:1973fx,Gross:1973id}: the coupling constant $g$ becomes smaller when the typical energy scale is getting large. 
%In other words, the quarks and gluons can be regarded as asymptotically free particles when these typical length scales are  short. This explains the fact that the role of the strong interaction in the deep inelastic scattering between the proton and the high-energy injected electron.
Generally, in field theories, when one takes into account perturbative loop-corrections, there are some divergences in the calculations. To remove those divergences and obtain some physically meaningful results, the parameters in the Lagrangian should be no more constants but depend on the typical energy scale at which such divergences are removed by renormalizing the parameters. This is also the case for the interaction parameter $g$ introduced in Eq.~(\ref{eq:CD}) in QCD. It is known that the interaction strength $g$ satisfies the following renormalization group equation to its leading order:
\begin{align}
\label{eq:RGQCD}
\mu \frac{\d g}{\d \mu} = -\frac{b_0}{(4\pi)^2}g^3 +\mathcal O(g^5),
\end{align}
with $\mu$ being the renormalization scale and $
b_0\equiv  \frac{11}{3} N_{\rm c}-\frac{2}{3}N_{\rm f}$. Since Eq.~(\ref{eq:RGQCD}) can be solved by the separation of variables method at this order, by defining $\alpha_s\equiv g^2/(4\pi)$, one finds the solution as
%\begin{align}
%\alpha_s (\mu) = \frac{\alpha_s}{ 1+\dis  \frac{b_0 \alpha_s}{2\pi} \log (\mu/\mu_0)},
%\end{align}
%where we defined 
%\begin{align}
%\alpha_s\equiv \frac{g^2}{4\pi}.
%\end{align}
%By defining
%\begin{align}
%1=  \frac{b_0 \alpha_s}{2\pi} \log (\Lambda_{\rm QCD}/\mu_0 ),
%\end{align}
%one can write it into 
\begin{align}
\alpha_s (\mu) = \frac{2\pi}{b_0 \log (\mu/\Lambda_{\rm QCD})},
\end{align}
where $\Lambda_{\rm QCD}\approx 200\ {\rm MeV}$ is the typical energy scale of QCD, and several experiments determines its value \cite{Tanabashi:2018oca}. This solution shows that the interaction strength $\alpha_s(\mu)$ runs towards a smaller value as $\propto 1/\log(\mu)$ when the typical energy scale of the system $\mu$ increases. %This asymptotic-freedom behavior matches the observation in the high-energy collisions between the electron and the proton in the 1960's, and QCD becomes to be believed as the fundamental theory for the strong interaction. 

\subsection{Symmetries}
\label{sec:sym}
The internal symmetries of QCD can be summarized as 
\begin{align}
\label{eq:Gsym}
\mG = {\rm SU}(N_{\rm c})_{\rm C} \times {\rm SU}(N_{\rm f})_{\rm L} \times{\rm SU}(N_{\rm f})_{\rm R} \times {\rm U}(1)_{\rm B},
\end{align}
where each symmetry group is defined as follows. Note here that the $\U(1)$ axial symmetry is not included in $\mG$ from the reason given at the end of this subsection.

\subsubsection{Color gauge symmetry}
The color gauge symmetry ${\rm SU}(N_{\rm c})_{\rm C}$ is the invariance under the following gauge transformation in color ${\rm SU}(N_{\rm c})$ space:
\begin{align}
\label{eq:color trans}
q\rightarrow V_{\rm C} q,\quad A_\mu \ra V_{\rm C} A_\mu V_{\rm C}^\dagger +\frac{\i}{g} (\d_\mu V_{\rm C}) V_{\rm C}^\dagger, \quad V_{\rm C} \equiv {\rm e} ^{-\i \theta^a (x) t^a}.
\end{align}
Here $\theta^a(x)$ is a local transformation parameter that depends on the coordinates (for the definition of $t^a$, see Eq.~(\ref{eq:talgebra})).  

To see the invariance of QCD Lagrangian under Eq.~(\ref{eq:color trans}), we first calculate the transformation laws for $ D_\mu q$ and the field strength $G_{\mu\nu}$,
\begin{align}
\label{eq:Dtranscolor}
D_\mu q  &\ra  V_{\rm C} (D_\mu q),\\
\label{eq:Gtranscolor}
G_{\mu\nu} & \rightarrow V_{\rm C} G_{\mu\nu} V_{\rm C}^\dagger.
\end{align}
Here, one can derive the second transformation by using the first one. (See Eqs.~(\ref{eq:CD}) and (\ref{eq:Gmunu}) for the definitions of the covariant derivative $D_{\mu}$ and $G_{\mu\nu}$, respectively.) From Eqs.~(\ref{eq:Dtranscolor}) and (\ref{eq:Gtranscolor}), one can confirm that $\mathcal L_{\rm QCD}$ is invariant under the color gauge transformation~(\ref{eq:color trans}). 

\subsubsection{Chiral symmetry}
The chiral symmetry ${\rm SU}(N_{\rm f})_{\rm L} \times{\rm SU}(N_{\rm f})_{\rm R}$ is the invariance under the independent rotation of right-handed and left-handed components of the quark field in the flavor space. Here, the right-handed quark $q_{\rm R}$ and the left-handed quark $q_{\rm L}$ are defined by
\begin{align}
\label{eq:chiral rep}
q_{\rm L} \equiv \frac{1-\gamma_5}{2} q,\quad 
q_{\rm R} \equiv \frac{1+\gamma_5}{2} q.
\end{align}
These transformations under the chiral transformation are
\begin{align}
\label{eq:chiral}
q_{\rm L}\rightarrow V_{\rm L} q_{\rm L},\quad 
q_{\rm L}\rightarrow V_{\rm R} q_{\rm R},\quad V_{\rm L,R}\equiv \e^{-\i \theta^a_{\rm L,R} \lambda^a}.
\end{align}
Here, $\lambda^a$ being the generator of ${\rm SU}(N_{\rm f})_{\rm L,R}$.
% which satisfies the same algebra of $t^a$ given in Eq.~(\ref{eq:talgebra}). 
Note that the transformation parameters $\theta^a_{\rm L,R}$ do not depend on $x^\mu$ (generally such a transformation is called a global symmetry), unlike $\theta^a(x)$ introduced in Eq.~(\ref{eq:color trans}) for the color gauge symmetry. 

Let us see the chiral transformation~(\ref{eq:chiral}) of the QCD Lagrangian~(\ref{eq:LQCD}). We first decompose the quark fields into
\begin{align}
q = q_{\rm R} + q_{\rm L},
\end{align}
and rewrite the quark sector of the QCD Lagrangian~(\ref{eq:Lquark}) into
\begin{align}
\label{eq:masstermRL}
\mathcal L_{\rm quark} = \bar q_{\rm L} \i \gamma^\mu D_\mu q_{\rm L} +  \bar q_{\rm R} \i \gamma^\mu D_\mu q_{\rm R} - \bar q_{\rm L} \hat m q_{\rm R} - \bar q_{\rm R} \hat m q_{\rm L}.
\end{align}
Here, $\hat m$ denotes the matrix in the flavor space, and its components are given by $m_i $. In Eq.~(\ref{eq:masstermRL}), the first two terms are invariant under the chiral transformation, whereas the last two mass terms are not invariant except for the particular {\it vector}-like choice of the transformation parameters $\lambda^a_{\rm L}=\lambda^a_{\rm R}$.\footnote{{\it Vector} transformations are in-phase between left- and right-handed quarks. Meanwhile, {\it Axial} transformations are  opposite-phase rotation with different chirality. } 
Therefore, the chiral symmetry is exact only when one sets the quark masses $m_{\rm i}$ to zero (this massless limit is called the chiral limit). Practically, the mass of u quark, $m_{ \rm u} \approx 3\ {\rm MeV}$ and that of d quark, $m_{\rm d}  \approx  5\ {\rm MeV}$ are small compared to $\Lambda_{\rm QCD}$ introduced in Sec.~\ref{sec:Asymptotic}. In general, the chiral symmetry is an approximate symmetry of QCD, when $m_ i$ is small compared to any typical energy scales of the system.

\subsubsection{Baryon number symmetry}
The baryon number symmetry ${\rm U}(1)_{\rm B}$ is the invariance under
\begin{align}
\label{eq:baryon}
q\rightarrow {\rm e} ^{-\i \alpha_{\rm B} } q,
\end{align}
where $\alpha_{\rm B}$ is the global transformation parameter. The QCD Lagrangian is invariant under this transformation. 

\subsubsection{U(1) axial symmetry and its quantum anomaly}
\label{sec:U(1)A}
In addition to the symmetry group $\mG$, the {\it classical Lagrangian} of QCD has the approximate axial symmetry ${\rm U}(1)_{\rm A}$, which is the invariance under 
\begin{align}
q\rightarrow {\rm e} ^{-\i \alpha_{\rm A} \gamma _5} q,
\end{align}
where $\alpha_{\rm A}$ is the transformation parameter. This transformation is equivalent to
\begin{align}
\label{eq:axial chiral form}
q_{\rm L}\rightarrow {\rm e} ^{\i \alpha_{\rm A}}q_{\rm L},\quad
q_{\rm R}\rightarrow {\rm e} ^{-\i \alpha_{\rm A}}q_{\rm R}.
\end{align}
Similarly to the chiral symmetry, ${\rm U}(1)_{\rm A}$  can be regarded as an approximate symmetry of QCD at the {\it classical Lagrangian level} in the presence of small quark masses $ m_i $. However, it is known that the quantum effects break the ${\rm U}(1)$ axial symmetry \cite{tHooft:1976rip,Kobayashi:1970ji}, 
\begin{align}
\label{eq:QCDanomaly}
\U(1)_{\rm A}\stackrel{\rm anomaly}{\ra} \mathbb{Z}(2N_{\rm f})_{\rm A}.
\end{align}
Here, $\mathbb{Z}(2N_{\rm f})_{\rm A}$ corresponds to the symmetry under the parameter choices of the parameter $\alpha_{\rm A}=n \pi/N_{\rm f}$ $(n=1,...,2N_{\rm f})$.\footnote{Unlike the chiral anomaly in the background electromagnetic field in Eq.~(\ref{eq:Chiral Anomaly}), the divergence of the $\U(1)_{\rm A}$ current in the presence of the gluons, discussed here, is related to the topological charge called the instanton number of the non-Abelian gauge field. This {\it topological nature} leads to the invariance for some particular {\it discrete} choices of the parameters for $\U(1)_{\rm A}$ transformation, $\mathbb{Z}(2N_{\rm f})_{\rm A}$. See, e.g., Refs.~\cite{tHooft:1976rip,Nair:2005iw} for the details.} Because of the presence of this quantum anomaly effect, we distinguish ${\rm U}(1)_{\rm A}$ from the symmetry group $\mG$. We will discuss the residual part $\mathbb{Z}(2N_{\rm f})_{\rm A}$  in the context of the continuity between the nuclear/quark superfluid phases at the end of Sec.~\ref{sec:SBP}. 

\section{Phase structure}
\label{sec:PS}
We introduce the order parameters of QCD and discuss those symmetry breaking patterns in Sec.~\ref{sec:OPQCD}. Section~\ref{eq:BCSquark} is devoted to explaining why the diquark condensation is favored at high-density QCD. 
In Sec.~\ref{sec:SBP}, we classify the phases of Fig.~\ref{fig:phase} by using the order parameters.

\subsection{Order parameters}
\label{sec:OPQCD}
The symmetry group of QCD, $\mG$ given in Eq.~(\ref{eq:Gsym}) can be spontaneously broken in the ground states. This spontaneous symmetry breaking can be characterized by the two representative order parameters in QCD defined as follows.

 Note that in this subsection, we explicitly write the flavor and the color indices:
\begin{align}
i,j,k,...&={\rm u,d,s},...\quad ({\rm flavor});\\
A,B,C,...&={\rm r,g,b},...\quad ({\rm color}),
\end{align}
where color indices range over red, green, and blue,... (r,g,b,...) in general. The general transformation laws under all of the symmetries introduced in Sec.~\ref{sec:sym}, $\mG \times {\rm U}(1)_{\rm A}$ can be written as follows:
\begin{align}
\label{eq:GqL}
(q_{\rm L})^i_A &\rightarrow \e^{\i \alpha_{\rm A}} \e^{-\i \alpha_{\rm B}} (V_{\rm L})_{ij} (V_{\rm C})_{AB} (q_{\rm L})^j_B,\\
\label{eq:GqR}
(q_{\rm R})^i_A &\rightarrow \e^{-\i \alpha_{\rm A}} \e^{-\i \alpha_{\rm B}} (V_{\rm R})_{ij} (V_{\rm C})_{AB} (q_{\rm R})^j_B.
\end{align}
%(For the definition of the left-handed and right-quark fields $q_{\rm R,L}$, see Eq.~(\ref{eq:chiral rep}))

\subsubsection{Chiral order parameter}
One of the important order parameters in QCD is the chiral order parameter defined by
\begin{align}
\label{eq:chiralOP}
\Phi_{ij}\equiv \left< (\bar q_{\rm R})^j_A (q_{\rm L})_A^i  \right>,
\end{align}
which is the matrix in the flavor space. By using Eqs.~(\ref{eq:GqL}) and (\ref{eq:GqR}), one finds that this order parameter is transformed under  $\mG \times {\rm U}(1)_{\rm A}$ as 
\begin{align}
\label{eq:Phitra}
\Phi_{ij}\rightarrow \e^{-2 \i \alpha_{\rm A}} (V_{\rm L})_{ik} \Phi_{kl} (V_{\rm R}^\dagger)_{lj}. 
\end{align}
We note that the determinant of $\Phi$ characterizes the quantum anomaly \cite{Kobayashi:1970ji}. The transformation law under (\ref{eq:Phitra}):
\begin{align}
\label{eq:detPhi}
\det \Phi \rightarrow \e^{-2 \i \alpha_{\rm A}}  \det \Phi,
\end{align}
follows that $\det \Phi$ preserves the chiral symmetry $\SU(N_{\rm f})_{\rm L} \times \SU(N_{\rm f})_{\rm R}$ but breaks $\U(1)_{\rm A}$. Here, we use Eq.~(\ref{eq:Phitra}) and $\det V_{\rm L}=\det V_{\rm R}=1$ to derive Eq.~(\ref{eq:detPhi}).

The chiral order parameter breaks the chiral symmetry into the subgroup:
\begin{align}
{\rm SU}(N_{\rm f})_{\rm L} \times{\rm SU}(N_{\rm f})_{\rm R}\ra {\rm SU}(3)_{\rm L+R},
\end{align}
where ${\rm SU}(N_{\rm f})_{\rm L+R}$ is the invariance under the {\it vector}-like choice of the parameters, $V_{\rm L} =V_{\rm R}$. Therefore, the chiral order parameter is an order parameter characterizing the spontaneous breaking of the chiral symmetry. Note here that ${\rm U}(1)_{\rm A}$ is not spontaneously broken, because  the quantum effects have already broken this symmetry as we explained in Sec.~\ref{sec:sym}.

\subsubsection{Diquark condensate}
The second order parameter in QCD is the diquark condensate. Each of the right-hand and left-handed quark fields is defined by
\begin{align}
\label{eq:dL}
(d_{\rm L}^\dagger)_{A i} \equiv \eps_{ABC} \eps_{ijk} \left<  (q_{\rm L})^j_B \mathcal O_{\rm C} (q_{\rm L})^k_C \right>,\\
\label{eq:dR}
(d_{\rm R}^\dagger)_{A i} \equiv \eps_{ABC} \eps_{ijk} \left<  (q_{\rm R})^j_B \mathcal O_{\rm C} (q_{\rm R})^k_C \right>.
\end{align}
Here, two of the quark fields in each of $d_{\rm R,L}^\dagger$ are chosen to be antisymmetric under the exchanges of the color and flavor, respectively (and also the spin which is not explicit in these expressions). The labels of quarks in the right-hand sides will be confirmed in Sec.~\ref{eq:BCSquark}. Besides, $\mathcal O_{\rm C}$ denotes the charge conjugation operator
\begin{align}
\mathcal O_{\rm C}\equiv \i \gamma^2 \gamma^0,
\end{align}
which is also necessary for the diquark condensate being a Lorentz scalar. By using Eqs.~(\ref{eq:GqL}) and (\ref{eq:GqR}),  the diquark condensates are transformed by $\mG \times {\rm U}(1)_{\rm A}$ as follows:
\begin{align}
\label{eq:dtrans}
(d_{\rm L})_{A i} &\ra \e^{2\i \alpha_{\rm A}} \e^{-2\i \alpha_{\rm B}} (V_{\rm L})_{ij} (V_{\rm C})_{AB} (d_{\rm L})_{B j},\\
(d_{\rm R})_{A i} &\ra \e^{-2\i \alpha_{\rm A}} \e^{-2\i \alpha_{\rm B}} (V_{\rm R})_{ij} (V_{\rm C})_{AB} (d_{\rm L})_{B j}.
\end{align}

Let us consider the realistic colors and focus our attention on the light quarks, and set $N_{\rm C}=N_{\rm f}=3$. The diquark condensate breaks the color gauge symmetry and the chiral symmetry into their {\it mixed} subgroup:
\begin{align}
\label{eq:CFL}
{\rm SU}(3)_{\rm C} \times {\rm SU}(3)_{\rm L} \times{\rm SU}(3)_{\rm R}\ra {\rm SU}(3)_{\rm C+L+R}.
\end{align}
Here, ${\rm SU}(3)_{\rm C+L+R}$ symmetry is the invariance under the particular combination of the gauge transformation and the chiral transformation, $V_{\rm C}^\dagger =V_{\rm L} =V_{\rm R}$. One can confirm this invariance by using the transformation law of the diquark condensate (\ref{eq:dtrans}). The resulting phase is invariant under {\it simultaneous} color and flavor transformations. For this reason, it is called the color flavor locked phase \cite{Alford:1998mk}. From Eq.~(\ref{eq:CFL}), the diquark condensate is an order parameter characterizing the color superconductivity, which is the ``spontaneous breaking'' of the color gauge symmetry. %The color superconductivity is an analogy to the normal superconductivity of metals, i.e., the spontaneously symmetry breaking of the $\U(1)_{\rm em}$ gauge symmetry. 
In addition to this color gauge symmetry breaking, the diquark condensate also breaks the baryon symmetry,
\begin{align}
\label{eq:baryonSSB}
{\rm U}(1)_{\rm B} \rightarrow \Z(2)_{\rm B},
\end{align}
where $\Z(2)_{\rm B}$ is the discrete symmetry, which corresponds to the choices of the parameter, $\alpha_{\rm B}=0,\pi$ in Eq.~(\ref{eq:baryon}). %In fact, $\phi$ changes its value under $\U(1)_{\rm B}$ except for  $\Z(2)_{\rm B}$. 
Therefore, the diquark condensate is also an order parameter for the superfluidity characterized by the spontaneously breaking of the ${\rm U}(1)_{\rm B}$ symmetry.

\subsection{Formation of diquark condensate in high-density QCD}
\label{eq:BCSquark}
In the high-density QCD ($\mu_{\rm B} \gg \Lambda_{\rm QCD},T$), quarks form a diquark condensate. In this regime, the degenerate quarks near the Fermi surface can be described by the weakly coupled QCD due to the asymptotic freedom. The interaction between two quarks is dominated by the one-gluon exchange which is proportional to a product of the $\SU(3)$ interaction vertices,
\begin{align}
\label{eq:Filtz}
(t^a)_{AB} (t^a)_{CD} = \frac{1}{6} (\delta_{AB}\delta_{CD}+\delta_{AC}\delta_{BD}) -\frac{1}{3} (\delta_{AB}\delta_{CD}-\delta_{AC}\delta_{BD}).
\end{align} 
Here, the first/last two terms represent the repulsive/attractive interaction between the quarks whose color labels are symmetric/antisymmetric  under the exchange of $A$ and $C$ or $B$ and $D$, respectively. Degenerate fermionic systems with a 2-body (4-point) attractive interaction have the instabilities towards the formation of the Cooper pairs \cite{Cooper:1956zz,Bardeen:1957kj}. According to this BCS mechanism, quarks form Cooper pairs in the attractive channels, and the finite diquark condensate is realized as the ground state of high-density QCD \cite{Bailin:1983bm,Iwasaki:1994ij,Alford:1997zt,Rapp:1997zu}.

%The anti-symmetric combination of the spins is energetically favorable. This is because an isotropy of the condensate allows a better use of the Fermi surface. Since the condensation is anti-symmetric in color indices, Pauli principle requires the anti-symmetric flavor induces. 

%The antisymmetric structure of spins is energetically favorable in general. This is because the isotropy of the condensate allows better use of the Fermi surface. Since the color indices are antisymmetric, the flavor should have the same structure due to the Pauli principle. 

The color, flavor, and spin structure of the diquark condensate are determined as follows. The color indices should be antisymmetric to have an attractive interaction. The spin indices should also be antisymmetric so that the total spin of the diquark condensate is zero. This isotropic combination is energetically favorable in general because it allows efficient use of the Fermi surface. Finally, from the Pauli principle, the flavor indices should also be antisymmetric. 

\subsection{Symmetry breaking patterns}
\label{sec:SBP}

Here, we will look at the phases classified by the chiral order parameter $\Phi_{ij}$ and the diquark condensate $d_{\rm R,L}$ \cite{Hatsuda:2006ps,Yamamoto:2007ah}. Let us consider $N_{\rm C} =N_{\rm f}=3$ and the most symmetric ground state:
\begin{align}
\label{eq:chiral condensate 3flavor}
\Phi_{ij} &= \sigma\delta_{ij},\\
\label{eq:diquark 3flavor}
(d_{\rm L})_{A i} &= -(d_{\rm R})_{A i} =\delta_{A i} \phi .
\end{align}
Here, $\sigma$ is called the chiral condensate.

%In Eq.~(\ref{eq:diquark 3flavor}), particular combinations between color and flavor labels are finite, e.g., red and up. This phenomenon is called the color flavor locking \cite{Alford:1998mk}. 

\begin{table}[t]
\centering\includegraphics[bb=0 0 624 198,height=4.3cm]{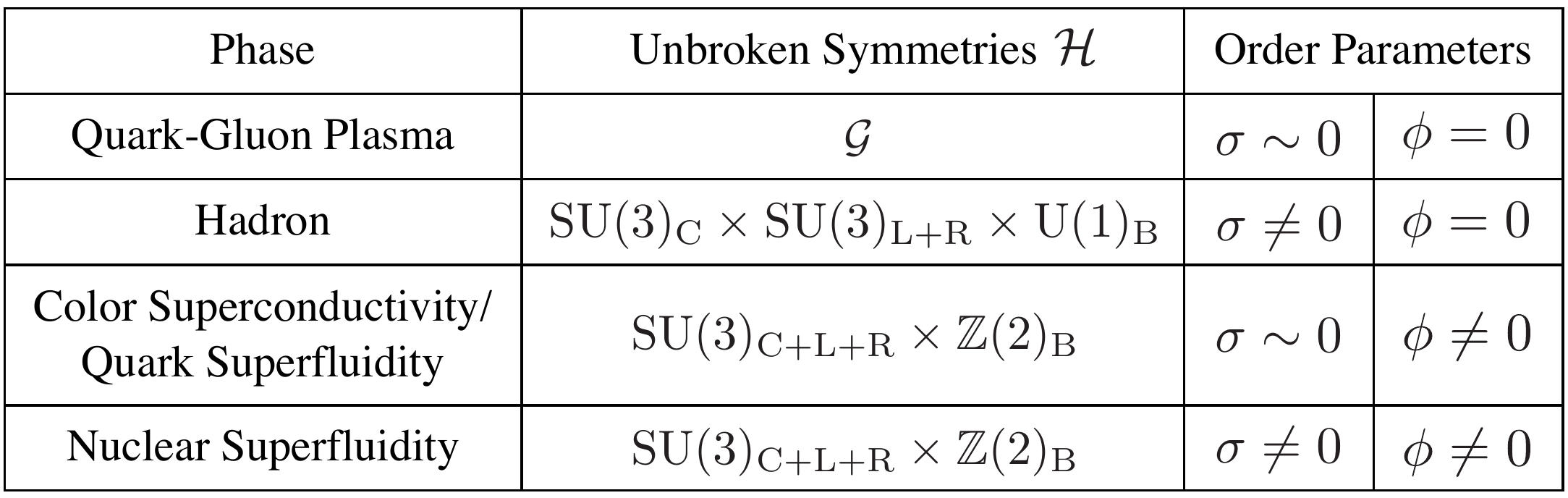}
%\vspace{4.3cm}
\caption{Symmetry breaking patterns of QCD phase diagram \cite{Hatsuda:2006ps}}
\label{tab:sbp}
\end{table}

In principle, there are four possible phases characterized by $\sigma$ and $\phi$, as summarized in Table \ref{tab:sbp}. Here, the symmetry breaking patterns of the individual phases are also listed ($\mathcal H$ denotes the unbroken symmetry of $\mathcal G$). The chiral condensate $\sigma$ classifies the quark-gluon plasma phase and the hadron phase. It is approximately zero (of an order of the quark mass) in the quark-gluon plasma phase and is finite in the hadron phase. On the other hand, any symmetries cannot distinguish the two superfluid phases, so that these two phases can be continuously connected.  This conjecture is called the hadron-quark continuity \cite{Schafer:1998ef}. %Nevertheless, we relate the nuclear and quark superfluid phases by the regions $\sigma \gtrsim \phi \neq$ and $\phi \gtrsim \sigma \neq0$, respectively.

Let us see the crossover of the superfluid phases on the aspects of the chiral symmetry and the $\U(1)_{\rm A}$ symmetry following the argument in Ref.~\cite{Hatsuda:2006ps}. We consider the order parameter, $(d_{\rm L})_{A i} (d_{\rm R}^\dagger)_{A j}=-\phi^2 \delta_{ij}$, which breaks these symmetries,
\begin{align}
(d_{\rm L})_{A i} (d_{\rm R}^\dagger)_{A j} \ra \e^{4\i \alpha_{\rm A}} (V_{\rm L})_{ik} (d_{\rm L})_{A k} (d_{\rm R}^\dagger)_{A l}  (V_{\rm R})_{lj}.
\end{align}
The quark superfluid phase with finite $\phi^2\neq0$ breaks the chiral symmetry but preserves $\Z(4)_{\rm A}$ ($\alpha_{\rm A} = 0,\pi/2,\pi,3\pi/2 $) of $\U(1)_{\rm A}$ as a subgroup. On the other hand, in the nuclear superfluid phase, $\sigma\neq 0$ and $\phi^2\neq 0$ break the chiral symmetry but preserves $\Z(2)_{\rm A}$ ($\alpha_{\rm A} = 0,\pi$). At a glance, the residual discrete $\U(1)_{\rm A}$ symmetries look different. However, the quantum anomaly has already broken $\U(1)_{\rm A}$ into $\Z(6)_{\rm A}$, which includes $\Z(2)_{\rm A}$ but not $\Z(4)_{\rm A}$. The residual $\U(1)_{\rm A}$ symmetry is the same $\Z(2)_{\rm A}$ in both phases, in the presence of the quantum anomaly. Not only the chiral symmetry but also $\U(1)_{\rm A}$ symmetry cannot classify the nuclear/quark superfluid phases.

%Let us compare the remaining symmetries of $\U(1)_{\rm A}$ with different ground states characterized by $\sigma$ and $(d_{\rm L})_{A i} (d_{\rm R}^\dagger)_{A j}=-\phi^2 \delta_{ij}$. Here, we chose the latter order parameter rather than $\phi$ because of the gauge invariance. From Eqs.~(\ref{eq:Phitra}) and 
%\begin{align}(d_{\rm L})_{A i} (d_{\rm R}^\dagger)_{A j} \ra \e^{4\i \alpha_{\rm A}} (V_{\rm L})_{ik} (d_{\rm L})_{A k} (d_{\rm R}^\dagger)_{A l}  (V_{\rm R})_{lj},\end{align}
%one can see that $\sigma$ and $\phi^2$ breaks the $\U(1)_{\rm A}$ into $\Z(2)_{\rm A}$ ($\alpha_{\rm A} = 0,\pi$) and $\Z(4)_{\rm A}$ ($\alpha_{\rm A} = 0,\pi/2,\pi,3\pi/2 $), respectively. At a glance, these two remaining symmetries are different.  Nevertheless, the presence of the quantum anomaly leads to $\Z(2)_{\rm A}$ in each case. In order to understand this, we remind that $\U(1)_{\rm A}$ has been already broken by the quantum anomaly as Eq.~(\ref{eq:QCDanomaly}). Therefore,  when we consider $\phi^2$ as an order parameter, we should include this quantum-anomaly effect and consider the common part between $\Z(4)_{\rm A}$ and $\Z(6)_{\rm A}$ at $N_{\rm f}=3$, which is nothing but $\Z(2)_{\rm A}$. From the above argument, the ground states in the nuclear/quark superfluid phases cannot be classified by the symmetry breaking pattern of $\U(1)_{\rm A}$ as well as the chiral symmetry \cite{Hatsuda:2006ps}. 

\section{Chiral magnetic effect}
\label{sec:CME}

So far, we have discussed the static properties of QCD. In this section, we discuss its dynamics. Chiral transport phenomena are the novel macroscopic transport phenomena induced by the chirality of quarks in QCD. As an example, we here discuss the so-called chiral magnetic effect (CME). In Sec.~\ref{sec:symCME}, we first demonstrate that the CME is a phenomenon specific to the presence of the chirality imbalance from the symmetry arguments. Next in Sec.~\ref{sec:dercationCME}, we derive the CME for noninteracting quarks under an external magnetic field. Finally, in Sec.~\ref{eq:CMWderivation}, we discuss the relevance of the CME to the low-energy hydrodynamic modes. In this section, we consider the chiral limit and set $ m_i =0$. 

\subsection{Symmetry argument}
\label{sec:symCME}
We here discuss the symmetry aspects of the CME. Let us write the electric current $\jv$ by using the electric field $\Ev$ and the magnetic field $\Bv$. From the rotational symmetry of the system, we obtain
\begin{align}
\label{eq:jvexpand}
\jv = \sigma_{\rm E} \Ev + \sigma_{\rm B} \Bv,
\end{align}
where $\sigma_{\rm E}$ and $\sigma_{\rm B}$ are some constants. The first term is nothing but the Ohmic law, which is allowed in the usual matter without chirality imbalance. On the other hand, the second term is allowed when $\sigma_{\rm B}$ is proportional to the chirality imbalance \begin{align}
\mu_5 =\frac{\mu_{\rm R}-\mu_{\rm L}}{2}.
\end{align}
The parity transformation leads to
\begin{align}
\jv\ra -\jv,\quad \Ev \ra -\Ev,\quad \Bv \ra \Bv,
\end{align}
and 
\begin{align}
\mu_5 \ra -\mu_5.
\end{align}
One can understand the last transformation by using the fact that the parity transformation exchanges the chirality,
\begin{align}
\mu_{\rm R} \ra \mu_{\rm L},\quad \mu_{\rm L}\ra \mu_{\rm R}.
\end{align}
As we will see in Sec.~\ref{sec:dercationCME} (in particular in Eq.~(\ref{eq:CMEgeneral})), the CME is the generation of such second term with \cite{Kharzeev:2007jp,Fukushima:2008xe,Nielsen:1983rb,Vilenkin:1980fu}
\begin{align}
\sigma_{\rm B} = \frac{\mu_5}{2\pi^2}.
\end{align}

\subsection{Derivation}
\label{sec:dercationCME}
We here first derive the chiral anomaly in the presence of the background electromagnetic fields and the CME. In this subsection, we first neglect the gluons and discuss these effects later. The discussion here follows the original argument in Ref.~\cite{Nielsen:1983rb} and also the recent lecture notes \cite{Landsteiner:2016led,Hidaka}. 

\subsubsection{Landau Level}
Let us consider the Weyl equation for the left-handed quark $ q _{\rm L}$,
%\footnote{As is clear from the derivation itself, the following discussion is not only the case for the quark field but any relativistic fermions.}
\begin{align}
\label{eq:WeylL}
\i \bar \sigma^\mu D_\mu  q _{\rm L} =0,
\end{align} 
where $\bar \sigma^\mu=(1,-\sigma^i)$, $D_\mu = \d_\mu - \i A_\mu$, and $A^\mu$ is the $\U(1)_{\rm em}$ gauge field. The Weyl equation can be written as
\begin{align}
\label{eq:Weyl 2}
\i(\d_t -\sigma^i D_i)   q _{\rm L}=0.
\end{align}
We set the magnetic field in the $z$ direction, $\Bv=(0,0,B)$, and take the particular 
\begin{align}
(A_x,A_y,A_z)=(0,Bx,0).
\end{align}
We write the 2-component field $ q _{\rm L}$ as 
\begin{align}
\label{eq:qLform}
 q _{\rm L} = \e^{-\i(\omega t-k_y y-k_z z) }
\left(
\begin{array}{c}
 q _-\\
 q _+
\end{array}
\right),
\end{align}
with $\omega$ and $k_i$ ($i=y,z$) being the energy and the momentum of the quark, respectively. Then, by substituting Eq.~(\ref{eq:qLform}) into Eq.~(\ref{eq:Weyl 2}), we find 
\begin{align}
\label{eq:LL}
\left(
\begin{array}{cc}
\omega+k_z & -\i (\d_x -Bx +k_y )\\
-\i (\d_x +Bx -k_y )&\omega -k_z
\end{array}
\right)
\left(
\begin{array}{c}
 q _-\\
 q _+
\end{array}
\right)=0.
\end{align}
We can readily obtain the following solution of this equation:
\begin{align}
\omega=-k_z;\quad  q _-\propto \exp\left[ -\frac{(Bx-k_y)^2}{2B} \right],\quad  q _+=0,
\end{align}
whereas a similar form,
\begin{align}
\omega=k_z;\quad  q _+\propto \exp\left[ \frac{(Bx-k_y)^2}{2B} \right],\quad  q _-=0,
\end{align}
is not a solution because we cannnot normalize the eigenvector in a particular way at $x\rightarrow \infty$. On the other hand, the Weyl equation for the right-handed quark is 
\begin{align}
\i \sigma^\mu D_\mu  q _{\rm R} =0,
\end{align}
with $\sigma^\mu=(1,\sigma^i)$ yields the solution for $\omega=k_z$ with the normalized eigenvector. Therefore, we obtain the zero eigenenergies of the Dirac fermion $q$ including both of the Weyl fermions $ q _{\rm L}$ and $ q _{\rm R}$,
\begin{align}
\label{eq:LLL}
\omega=\pm k_z,
\end{align}
where the signs correspond to different chirality of the quarks. These gapless eigenmodes are called the lowest Landau level. 

In general, the dispersion relation for Eq.~(\ref{eq:LL}) can be calculated  as
\begin{align}
\label{eq:E}
\omega^2 = k_z^2 +B(2n+1-s), 
\end{align}
where $n=0,1,2,...$ and $s=\pm1$ is the eigenvalue of $\sigma^z$. The lowest landau level corresponds to $(n,s)=(0,1)$. Note that Eq.~(\ref{eq:E}) does not depend on $k_y$, so that any eigenvalues are degenerated by all of the degrees of freedom that $k_y$ can take in our gauge choice. This is called the Landau degeneracy, and it is given by $B/(2\pi)$ per unit area.\footnote{Let us calculate the Landau degeneracy. We consider the $xy$-plane bounded by $0<x<L_x,\ 0<y<L_y$. The momentum in $y$ direction can take $k_y=2\pi n_y/L_y$, ($n_y \in \Z$). However, the center of the wave function $x_{\rm c}\equiv B/k_y$ must be inside the range of $x$. This yields the upper limit of $n_y$ such that 
\begin{align}
0<\frac{B}{k_y}<L_x \Leftrightarrow 0<n_y<\frac{L_x L_y B}{2\pi}.
\end{align}
All of the modes included in this inequality contribute to the degeneracy. The number of such modes per unit area is: 
\begin{align}
\label{eq:Landau deg}
N_y = \frac{B}{2\pi} .
\end{align}
}

\subsubsection{Chiral anomaly}
\begin{figure}[t]
\begin{tabular}{cc}
\begin{minipage}{1.0\textwidth}
\subfigure[$E=0$]{ \includegraphics[bb=0 0 595 595,height=7cm]{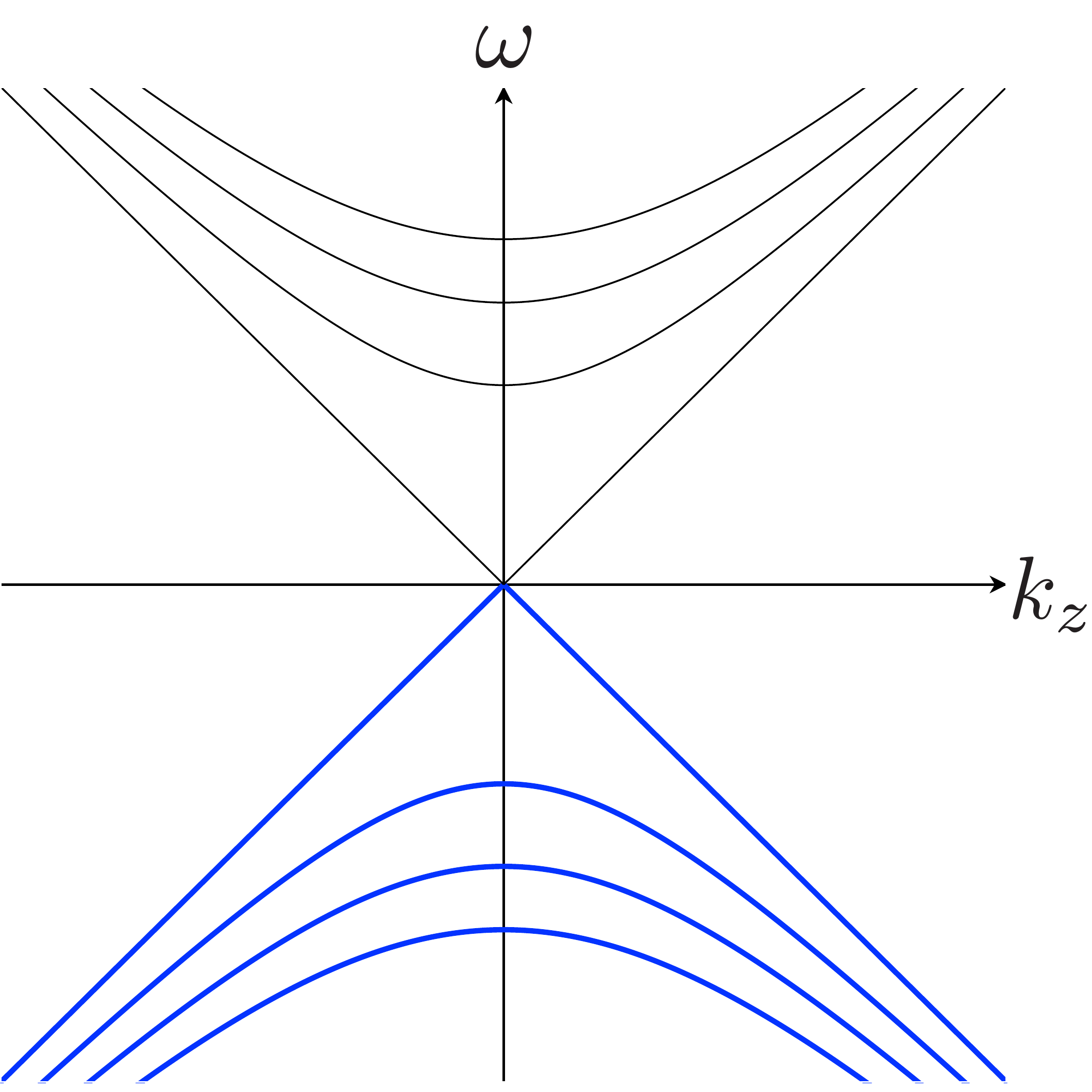}\label{fig:OSa}}~
\subfigure[$E\neq0$]{ \includegraphics[bb=0 0 595 595,height=7cm]
{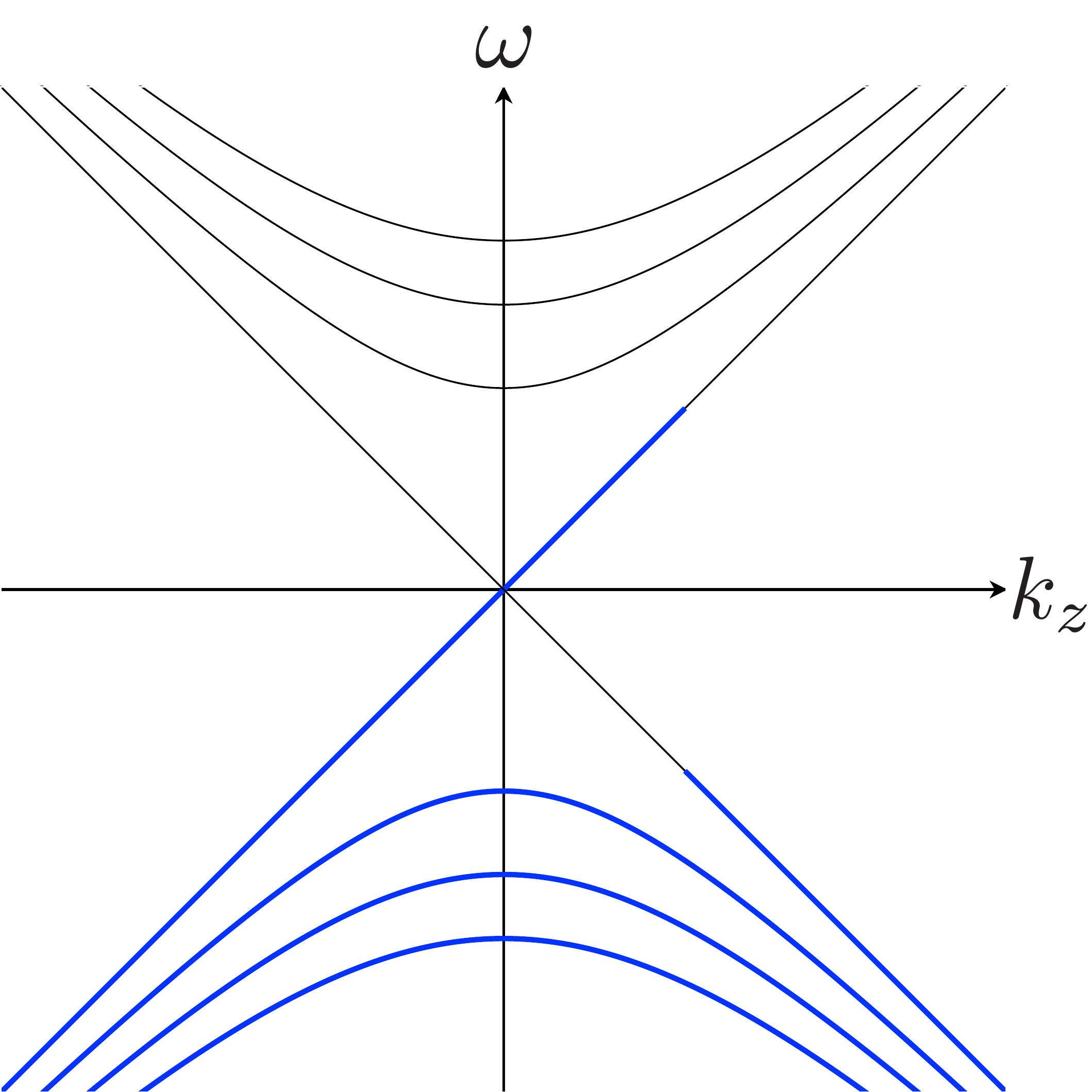}\label{fig:OSb}}
\caption{Occupied staes of Dirac fermions in a static magnetic in the presence/absence of the electric field $E$.}
\label{fig:Pin5s}
\end{minipage}~
\end{tabular}
\end{figure}
We here derive the chiral anomaly. Figure.~\ref{fig:OSa} shows the spectrum in a static magnetic field, i.e., the dispersion relation (\ref{eq:E}). Here, the branches crossing the origin correspond to the lowest Landau level (\ref{eq:LLL}), and the other curves correspond to the higher Landau levels.

Next, we apply an electric field in the $z$-direction, $\Ev=(0,0,E)$.  Then, the electric field accelerates the particles, 
\begin{align}
\frac{\d k_z}{\d t} = E.
\end{align}
When we apply the electric field by time $\tau$, the fermions acquire finite momentum,
\begin{align}
\Delta k_z = E\tau.
\end{align}
We illustrate the occupied state after applying the electric field  $E$ as Fig.~\ref{fig:OSb}. We can interpret here that some of the minus-energy particles are excited to get positive energy. In this accelerating process, the charges for the right-handed and left-handed fermions in volume $V$ are varied by
\begin{align}
\Delta Q_{\rm R} &= \frac{BV}{2\pi} \int^{\Delta k_z}_0 \frac{\dd k_z }{2\pi}=\frac{\tau VEB}{4\pi^2},\\
\Delta Q_{\rm L} &= -\frac{BV}{2\pi} \int^{\Delta k_z}_0 \frac{\dd k_z }{2\pi}=-\frac{\tau VEB}{4\pi^2}.
\end{align}
Note that the factor $B/(2\pi)$ corresponds to the Landau degeneracy~(\ref{eq:Landau deg}). Therefore, the total charge does not change $\Delta Q= Q_{\rm R}+Q_{\rm L} =0$, whereas the axial charge varies 
\begin{align}
\Delta Q_5 =\Delta Q_{\rm R} - Q_{\rm L} = \frac{\tau VEB}{2\pi^2}.
\end{align}
We can interpret this equation as an additional source of the classical conservation law associated with the axial charge symmetry per unit time and volume. Thus, we obtain
\begin{align}
\d_\mu j^\mu_5 =  \frac{EB}{2\pi^2}.
\end{align}
This equation in static electric and magnetic fields can be generalized into,
\begin{align}
\label{eq:Chiral Anomaly}
\d_\mu j^\mu_5 =  C \Ev\cdot \Bv,
\end{align}
where
\begin{align}
\label{eq:C}
C\equiv \frac{1}{2\pi^2}.
\end{align}
Equation (\ref{eq:Chiral Anomaly}) is the same relation as that of the triangle anomalies in quantum field theories \cite{Adler:1969gk,Bell:1969ts}. 

We make some remarks on this triangle anomaly relation. It is exact at all orders of perturbation theory according to the Adler-Bardeen's theorem \cite{Adler:1969er}. Moreover, this is an ``anomalous'' Ward-Takahashi identity, which does not depend on any perturbative calculations \cite{Fujikawa:1979ay}. Therefore, the presence of gluons does not affect Eq.~(\ref{eq:Chiral Anomaly}).

\subsubsection{Chiral magnetic effect}
Let us evaluate the currents in the presence of each chemical potential of left- or right-handed fermions, $\mu_{\rm R,L}$.  Using the Fermi-Dirac distribution function, we obtain
\begin{align}
j^z_{\rm R} &= \frac{B}{2\pi}\int^{\infty}_0 \frac{\dd k_z}{2\pi} \left(  \frac{1}{1+\e ^{\frac{k_z-\mu_{\rm R}}{T}}} - \frac{1}{1+\e ^{\frac{k_z+\mu_{\rm R}}{T}}} \right) = \frac{\mu_{\rm R}B}{4\pi^2},\\
j^z _{\rm L} &= -\frac{B}{2\pi}\int^{\infty}_0 \frac{\dd k_z}{2\pi} \left(  \frac{1}{1+\e ^{\frac{k_z-\mu_{\rm L}}{T}}} - \frac{1}{1+\e ^{\frac{k_z+\mu_{\rm L}}{T}}} \right) = -\frac{\mu_{\rm L}B}{4\pi^2}.
\end{align}
Here the first and second terms of each equation correspond to the particle and anti-particle contributions, respectively. Note also that only the lowest Landau level contributes to the currents. By writing these in the vector form, we get 
\begin{align}
\jv_{\rm R} = \frac{\mu_{\rm R}\Bv}{4\pi^2},\\
\jv_{\rm L} = -\frac{\mu_{\rm L}\Bv}{4\pi^2}.
\end{align}
Thus, the total and axial currents are
\begin{align}
\label{eq:CMEgeneral}
\jv\equiv \jv_{\rm R} +\jv_{\rm L} = \frac{\mu_5 \Bv}{2\pi^2}, \\
\jv_5\equiv \jv_{\rm R} -\jv_{\rm L} =  \frac{\mu \Bv}{2\pi^2}, 
\end{align}
where $\mu_5\equiv (\mu_{\rm R}-\mu_{\rm L})/2$ and $\mu\equiv (\mu_{\rm R}+\mu_{\rm L})/2$. The creation of the vector current $\jv$ and the axial current $\jv_5$ are called the chiral magnetic effect (CME) \cite{Kharzeev:2007jp,Fukushima:2008xe,Nielsen:1983rb,Vilenkin:1980fu} and the chiral separation effect (CSE) \cite{Son:2004tq,Metlitski:2005pr}, respectively. As we will see in Chap.~\ref{sec:LangevinCME}, we can also derive the CME and CSE in the use of the anomalous commutation relation (\ref{eq:PBn5n}), which is equivalent to the triangle anomaly relation (\ref{eq:Chiral Anomaly}).

\subsection{Chiral magnetic wave}
\label{eq:CMWderivation}
We here show the presence of the CME and CSE leads to the collective gapless modes called the chiral magnetic wave (CMW) \cite{Kharzeev:2010gd,Newman:2005hd}. In the absence of the electric field, the conservation laws for the electric charge and the axial charge are
\begin{align}
\label{eq:dj}
\frac{\d n}{\d t} +\na \cdot \jv =0,\\
\label{eq:dj5}
\frac{\d n_5}{\d t} +\na \cdot \jv_5 =0.
\end{align}
The vector and the axial currents are generated by the CME and the CSE as 
\begin{align}
\label{eq:CMEn5}
\jv = \frac{\mu_5 \Bv }{2\pi^2} = \frac{n_5 \Bv }{2\pi^2\chi_5}, \\
\label{eq:CSEn}
\jv_5 =  \frac{\mu \Bv }{2\pi^2} = \frac{n \Bv }{2\pi^2\chi}. 
\end{align}
Here we write the conserved charge densities $n$ and $n_5$ into
$n=\chi \mu$ and $n_5=\chi_5 \mu_5$ by using the the susceptibilities 
\begin{align}
\chi\equiv \frac{\d n}{\d \mu}, \quad 
\chi_5 \equiv \frac{\d n_5}{\d \mu_5}.
\end{align}
We substitute Eqs.~(\ref{eq:CMEn5}) and (\ref{eq:CSEn}) into Eqs.~(\ref{eq:dj}) and (\ref{eq:dj5}), respectively. Then, we obtain
\begin{align}
\label{eq:dj2}
\frac{\d n}{\d t} +  \frac{\Bv}{2\pi^2\chi_5}\cdot  \na n_5  =0,\\
\label{eq:dj52}
\frac{\d n_5}{\d t} + \frac{\Bv}{2\pi^2\chi}\cdot  \na n  =0.
\end{align}
By eliminating $n_5$ from these equations, we find a wave function
\begin{align}
\label{eq:waveeqCMW}
\frac{\d^2 n}{\d t^2} = -  \frac{\Bv}{2\pi^2\chi_5}\cdot  \na \frac{\d n_5}{\d t}
=   \frac{\Bv}{4\pi^2\chi \chi_5}\cdot  \na ( \Bv \cdot \na n)
=   \frac{\Bv^2}{4\pi^2\chi \chi_5}\na_s^2 n,
\end{align}
where we define the projection of the derivative into the direction of the magnetic field, 
\begin{align}
\na_s \equiv \frac{\Bv \cdot \na}{|\Bv|}.
\end{align}
We can interpret Eq.~(\ref{eq:waveeqCMW}) as a propagating mode due to the fluctuation of charge density and axial charge density. This collective wave is called the CMW.

\newpage \thispagestyle{empty}

\chapter{Theory of dynamic critical phenomena}
\label{chap:TDCP}
In this chapter, we give an overview of the theory of dynamic critical phenomena. In Sec.~\ref{sec:DUCgeneral}, we first review the dynamic universality class and explain the idea of dynamic critical phenomena. In Sec.~\ref{eq:FormDCP}, we formulate the static/dynamic low-energy effective theories of a general critical system. In Sec.~\ref{sec:StaticRGgeneral}, we also illustrate the RG analysis of the effective theories.

\section{Dynamic universality class}
\label{sec:DUCgeneral}
Near a second-order phase transition or a critical point, the large correlation length of an order parameter, $\xi$, leads to unusual hydrodynamics \cite{Hohenberg:1977ym}. For example, the typical time scale of diffusion becomes larger when the nuclear liquid-gas critical point is approached  \cite{Kawasaki,Halperin1974b,Siggia1976}. Remarkably, this critical slowing down is independent of the microscopic details of the system. In fact, several molecules, such as water, carbon dioxide, xenon, etc., exhibit experimentally the same dynamic critical behavior \cite{Swinney1973}. Moreover, as we mentioned in the introduction, relativistic fluid near the high-temperature QCD critical point also displays the same dynamic critical phenomenon as above \cite{Fujii:2003bz,Fujii:2004jt,Son:2004iv,Minami:2011un}. This shows the universality between the quark-gluon system at $T\sim\Lambda_{\rm QCD}\sim 200\ {\rm MeV}$ and the molecular systems at $T\sim300$ K $\sim30$ meV.

\begin{figure}[t]
\includegraphics[bb=0 0 1220 450,height=5cm]{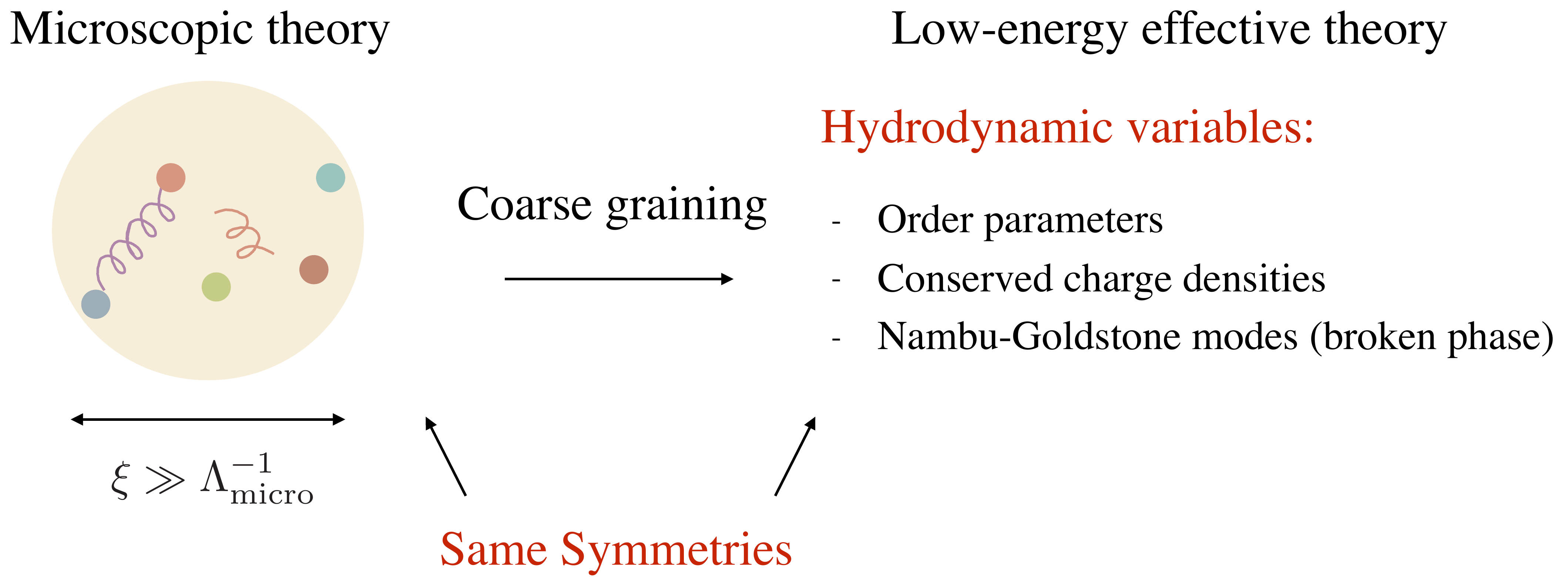}\label{fig:DUC}
\caption{Idea of the dynamic universality class}
\end{figure}

Let us look into the details of the universality of critical phenomena. We here consider the general system near a second-order phase transition or a critical point. As schematically illustrated in Fig.~\ref{fig:DUC}, the correlation length of the order parameter, $\xi$, is much larger than the typical microscopic length scale, $\Lambda_{\rm micro}^{-1}.$\footnote{Note here that we consider the system not {\it directly at} the critical point $\xi\ra \infty$, but sufficiently {\it near} the critical point where $\xi\gg\Lambda_{\rm micro}^{-1}$ ($\Lambda_{\rm micro}\neq 0$). Otherwise, the hydrodynamic description breaks down. In other words, we extrapolate the calculation on $\xi\gg\Lambda_{\rm micro}^{-1}$ to the critical point perturbatively.} Therefore, to describe the dynamic critical phenomena characterized by $\xi$, we can coarse grain or integrate out microscopic degrees of freedom, such as quarks and gluons in QCD. Then, it is sufficient to use the low-energy effective theory of the hydrodynamic variables at large-length and long-time scales of the system. Typical hydrodynamic variables are order parameters (e.g., magnetization), conserved-charge densities (e.g., energy and momentum densities), and Nambu-Goldstone modes associated with spontaneously symmetry breaking (e.g., superfluid phonon). Symmetries of the low-energy effective theory must be the same as those of the microscopic theory before coarse graining or integrating out. We can classify dynamic critical phenomena based on the hydrodynamic variables and the symmetries of the systems. The classification of dynamic critical phenomena is called the dynamic universality class \cite{Hohenberg:1977ym}.  

We summarize the conventional classification of dynamic critical phenomena studied by Hohenberg and Halperin \cite{Hohenberg:1977ym} in Table \ref{tab:DUC}. We here show, for each of the universality classes, internal symmetry, the property of order parameter (whether it is conserved or not) and conserved charge (the symmetry related by the Noether's theorem), and a typical system.

\begin{table}[t]
\centering\includegraphics[bb=0 0 666 312,height=6.5cm]{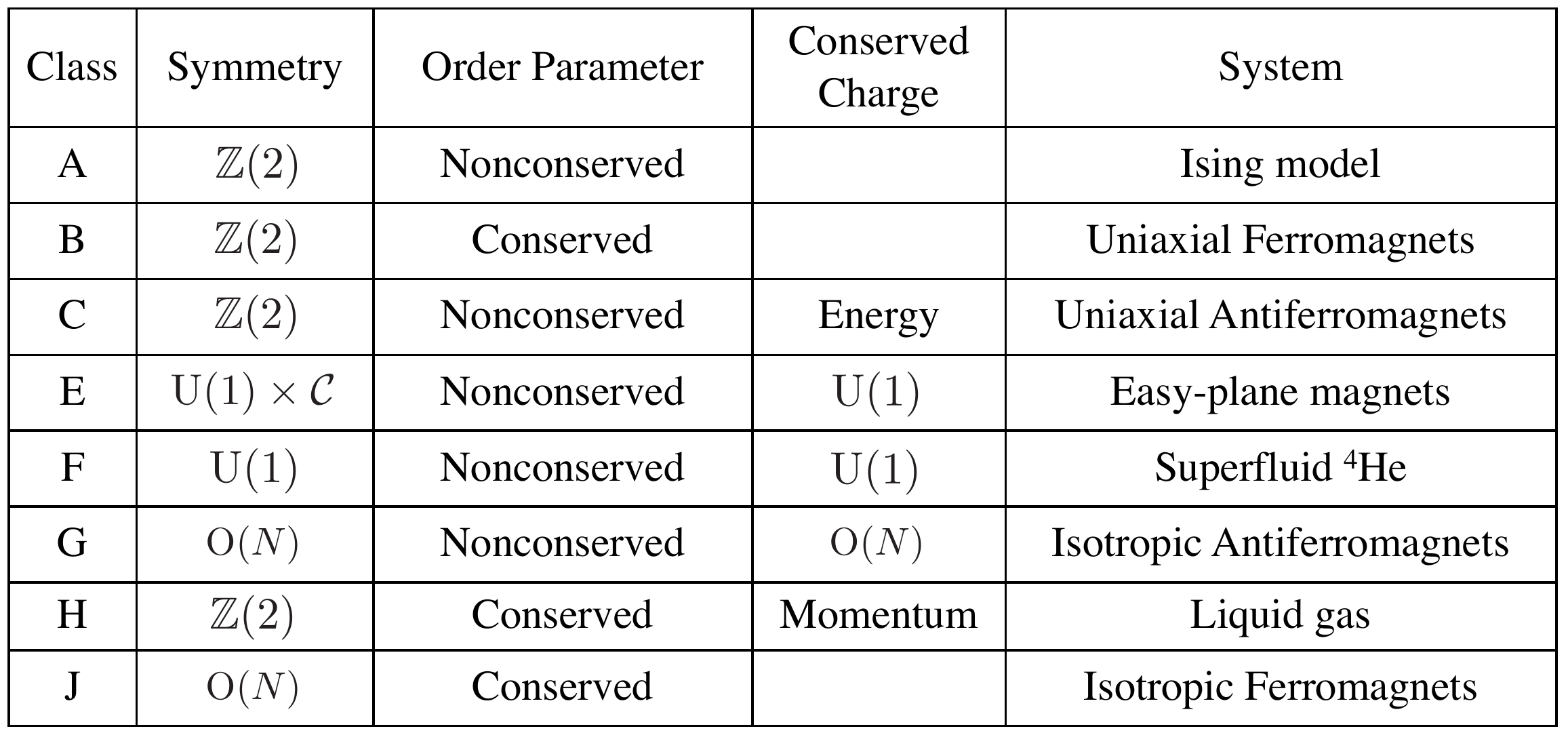}
  \caption{Conventional dynamic universality class}
 \label{tab:DUC}
\end{table}
%\end{center}

\section{Formulation}
\label{eq:FormDCP}

In this section, we give more details about the low-energy effective theory of critical phenomena. In Secs.~\ref{sec:GLgeneral} and \ref{sec:general Langevin theory}, we formulate the Ginzburg-Landau theory/Langevin theory to describe static/critical critical phenomena, respectively.
%\footnote{In the dynamic part, we will study the time evolution of the system {\it near equilibrium}, such that the dynamics can be described based on the static property of the system. This assumption is necessary, because the original criticality in the thermodynamic free energy may not be well-defined otherwise.} 
 In Sec.~\ref{sec:DPTgeneral}, we demonstrate how to obtain the field theory equivalent to the Langevin theory derived in Sec.~\ref{sec:general Langevin theory}. This field theory will help us to apply the renormalization group approach in Sec.~\ref{sec:dynamics}.

\subsection{Ginzburg-Landau theory}
\label{sec:GLgeneral}
The static critical phenomena and the static universality class can be determined only by the order parameters and the symmetries of the system. This notion is called the Ginzburg-Landau theory, and its idea can be summarized as follows.

Since a second-order phase transition or a critical point is/can be regarded as a continuous transition, we can consider the parameter region, e.g., temperature, with a small order parameter. Then, we can expand the free energy of the system with respect to the order parameter. Moreover, as we are interested in the long-range behavior of the system characterized by $\xi$, we can also expand this Ginzburg-Landau free energy with respect to the derivative. Here we can neglect higher-order terms because such terms are suppressed by $O(\na/\Lambda)$. Here $\Lambda$ is the momentum cutoff, which gives the upper limit of the applicability of the effective theory.\footnote{Practically, we are constructing the low-energy effective theory by integrating out the microscopic degrees of freedom whose energy scale is larger than the typical energy scale of the critical phenomena, i.e., the inverse correlation length $\xi^{-1}$. It follows that $\Lambda$ should be chosen as to satisfy $\xi^{-1}\ll \Lambda$.} See, e.g., Refs.~\cite{Polchinski:1992ed,Kaplan:1995uv,Kaplan2005} for the general construction of effective field theories. 

We generalize the above expansion of the order parameter to all of the hydrodynamic variables. In the Ginzburg-Landau free energy, we write down all consistent terms with the symmetries of the system from the double expansion of the derivative and the hydrodynamic variables. Let us write one of the hydrodynamic variables of the system as $\psi_I$ ($I,J,K=1,...,n$). Then, we can formally write the free energy of the system in the following form:\begin{align}
\label{eq:Fgeneral}
F[\psi] &= \int \dd \rv \left [ \frac{1}{2}\psi_I\beta_{IJ}(\na) \psi_J + \frac{1}{3!}\beta_{IJK} (\na) \psi_I \psi_J \psi_K\right.\notag\\
&\left.\qquad\qquad\qquad\qquad\qquad\qquad+ \frac{1}{4!}\beta_{IJKL} (\na)\psi_I \psi_J \psi_K \psi_L+\cdots\right].
\end{align}
Here $\beta_{IJ}$, $\beta_{IJK}$, $\beta_{IJKL}$ are some functions of the spatial derivative. These coefficients can also be expanded with respect to the derivative. Summation over repeated indices $I, J, K$, and $L$ is implied.

%Typically, the higher order terms with respective each of the derivative or the hydrodynamic variables will be found irrelevant in the spirit of the so-called Renormalization Group (RG), as we will see later. Nevertheless, as we will see the details, from the so-called Renormalization Group (RG) treatment, the relevant higher order terms are few but can but they play an important role for the fluctuations enhanced when the system approaches the critical point. 

Once we determine the functional of the free energy, we can calculate any correlation functions among the hydrodynamic variables by using the following definition of the expectation value:
\begin{align}
\label{eq:expect}
\displaystyle \left\langle\mathcal{O}[\psi]\right\rangle
=\frac{\displaystyle\int\prod_{I} \mathcal{D}\psi_{I}\mathcal{O}[\psi]e^{-\beta F[\psi]}}{\displaystyle\int\prod_{I} \mathcal{D}\psi_{I}e^{-\beta F[\psi]}}.
\end{align}
Here, $\beta\equiv T^{-1}$ is the inverse temperature, and $\mathcal O[\psi]$ is some product of the hydrodynamic variables.
We can also calculate thermodynamic quantities related to these static correlation functions. Static critical exponents characterize the critical phenomena of these thermodynamic quantities. Systems with the same critical exponents belong to the same universality class. As we will see later, we can uniquely determine static critical exponents by using the renormalization group. Thus, the static universality class is only determined by the order parameter and the symmetries, which constrains the possible terms of the Ginzburg-Landau free energy.\footnote{
You may logically think that not only the order parameter but also conserved charge densities can affect the {\it static} universality class. Such variables are normalized so that those mass terms in the Ginzburg-Landau free energy become unity (unlike the order parameter field) under the renormalization group. Due to this fact, the loop corrections on the fixed point structure solely determined by the order parameters tend to be irrelevant. For example, it is shown that the inclusion of the energy density does not change the static universality class governed by the Wilson-Fisher fixed-point \cite{HHSI,HHSII}.      
%On the other hand, the inclusion of the massless gauge field changes the existence of the fixed point, so that the order of the phase transition has been modified to the first order. 
}

\subsection{Langevin theory}
\label{sec:general Langevin theory}
The Langevin theory is the low-energy effective theory at large-length and long-time scales, including macroscopic dissipation effects. The time evolution of the hydrodynamic variable $\psi_I$ is described by the following Langevin equation \cite  {Chaikin,Forster,Mazenko,Tauber}:
\begin{align}
\label{eq:Nonlinear Langevin}
\frac{\d \psi_I(t,\rv)}{\d t} = -\gamma_{IJ}(\na) \frac{\delta F[\psi]}{\delta \psi_J (t,\rv)}  -\int \dd \rv' \left [    \psi_I(t,\rv), \psi_J(t,\rv') \right] \frac{\delta F[\psi]}{\delta \psi_J (t,\rv')} + \xi_I (t,\rv). 
\end{align}
These three terms in the right-hand side are called the dissipative term, the reversible term,%
\footnote{Strictly speaking, there is another contribution in the reversible term at finite temperature,
\begin{align}
T \frac{\delta }{\delta \psi_J (t,\rv')} \left [    \psi_I(t,\rv), \psi_J(t,\rv') \right]. 
\end{align}
This term is required so that the Fokker-Plank equation corresponding to Eq.~(\ref{eq:Nonlinear Langevin}) yields the equilibrium distribution $\propto \e^{-\beta F}$ as a steady-state solution (for more details, see, e.g., Refs.~\cite{Chaikin,Mazenko,Tauber}). Nevertheless, this term is found to be zero or unimportant in most cases. 
}
 and the noise term, respectively. 

\subsubsection{Dissipative term}
The first term of Eq.~(\ref{eq:Nonlinear Langevin}) describes the relaxation process of the hydrodynamic variables to its equilibrium value. %, such that the time evolution can be determined by the Ginzburg-Landau Free energy $F[\psi]$. 
We consider the system slightly apart from the equilibrium. It follows that we can interpret this first term as a leading term of the expansion with respect to the variation of the free energy, $\Psi_I\equiv \delta F/\delta \psi_I$,
\begin{align}
\label{eq:fexpansion}
f_I[\Psi]=f_I[0]+\gamma_{IJ}\Psi_J+\cdots, 
\end{align}
where, $\Psi_I=0$ at the equilibrium. Note that $f_I[0]$ and $\Psi_J$ are some functions of $\psi_I$, in general. The leading term $f_I[0]$ is the 0-th order of the perturbation which may be regarded as an equilibrium. On the other hand, the left-hand side of Eq.~(\ref{eq:Nonlinear Langevin}), $\d \psi_I /\d t$ vanish at this 0-th order. It follows that we can set $f_I[0]=0$.
%\footnote{Logically, one can add any variables which vanishes at equilibrium to the right hand side of Eq.~(\ref{eq:fexpansion}). Actually, the noise term satisfies this condition.} 
%By redefining the hydrodynamic variables as those fluctuations from the variables at equilibrium, to say $\delta x_i\equiv x_i-\left< x_i \right>$ and write $\delta x_i \ra x_i$, we can set $f_i[0]=0$. 
%In addition to this expansion with respect to $\Psi$, 
We also expand the coefficient $\gamma_{IJ}$ with respect to the derivative as we are interested in the long-range behavior of the system, 
\begin{align}
\label{eq:gammaexpansion}
\gamma_{ij}(\na) = \gamma^{(0)}_{ij} + \gamma^{(2)}_{ij} \na^2+\cdots.
\end{align}
 
\subsubsection{Reversible term}
The second term describes the reversible dynamics at a macroscopic scale. In the expression of this second term, $\left[A(t,\rv),B(t,\rv')\right]$ denotes the Poisson bracket, which can be postulated from the symmetry algebra. One can obtain the expression of the Poisson bracket by computing the microscopic commutation relation of the operators corresponding to the hydrodynamic variables. We can interpret this term as the classical limit of the Heisenberg equation,
\begin{align}
\int \dd \rv'  \left[  \psi_I(t,\rv),\psi_J(t,\rv') \right] \frac{\delta F[\psi]}{\delta \psi_J(t,\rv')}
=&\left[\psi_I(t,\rv),F[\psi]\right].
\end{align}
One may show this relation, order by order of the expansions in Eq.~(\ref{eq:Fgeneral}).

One distinct difference from the first dissipative term is that this second reversible term does not contribute to the time derivative of the free energy:
\begin{align}
\frac{\dd \ex{F[\psi]}}{\dd t}&=\ex{\frac{\delta F[\psi]}{\delta \psi_I }\frac{\d \psi_I}{\d t}}\notag\\
&=-\gamma_{IJ}\ex{\frac{\delta F}{\delta \psi_I }\frac{\delta F}{\delta \psi_J}}-\int \dd \rv' \ex{\left[ \psi_I ,\psi_{J}\right]\frac{\delta F}{\delta \psi_I }\frac{\delta F}{\delta \psi_J}}+\ex{\frac{\delta F}{\delta \psi_I }\xi}
\notag\\
&=-\gamma_{IJ}\ex{\frac{\delta F}{\delta \psi_I }\frac{\delta F}{\delta \psi_J}}.
\end{align}
Therefore, the second term describes the time evolution of the system without increasing the entropy. Intuitively, the dissipative term breaks the time-reversal symmetry, whereas the reversible term does not.

\subsubsection{Noise term}
The third term is the random driving force, which originates from the underlying microscopic degrees of freedom. Those statistical properties should be dictated by the nature of equilibrium as follows: 
\begin{align}
\label{eq:xiexp}
\left<\xi_I(t,\rv) \right>&=0,\\
\label{eq:noisecorgeneral}
\left< \xi_I(t,\rv) \xi_J(t',\rv') \right>&=2T\gamma_{IJ}(\na)  \delta (t-t') \delta^d(\rv-\rv').
\end{align}
Equation~(\ref{eq:xiexp}) stems from the requirement that $\left< \psi_I (t,\rv) \right>$ should not be affected by the presence of $\xi_{\rm I}$. The matrix $\gamma_{IJ}$ in the random-force correlation~(\ref{eq:noisecorgeneral}) should be the same as that in the dissipative term (\ref{eq:gammaexpansion}). The relation between the dissipation effects and the random force is called the (second) fluctuation-dissipation relation.\footnote{In the context of the Brownian motion, this relation is due to the fact that both fluctuation and dissipation of the Brownian particles come from the same origin: random impact of surrounding molecules.}
%Besides, hydrodynamic variables $\psi_I$ and noise variables $\xi_J$ are independent so that
%\begin{align}
%\label{eq:xixiexp}
%\left <\psi_I(t,\rv) \xi_J(t,\rv) \right>&=0,
%\end{align}
%in the use of Eq.~(\ref{eq:xiexp}).

%\subsubsection{Notes}
%We here note that, in the above construction, the Ginzburg-Landau theory and the Langevin theory can be derived uniquely  in the use of the systematical expansions and the symmetry algebra. In this sense, the low-energy effective theory is perfectly determined, once we assume the hydrodynamic variables and the symmetries of the system.  

\subsection{Dynamic perturbation theory}
\label{sec:DPTgeneral}
We here convert the Langevin theory derived in Sec.~\ref{sec:general Langevin theory} into the path-integral formulation of a field theory. This formulation is called the Martin-Siggia-Rose-Janssen-de Dominicis (MSRJD) formalism \cite{Martin:1973zz,Janssen:1976,DeDominicis:1978} and helps us to apply the renormalization group method systematically. In this subsection, we follow the derivation in Ref.~\cite{Tauber}. 

We first formally write the Langevin equation~(\ref{eq:Nonlinear Langevin}) and the fluctuation-dissipation relation~(\ref{eq:noisecorgeneral}) (in the unit of $T=1$) in the following form:
\begin{align}
\label{eq:generalLangevin}
 \frac{\partial \psi_I(\rv{,t})}{\partial t}
 &=\mathcal{F}_I[\psi]+ \xi_I(\rv,t) ,\\
\label{eq:generalFDR}
\langle \xi_I(\rv,t)\xi_J(\rv',t')\rangle&=2\gamma_{IJ}(\na)\delta^d(\rv-\rv')\delta(t-t').
\end{align}
Here, $\mathcal{F}_I$ may involve all the hydrodynamic variables $\{ \psi_I \}$, and the noises $\{ \xi_I \}$ are assumed to obey the Gaussian white noise. 

We consider the correlation functions for the hydrodynamic variables under various configurations of the noise variables,
\begin{align}
\label{eq:Opsibar}
\left< \mathcal O[\bar \psi] \right>_\xi = 
 \mathcal{N} \int \mathcal D \xi \mathcal O[\bar \psi]  \exp \left[ 
 - \frac{1}{4} \int \dd t \int \dd \rv \, \xi_I \gamma_{IJ}^{-1} \xi_J 
 \right],
\end{align}
where $\bar\psi_I$ denotes the formal solution of the Langevin equations, and $\mathcal{N}$ is a normalization factor. We can reproduce Eq.~(\ref{eq:generalFDR}) by using the distribution~(\ref{eq:Opsibar}). To carry out the path integral of $\xi$ in Eq.~(\ref{eq:Opsibar}) in the presence of the noise-dependent variable $\mathcal{O}[\bar\psi]$, we insert the following identity into the right-hand side of Eq.~(\ref{eq:Opsibar}):
\begin{align}
\label{eq:identMSR}
1= \int \mathcal D \psi \prod_{I}\delta(\psi_I-\bar \psi_I) = \int \mathcal D \psi\,   \prod_I \prod_{\rv,t}  \delta \left( \frac{\partial \psi_I}{\partial t} - \mathcal{F}_I[\psi]-\xi_J \right).
\end{align}
We here omit the Jacobian $\det \left( \partial_t - \delta \mathcal F / \delta \psi \right)$ in the right-hand side. We can justify this by getting rid of some unnecessary graphs containing the so-called \textit{closed response loops} in diagrammatic calculations (see Ref.~\cite{Tauber} for more details). Then, we can replace $\mathcal{O}[\bar \psi]$ by $\mathcal{O}[\psi]$ and integrate over the noise variables of Eq.~(\ref{eq:Opsibar}). Finally, we get
\begin{align}
\label{eq:Opsi}
\left< \mathcal O[\psi] \right> = \mathcal{N}'\int \i \mathcal D \tilde \psi \int \mathcal D \psi\, \mathcal O [\psi]
\exp \left( - S[\tilde\psi,\psi] \right).
\end{align}
Here, $\mathcal{N}'$ is a normalization factor, and we use the Fourier representation of the delta function. We introduce the pure imaginary auxiliary field $\tilde \psi_I$ called the response field for each hydrodynamic variable $\psi_I $. The MSRJD effective action $S[\tilde\psi,\psi]$ is given by %which was originally the Langevin theory given by (\ref{eq:generalLangevin}) and (\ref{eq:generalFDR}):
\begin{align}
\label{eq:MSR}
S[\tilde\psi,\psi]= \int \dd t \int \dd \rv\, \left[ \tilde\psi_I \left( \frac{\partial \psi_I}{\partial t}  -\mathcal{F}_I[\psi] \right)- \tilde\psi_J \gamma_{IJ}(\na)\tilde\psi_I \right].
\end{align}
By the use of the path-integral technique, we can calculate any correlation functions (\ref{eq:Opsi}) among the hydrodynamic variables $\psi_I$ and the response fields $\tilde \psi_I$.

\section{Renormalization group analysis}
\label{sec:StaticRGgeneral}
The renormalization group (RG) is a powerful method to solve the theory near a second-order phase transition or a critical point. In Sec.~\ref{sec:staticsgeneral}, we first review the static RG for the Ginzburg-Landau free energy following Ref.~\cite{Chaikin}. Next, in Sec.~\ref{sec:dynamics}, we move to the dynamic RG based on the MSRJD action and make remarks on its difference from the static one.  

\subsection{Statics}

\label{sec:staticsgeneral}
 
\subsubsection{RG transformation}
We perform static RG transformation on the path integral of the partition function:
\begin{align}
\label{eq:staticpartition}
Z&=\int \mathcal{D}\psi e^{-\beta F_{\Lambda}[\psi]},
\end{align}
with $F_{\Lambda}[\psi]$ being a Ginzburg-Landau free energy. The static RG consists of the following steps:

(i) Integrating over $\psi(\qv)$ in the  momentum shell:
\begin{align}
\label{eq:shell}
\Lambda/b<q<\Lambda;
\end{align}

(ii) Rescaling the momentum scale and all of the fields:
\begin{align}
\label{eq:rescalestatic}
\qv &\ra \qv ' = b \qv;\\
\label{eq:rescalestatic2}
\psi(\qv)&\ra  \psi'(\qv') = [\zeta(b)]^{-1}\psi(\qv).
\end{align}
Here, $q\equiv |\qv|$, $\Lambda$ is the momentum cutoff, $b>1$ is the renormalization scale. We will determine the explicit form of the scaling function $\zeta(b)$ in a moment. At the second-order phase transition or the critical point, the correlation length diverges, and the typical length-scale disappears. Therefore, we expect the RG invariance of the system. This emergent symmetry brings strong constraints on the calculation.

%This is also because the higher momentum degrees of freedom is not important for the long-range behavior of the system. We will use this condition from the hydrodynamic regime and solve the theory including the nonlinear Gaussian terms. 

\subsubsection{RG of the Ginzburg-Landau free energy}
To apply the RG to $F_\Lambda[\psi]$, we decompose the field $\psi(\qv)$ into the low momentum part $\psi^<$ and the high momentum part $\psi^>$:
\begin{align}
\label{eq:fielddecomp}
\psi(\qv) = \psi^{<}(\qv) + \psi^{>}(\qv),
\end{align}
where 
\begin{align}
\label{eq:high}
\psi^<(\q)&=\left\{\begin{array}{cc}
\psi(\q)&(0<q<\Lambda/b);\\
0&(\Lambda/b<q<\Lambda),
\end{array}\right.\\
\label{eq:low}
\psi^>(\q)&=\left\{\begin{array}{cc}
0&(0<q<\Lambda/b);\\
\psi(\q)&(\Lambda/b<q<\Lambda).
\end{array}\right.
\end{align}
We first integrate over the high-momentum field on the partition function:
\begin{align}
Z
&=\int \mathcal{D}\psi^<(\q)\mathcal{D}\psi^>(\q) e^{-\beta F_{\Lambda}[\psi^<+\psi^>]}\notag\\
&=\int \mathcal{D}\psi^<(\q) e^{-\beta F_{\Lambda/b}[\psi^<]}.
\end{align}
Here, $\beta F_{\Lambda/b}[\psi^<(\q)]$ is the effective action for $\psi^{\rm <}$, after integrating out $\psi^>$:
\begin{align}
e^{-\beta F_{\Lambda/b}[\psi^<]}\equiv \int\mathcal{D}\psi^>(\q) e^{-\beta F_{\Lambda}[\psi^< +\psi^>]}.
\end{align}
Next, we rescale the momentum $\qv \ra \qv ' = b \qv$ and the low momentum field $\psi^<(\q)$ into the original field $ \psi' (\qv) $ without the label ``$<$'' or ``$>,$''
\begin{align}
\psi^<(\qv)\ra \psi' (\qv) = [\zeta (b)]^{-1} \psi^<(\qv) .
\end{align}
A series of these operations is the static RG transformation. 

\subsubsection{Scaling factor $\zeta(b)$}

Let us determine the explicit form of $\zeta(b)$ from the RG invariance of the static correlation function.

From the translational symmetry of the system, we can generally write the static correlation function of the fields as
\begin{align}
\label{eq:psipsi}
\ex{\psi(\q_{1})\psi(\q_{2})}=C(\q_{1})(2\pi)^{d}\delta(\q_{1}+\q_{2}).
\end{align}
Here, the expectation value is defined in Eq.~(\ref{eq:expect}) and $C(\q)$ is some function of $\qv$. On the other hand, the correlation function after the RG transformation is 
\begin{align}
\label{eq:psi'psi'corr}
\ex{\psi'(\q'_{1})\psi'(\q'_{2})}&=C(\q'_{1})(2\pi)^{d}\delta(\q'_{1}+\q'_{2})=\zeta^{-2}(b)\ex{\psi(\q_{1})\psi(\q_{2})}.
\end{align}
We plug Eq.~(\ref{eq:psipsi}) into Eq.~(\ref{eq:psi'psi'corr}) and find 
\begin{align}
C(\q'_{1})(2\pi)^{d}\delta(\q'_{1}+\q'_{2})&=\zeta^{-2}(b)C(\q_{1})(2\pi)^{d}\delta(\q_{1}+\q_{2}),\notag\\
C(b\q_{1})b^{-d}&=\zeta^{-2}(b)C(\q_{1}).
\end{align}
At the critical point $\xi\rightarrow \infty$, the anomalous dimension $\eta$ characterizes the critical behavior of the correlation, $C(\qv)=q^{2-\eta}$. We thus obtain the expression for $\zeta(b)$ as
\begin{align}
\label{eq:zeta}
\zeta(b) = b^{\frac{d+2-\eta}{2}}.
\end{align}
In particular, $\eta=0$ for the Gaussian distribution. 

\subsubsection{Higher order terms}
We here discuss the RG of the following general term:
\begin{align}
\beta F^{p}_{\Lambda}[\psi]&=\int \dd^{d}\rv  u_{p}(\na)  \psi^{p}(\bm{\rv}) \notag\\
\label{eq:Fp}
&=\int \prod_{i=1}^{p}\frac{\dd^{d}\q_{i}}{(2\pi)^{3}} u_{p}(\qv_1,...,\qv_p) \psi(\q_{1})\cdots\psi(\q_{p})\delta\left(\q_{1}+\cdots+\q_{p}\right).
\end{align}
In particular, the terms with $p\leq 2$ are Gaussian terms; the terms with $p> 2$ are the non-Gaussian terms. 

We can expand $u_{p}(\na)$ by using the rotational symmetry of the system as
\begin{align}
u_{p}(\na) = u_p + u_p' \na^2 +\cdots. 
\end{align}
Here, $u_p$ and $u_p'$ are some constants. The parity symmetry under $\na\ra -\na$ prohibits the term proportional to $\na$. The coefficient of the higher-order derivative acquires an additional scaling factor, $b^{-2}$, after rescaling the momentum and the fields. Therefore, higher-order derivative terms with fixed $p$ are relatively suppressed under the RG procedure. In particular, we set $u_p'=0$ for $p>2$ from now on. 

By applying the RG transformation to Eq.~(\ref{eq:Fp}) we obtain
\begin{align}
\beta F^{p}_{\Lambda}[\psi]&=u_{p}b^{-pd}\zeta^{p}(b)b^{d}\int \prod_{i=1}^{p}\frac{\dd^{d}\q'_{i}}{(2\pi)^{3}} \psi'(\q'_{1})\cdots\psi'(\q'_{p})\delta\left(\q'_{1}+\cdots+\q'_{p}\right).
\end{align}
The new parameter $u_p'$ after the single RG step satisfies
\begin{align}
u'_{p}=b^{-pd}\zeta^{p}(b)b^{d}u_{p}\equiv b^{\lambda_{p}}u_{p}.
\end{align}
Here, $\lambda_p \equiv p+d-p(d+\eta)/2$. We can classify terms depending on the value of $\lambda_p$, 
\begin{align}
\lambda_p = \left\{ 
\begin{array}{cc}
\label{eq:relevant}
>0& {\rm relevant},\\
=0& {\rm marginal},\\
<0& {\rm irrelevant}.
\end{array}
\right.
\end{align}
When $u_p$ is a relevant parameter, it becomes larger under the RG transformation; when $u_p$ is an irrelevant parameter, it becomes smaller under the RG transformation. Therefore, we can neglect all the irrelevant parameters from the beginning. We can define the above critical dimension $d_{\rm c}(p)$ such that $u_p$ is irrelevant when $d>d_{\rm c}$. In particular, we find $d_{\rm c}(3)=6$ and $d_{\rm c}(4)=4$.

\subsubsection{RG equation}
We here show an outline of the RG of a scalar field theory (see Appendix~\ref{chap:staticRG} for details). The Ginzburg-Landau free energy is given by
\begin{align}
\label{eq:GLW}
\beta F[\psi] = \int \dd \rv \left[ \frac{r}{2} \psi^2 + \frac{1}{2}(\na \psi)^2 +u\psi^4+h\psi  \right]
\end{align}
with $\psi$ being a scalar field. It is worth noting that Eq.~(\ref{eq:GLW}) has $\Z(2)$ symmetry under the following transformation:
\begin{align}
\psi\ra-\psi.
\end{align}
This is the same symmetry as that of the Ising model, so that we can map the scaler field $\psi$ into the magnetization $m$ and expect the same static critical phenomena. 

According to the discussion just under Eq.~(\ref{eq:relevant}), relevant terms at $d=4$ are only the Gaussian terms. On the other hand, when $d$ is slightly smaller than 4, $u$ grows a little bit under the RG transformation. Therefore, by working at 
\begin{align}
\label{eq:epsilonexpansion}
d=4-\epsilon
\end{align}
with the small $\epsilon$ ($0<\epsilon\ll1$), we can construct the perturbation theory based on the expansion with respect to $u$ (or $\epsilon$). Starting from this assumption, we will eventually find $u=\mathcal O(\epsilon)$ and check the consistency of this perturbation scheme. 

%Starting from the parameter at the $l$-th steps for the RG step, $r_l$ and $u_l$, we can calculate the parameters at $l+1$-th steps:
We can evaluate the parameters $r$ and $u$ under the RG transformation. When we write these values after the RG as $r'$ and $u'$, we obtain 
\begin{align}
\label{eq:r're}
r'&=b^{2}\left[r
+6u\int^>_{\q}  \frac{1}{r+\qv^2}
+O(u^{2})\right],\\
\label{eq:u're}
u'&=b^{\epsilon}\left[ u-36u^{2}\int^>_{\q} \frac{1}{(r+\qv^2)^2}\right].
\end{align}
Here, the integrals $\int^>$ represent the integration over the high-momentum degrees of freedom $\psi^>$ (see Eq.~(\ref{eq:int>}) for the definition of this integral); overall factors $b^2$ and $b^\epsilon$ come from the rescaling. See Appendix \ref{chap:staticRG} for the derivation of Eqs.~(\ref{eq:r're}) and (\ref{eq:u're}), which correspond to Eqs.~(\ref{eq:r'}) and (\ref{eq:u'}), respectively. 

We work on the thin momentum shell by setting $b\equiv \e^l\simeq 1+l$ with $l\ll1$. Then, Eqs.~(\ref{eq:r're}) and (\ref{eq:u're}) reduce to the following RG equations (for the derivation, see Eqs.~(\ref{eq:RGEqr}) and (\ref{eq:RGEqu})):
\begin{align}
\label{eq:rbarRG}
\frac{\dd \bar r}{\dd l}&= 2\bar r +12 \bar u -12 \bar r\bar u,\\
\label{eq:ubarRG}
\frac{\dd \bar u}{\dd l}&=\epsilon \bar u-36\bar u^{2},
\end{align}
where 
\begin{align}
\label{eq:WFbody}
\bar r \equiv \frac{r}{\Lambda^2},\quad\bar u \equiv \frac{ u}{8\pi^2\Lambda^\epsilon},
\end{align}
are dimensionless parameters. These equations describe the parameters under the RG transformation at a general renormalization scale $l=\ln b$. The stable fixed-point solution is called the Wilson-Fisher fixed point (its derivation is given in Appendix~\ref{sec:WFappen}):
\begin{align}
\label{eq:WF}
\bar r =- \frac{\epsilon}{6}\quad \bar u = \frac{\epsilon}{36}.
\end{align}
We can calculate the critical exponents from these fixed point values.

\subsection{Dynamics}
\label{sec:dynamics}
We can also apply the dynamic RG analysis to the MSRJD action $S[\psi,\psi]$ given in Eq.~(\ref{eq:MSR}).% in an analogy to the static RG of the Ginzburg-Landau free energy. 
We carry out the RG transformation of $S[\psi,\psi]$ on the path integral defined in the following partition function: 
\begin{align}
Z=\int \mathcal D \i \tilde \psi \mathcal D \psi \e^{-S[\psi,\psi]}.
\end{align}
This is analogous to the static partition function~(\ref{eq:staticpartition}). In this subsection, we make some remark on the important differences from the static RG procedure. 

\subsubsection{Dynamic critical exponent} 
The first difference is the presence of time $t$ or frequency $\omega$. In addition to the rescaling of the space coordinate in Eq.~(\ref{eq:rescalestatic}), frequency (or time) is regarded as a parameter which scales by
\begin{align}
\omega\ra \omega'=b^z \omega,
\end{align} 
where $z$ is called the dynamic critical exponent which characterizes the dynamic universality class.

\subsubsection{Response fields}
The second difference is the presence of the response field $\tilde \psi$. In the presence of this additional field, the Green function in the field theory becomes a matrix form in ($\psi,\tilde \psi$) space. In particular, the bilinear term of $\tilde \psi$ in the MSRJD action yields the so-called noise vortex. We will look into these points explicitly in our main analysis in Sec.~\ref{sec:DPTCME}.

There is also an extension of the RG procedure. In addition to the  rescaling for the hydrodynamic field $\psi$, Eq.~(\ref{eq:rescalestatic2}), we also scale the response field as
\begin{align}
\label{eq:dynamicrescalegeneral}
\tilde \psi(\rv) &\ra b^{\tilde a}\tilde \psi(\rv).
\end{align}
Here, note that the scaling factor $b^{\tilde a}$ is generally independent of that for the hydrodynamic variable, $\zeta(b)$, in Eq.~(\ref{eq:zeta}). We will see the rescaling of the response fields in our main calculation, Eqs.~(\ref{eq:rescalephitilde})--(\ref{eq:rescalen5tilde}).

\subsubsection{Frequency-dependent internal loops}
The third difference can be seen in the integrating out procedure of the higher-momentum fields. In addition to the same momentum integral as that of the static RG, Eq.~(\ref{eq:shell}), we carry out frequency integrals over $-\infty<\omega<\infty$ in the dynamic formulation. Integrals in frequency space are performed by using contour integrals. These integral pick up hydrodynamic poles of the system.%; these are characteristics of the dynamics. 

\chapter{Dynamic critical phenomena of the high-density QCD critical point}
\chaptermark{High-deinsity QCD critical point}
\label{chap:DCPHD}

In this chapter, we study the static and dynamic critical phenomena near the high-density QCD critical point. In Sec.~\ref{sec:HVsuper}, we summarize the hydrodynamic variables of the system by following the strategy in Sec.~\ref{sec:DUCgeneral}. In Sec.~\ref{sec:StaticsHDCP}, based on the general discussion in Sec.~\ref{sec:GLgeneral}, we study the static universality class of the system. In Sec.~\ref{sec:dynamicsHDCP}, we apply the formulation in Sec.~\ref{sec:general Langevin theory}  and study its dynamic universality class. We conclude with Sec.~\ref{sec:conclusionHDCP} with some discussions. 

\section{Hydrodynamic variables}
\label{sec:HVsuper}
The hydrodynamic variables of the system can be summarized as follows: 
\begin{table}[h]
\begin{tabular}{cll}
(i)& the chiral condensate, &\, $\sigma\equiv \bar q q- \left<\bar q q\right>$;\\
(ii)& the baryon number density, &$n_{\rm B}\equiv  \bar q \gamma^0 q- \left<\bar q \gamma^0 q\right>$;\\
(iii)& the energy-momentum densities, &\ \ $\varepsilon \equiv T^{00}-\left<T^{00} \right>$ and $\pi^i\equiv \left<T^{0i} \right>$;\\
(iv)&the superfluid phonon, & \ \ $\theta$.
\end{tabular}
\end{table}

Some remarks are in order here: the chiral condensate $\sigma$ is defined as the scaler channel of the general chiral order parameter (\ref{eq:chiralOP}), namely Eq.~(\ref{eq:chiral condensate 3flavor}); the baryon number density $n_{\rm B}$ is the conserved charge density for the baryon number symmetry under its transformation, Eq.~(\ref{eq:baryon}); the superfluid phonon $\theta$ is the Nambu-Goldstone mode associated with the spontaneous braking of the baryon number symmetry, Eq.~(\ref{eq:baryonSSB}). The superfluid phonon $\theta$ corresponds to the phase degrees of freedom of $\phi$ in Eq.~(\ref{eq:diquark 3flavor}). Among others, $\sigma$, $n_{\rm B}$, and $\varepsilon$ are defined as the fluctuations around the equilibrium. Note that the hydrodynamic variables in the system near the high-temperature QCD critical point are the same as (i)$ - $(iii) of the high-density critical point.

We note that the following variables are not hydrodynamic variables. The first one is the amplitude fluctuation of the diquark condensate, or the fluctuation of $|\phi|$ in Eq.~(\ref{eq:diquark 3flavor}). This is because, the high-density QCD critical point is characterized by the massless chiral condensate $\sigma$, where the diquark condensate is nonvanishing. The second ones are the Nambu-Goldstone modes associated with the chiral symmetry breaking in Eq.~(\ref{eq:CFL}), e.g., the pions. As we mentioned under Eq.~(\ref{eq:masstermRL}), the chiral symmetry is explicitly broken by the finite quark mass. It follows that the pions acquire finite masses and can be integrated out (or coarse grained) according to the general discussion in Sec.~\ref{sec:DUCgeneral}. The third ones are the gluons which are gapped due to the color Meissner effect in the color superconducting phase.% In a similar manner to the pions, the massive gluons are also absent in our effective theory. 

As we will give a discussion in Sec.~\ref{sec:conclusionHDCP}, $\varepsilon$ and $\pi^i$ do not affect the static and dynamic critical phenomena of the system. Therefore, we first neglect $\varepsilon$ and $\pi^i$ and consider only $\sigma$, $n_{\rm B},$ and $\theta $ as hydrodynamic variables in the following sections.

\section{Static critical phenomena}
\label{sec:StaticsHDCP}
We first construct the Ginzburg Landau theory by following the argument in Sec.~\ref{sec:GLgeneral}. The general Ginzburg-Landau free energy consistent with the symmetries of the system, such as the chiral symmetry, the $\U(1)_{\rm B}$ symmetry, and the discrete $\mathcal{ C} \mathcal{ P}\mathcal{ T}$ symmetries (charge conjugation, parity, and time-reversal symmetries, respectively) is given by
\begin{align}
\label{eq:GL}
F[\displaystyle \sigma,n_{\rm B},\theta]
=\int \dd \bm{r} \biggl[
\frac{a}{2}({\bm \nabla}\sigma)^{2}+b'{\bm \nabla}\sigma \cdot {\bm \nabla}n_{\rm B}+\frac{c}{2}({\bm \nabla}n_{\rm B})^{2}  +\frac{\rho}{2}({\bm \nabla}\theta)^{2}+V(\sigma,n_{\rm B} )  \biggr] \,, 
\end{align}
where
\begin{align}
\label{eq:VHDCP}
V(\displaystyle \sigma,n_{\rm B})=\frac{A}{2}\sigma^{2}+B\sigma n_{\rm B}+\frac{C'}{2}n_{\rm B}^{2}  +\cdots .
\end{align}
Here $a,b',c,\rho,A,B,C'$ are the expansion parameters and 
``$\cdots$'' denotes higher order terms. The prime notation is used to distinguish from other characters introduced so far. Note that $b'$ and $B$ can be finite only when $m_{\rm q}\neq 0$ and $\mu_{\rm B}\neq 0$, otherwise the term $\propto \sigma n_{\rm B}$ is prohibited by the chiral symmetry and the charge conjugate symmetry.

An important observation here is that the superfluid phonon $\theta$ is decoupled from the other hydrodynamic variables $\sigma$ and $n_{\rm B}$ at the Gaussian level. This is because  the time-reversal symmetry prohibits the mixing between $\theta$ and $\sigma$ (or $n_{\rm B}$). Beyond the mean-field level, one can find that the nonlinear couplings between $\sigma$ and $\theta$, e.g., $\sigma^2(\na \theta)^2$ are irrelevant at the Wilson-Fisher fixed point at $d=3$.%
\footnote{Note that the couplings between $\theta$ and $\sigma$ or $n_{\rm B}$ contain the derivative due to the $\U(1)_{\rm B}$ symmetry of the system, i.e., the invariance under $\theta \ra \theta +\alpha_{\rm B}$.} Therefore, the static property of the system is only determined by $\sigma$ and $n_{\rm B}$, even when one takes into account higher-order terms. It follows that the static universality class of the high-density critical point is the same as that of the high-temperature critical point. In particular, we obtain the following static correlation functions:
\begin{align}
\label{eq:corr}
\displaystyle \left\langle\sigma(\bm{r})\sigma({\bm 0})\right\rangle&=\frac{1}{4\pi r} e^{-r/\xi}\,, \\
\label{eq:chi}
\chi_{\rm B} &\equiv \frac{\d n_{\rm B}}{\d \mu_{\rm B}} 
= \frac{1}{V T}\left\langle n_{\rm B}^{2}\right\rangle_{\bm{q}\rightarrow {\bm 0}}=\frac{A}{\Delta}\,,
\end{align}
where $V$ and $T$ being spatial volume and temperature of the system, respectively. We also define
\begin{align}
\label{eq:Delta}
\xi \sim \Delta^{-\frac{1}{2}},\quad
\Delta\equiv AC'-B^2.
\end{align}
The critical point is characterized by $\Delta\ra 0$, so that the correlation length of $\sigma$, $\xi$, diverges. Because of the mixing between $\sigma$ and $n_{\rm B}$ mentioned under Eq.~(\ref{eq:VHDCP}), $\chi_{\rm B}$ and the susceptibility for the chiral condensate, $\chi_{m_{\rm q}}$, diverge as 
\begin{align}
\label{eq:suscptibilities}
\chi_{\rm B} & \sim \xi^{2-\eta},\\
\chi_{m_{\rm q}} &\equiv \frac{\d \sigma }{\d m_{\rm q}} \sim \xi^{2-\eta}.
\end{align}
Here, the anomalous dimension $\eta$ is determined by the static universality class of the system $(\eta\simeq 0.04$ including non-Gaussian effects \cite{Chaikin}).

We here give the reason why the high-density QCD critical point belongs to the same static universality class as that of the Ising model. Since $\sigma$ and $ n_{\rm B}$ are mixed in the potential, it is useful to change the basis in the $(\sigma,n_{\rm B})$ plane. We chose the basis such that the flat direction, $n'_{\rm B}\propto -B\sigma +A n_{\rm B}$, appears at the critical point $\Delta\ra 0$. After integrating out the gapped degrees of freedom orthogonal to $n'_{\rm B}$, we can obtain the free energy of the system as
\begin{align}
\label{eq:FnB}
F[\tilde n'_{\rm B}] =\int \dd \bm{r} \biggl[
\frac{a}{2}({\bm \nabla} n'_{\rm B})^{2}+\frac{A'}{2}n'^2_{\rm B}+ v n'^3_{\rm B} + u' n'^4_{\rm B} \biggr],
\end{align}
with some parameters $A'$, $v$, and $u'$. By taking the particular complete of square, we can map this functional into the same functional as that of the Ising model, namely, Eq.~(\ref{eq:GLW}).

\section{Dynamic critical phenomena}
\label{sec:dynamicsHDCP}
In this section, we study the dynamic critical phenomena of the system in the vicinity of the high-density QCD critical point. First, in Sec.~\ref{sec:LangevinHDCP}, we construct the nonlinear-Langevin equations of the high-density QCD critical point. Next, in Sec.~\ref{eq:HMHDCP}, we solve the linearized Langevin equation and obtain the hydrodynamic modes of the system. Then, in Sec.~\ref{sec:z}, we calculate the dynamic critical exponent, which determines the dynamic universality class.

\subsection{Langevin theory}
\label{sec:LangevinHDCP}
 We can write down the Langevin equation of the system by following Eq.~(\ref{eq:Nonlinear Langevin}) as
\begin{align}
\label{eq:Langesigma}
\frac{\d \sigma}{\d t}&=-\Gamma\frac{\delta F}{\delta\sigma}+\tilde{\lambda}\bm{\nabla}^{2}\frac{\delta F}{\delta n_{\rm B}}+\xi_{\sigma}\,,
\\
\label{eq:LangenB}
\frac{\d n_{\rm B}}{\d t}&=\tilde{\lambda}\bm{\nabla}^{2}\frac{\delta F}{\delta\sigma}+\lambda\bm{\nabla}^{2}\frac{\delta F}{\delta n_{\rm B}} -\int_V \displaystyle[n_{\rm B},\theta]\frac{\delta F}{\delta\theta}+\xi_{n_{\rm B}}\,,
\\
\label{eq:Langetheta}
\frac{\d \theta}{\d t }&= - \int_V [\theta, n _{\rm B}]\frac{\delta F}{\delta n_{\rm B}}-\zeta\frac{\delta F}{\delta\theta}+\xi_{\theta}\,.
\end{align}
Here, we use the simplified notation of the integral,
\begin{align}
\label{eq:simple}
\int_V [A,B]\frac{\delta F}{\delta B} \equiv 
\int \dd \rv' [A(t,\rv),B(t,\rv')]\frac{\delta F}{\delta B(t,\rv')}.
\end{align}
The kinetic coefficients $\Gamma$, $\lambda$, and $\tilde \lambda$ originate from the expansion~(\ref{eq:gammaexpansion}) and are related to the noise terms $\xi_{\sigma}$, $\xi_n$, and $\xi_{\theta}$ by the fluctuation-dissipation relations corresponding to Eq.~(\ref{eq:noisecorgeneral}), 
\begin{align}
\left<  \xi_\sigma (t,\rv)  \xi_\sigma (t,\rv')  \right>  &= 2T \Gamma  \delta (t-t') \delta^d(\rv -\rv'),\\
\left<  \xi_{n_{\rm B}} (t,\rv)  \xi_{n_{\rm B}} (t,\rv')  \right>  &= -2T \lambda  \na^2 \delta (t-t') \delta^d(\rv -\rv'),\\
\left<  \xi_\sigma (t,\rv)  \xi_{n_{\rm B}} (t,\rv')  \right>  & = -2T \tilde \lambda  \na^2 \delta (t-t') \delta^d(\rv -\rv'),\\
\left<  \xi_\theta (t,\rv)  \xi_\theta (t,\rv')  \right>  &= 2T \zeta  \delta (t-t') \delta^d(\rv -\rv').
\end{align}
Here, $d$ denotes the spatial dimension. 

We postulate the following Poisson bracket, 
\begin{align}
\label{eq:ntheta}
[\theta(t,\bm{r}),n_{\rm B}(t',\bm{r}')]&=\delta(t-t')\delta^d(\bm{r}-\bm{r}'),
\end{align} 
so that $n_{\rm B}$ and $\theta$ satisfy the canonical conjugate relation. The reason why these variables are canonical conjugate can be understood as follows. We start from the QCD Lagrangian at finite baryon chemical potential,
\begin{align}
\mathcal L'_{\rm QCD} &= 
\mathcal L_{\rm QCD} + \mu_{\rm B} \bar q \gamma^0 q.
\end{align}
We regard $\mu_{\rm B}$ as an auxiliary gauge field $a^\mu = (\mu_{\rm B},\zev)$, and rewrite the Lagrangian into 
\begin{align}
\mathcal L'_{\rm QCD} &= 
\mathcal L_{\rm QCD} + a_\mu \bar q \gamma^\mu q,
\end{align}
which has the gauge invariance under the local $\U(1)_{\rm B}$ transformation,
\begin{align}
\label{eq:gaugeHDCP}
a_\mu \ra a_\mu -\d_\mu \alpha_{\rm B},\quad \theta\ra \theta +\alpha_{\rm B}.
\end{align}
Next, we consider the effective Lagrangian ${\cal L}_{\rm eff}$ of the superfluid phonon $\theta$. Since ${\cal L}_{\rm eff}$ should also possess the local gauge symmetry (\ref{eq:gaugeHDCP}), the expression of ${\cal L}_{\rm eff}$ is dictated by the covariant derivative 
\begin{align}
D_\mu \theta = (\d_t \theta+a^0,\na \theta).
\end{align}
Thus, we find
\cite{Weinberg:1996kr, Son:2002zn}
\begin{align}
\label{eq:LeffHDCP}
{\cal L}_{\rm eff} = {\cal L}_{\rm eff}\left( \frac{\d \theta}{\d t} + \mu_{\rm B}, {\bm \nabla} \theta\right),
\end{align}
which yields the canonical conjugation relation between $n_{\rm B}$ and $\theta$, 
\begin{align}
n_{\rm B} \equiv \frac{\delta {\cal L}_{\rm eff}}{\delta \mu_{\rm B}} = \frac{\delta {\cal L}_{\rm eff}}{\delta \dot \theta},\quad  \dot\theta \equiv \dis \frac{ \d \theta}{ \d t}.
\end{align}

Some remarks on the Langevin equations (\ref{eq:Langesigma})--(\ref{eq:Langetheta}) are in order here. The derivative expansions for the first two terms of Eq.~(\ref{eq:LangenB}) start from $\na^2$ so that the total baryon number is conserved $\dot n_{\rm B}\ra0$, (${\bm q} \rightarrow {\bm 0}$). The coefficient of the second term of Eq.~(\ref{eq:Langesigma}) should have the same kinetic coefficient as that of the first term in Eq.~(\ref{eq:Langen}), according to the Onsager's principle (see Ref.~\cite{Landau:1980mil} for the derivation of the Onsager's principle).

By substituting Eq.~(\ref{eq:ntheta}) and the variation of the free energy to the linear level of hydrodynamic variables,
\begin{align}
\label{eq:sigmaconj}
\displaystyle \frac{\delta F}{\delta\sigma}&=(A-a\bm{\nabla}^{2})\sigma+(B-b'\bm{\nabla}^{2})n_{\rm B}, \\
\displaystyle \frac{\delta F}{\delta n_{\rm B}}&=(B-b'\bm{\nabla}^{2})\sigma+(C'-c\bm{\nabla}^{2})n_{\rm B}, \\
\label{eq:thetaconj}
\displaystyle \frac{\delta F}{\delta\theta}&=-\rho\bm{\nabla}^{2}\theta,
\end{align}
into Eqs.~(\ref{eq:Langesigma})--(\ref{eq:Langetheta}), we obtain the Langevin equations to the order of $O({\bm q}^2)$ in frequency-momentum space $(\omega, {\bm q})$,    
\begin{align}
\label{eq:Mx=0}
\mathcal {M} 
\left(\begin{array}{c}
\sigma\\
n\\
\theta
\end{array}\right)=0\,,
\end{align}
where 
\begin{align}
\mathcal{M} \equiv\left(\begin{array}{ccc}
\i\omega-\Gamma A-(\Gamma a+\tilde{\lambda}B)\bm{q}^{2} &-\Gamma B-(\Gamma b'+\tilde{\lambda}C')\bm{q}^{2}&0\\
-(\tilde{\lambda}A+\lambda B)\bm{q}^{2}& \i\omega-(\tilde{\lambda}B+\lambda C')\bm{q}^{2} & \rho\bm{q}^{2}\\
-B-b'\bm{q}^{2} &-C'-c\bm{q}^{2} &\i\omega-\zeta \rho \bm{q}^{2}
\end{array}\right).
\end{align}
We here omit the noise terms because these are not important in the following argument, where we only need the expressions of hydrodynamic modes.

\subsection{Hydrodynamic modes}
\label{eq:HMHDCP}
\label{sec:mode}
We obtain the hydrodynamic modes of the system by solving the proper equation, $\det \mathcal{M}=0$. 
This equation reduces to 
\begin{align}
\label{eq:prop}
\omega^{3}+\i(x_{1}+x_{2}\bm{q}^{2})\omega^{2}-y\bm{q}^{2}\omega-\i z\bm{q}^{2}=0\,,
\end{align}
where
\begin{align}
x_{1}&\equiv\Gamma A\,, \nonumber \\
x_{2}&\equiv\Gamma a+2\tilde{\lambda}B+\lambda C'+\zeta \rho\,, \nonumber \\
y&\equiv\Gamma\lambda\Delta+C'\rho+\Gamma\zeta A\rho\,, \nonumber \\
z&\equiv\Gamma \rho\Delta\,.
\end{align}
In Eq.~(\ref{eq:prop}), we ignore the higher-order terms of ${\bm q}$. At this order, the left-hand 
side of Eq.~(\ref{eq:prop}) can be factorized as
\begin{align}
& \ \ \ \left[\omega+\i x_{1}+\i\left(x_{2}-\frac{y}{x_{1}}+\frac{z}{x_{1}^{2}}\right)\bm{q}^{2} + O(\bm{q}^{3}) \right] 
\nonumber \\
& \times \left[\omega-\sqrt{\frac{z}{x_{1}}}|\bm{q}|+\frac{\i}{2}\left(\frac{y}{x_{1}}-\frac{z}{x_{1}^{2}}\right)\bm{q}^{2}
+ O(\bm{q}^{3}) \right]
\nonumber \\
& \times \left[\omega+\sqrt{\frac{z}{x_{1}}}|\bm{q}|+\frac{\i}{2}\left(\frac{y}{x_{1}}-\frac{z}{x_{1}^{2}}\right)\bm{q}^{2}
+ O(\bm{q}^{3}) \right]=0\,.
\end{align}
Form this factorized form, one can find three hydrodynamic modes of the system near the high-density QCD critical point: the relaxation mode and the pair of the superfluid phonons with the dispersion relations,
\begin{align}
\omega_{1}&=-\i\Gamma A+O(\bm{q}^{2})\,,\\
\label{eq:omega23}
\omega_{2,3}&=\pm c_{\rm s}|\bm{q}|+O(\bm{q}^{2})\,,
\end{align}
respectively. Here,
\begin{align}
\label{eq:csSF}
c_{\rm s}\equiv \displaystyle\sqrt{\frac{\rho}{\chi_{\rm B}}}
\end{align}
is the speed of the superfluid phonon, whose existence is dictated by the spontaneously symmetry breaking of the $\U(1)_{\rm B}$ symmetry, Eq.~(\ref{eq:baryonSSB}). We here find the critical slowing down of the speed of the superfluid phonon, $c_{\rm s} \rightarrow 0$. Note that we have shown the divergence of $\chi_{\rm B}$ when the critical point is approached $\xi\ra \infty$,  in Eq.~(\ref{eq:suscptibilities}).

\subsection{Dynamic critical exponent}
\label{sec:z}
We can estimate the dynamic critical exponent $z$ from $\omega \sim \xi^{-z}$ by using Eq.~(\ref{eq:omega23}),
\begin{align}
\label{eq:zdef}
c_{\rm s}\sim\xi^{1-z}.
\end{align}
Here, we use the matching condition at $\xi q \sim1$ between the critical regime ($\xi q\gg1$) and the hydrodynamic regime ($\xi q\ll1$).  %The value of $z$ characterizes the dynamic universality class of the system. 

To determine the value of $z$, we use the following $\xi$ dependences of $\rho$ and $\chi_{\rm B}$ near the high-density QCD critical point:
\begin{align}
\label{eq:paraxi}
\rho\sim\xi^{0}, \quad \chi_{\rm B}\sim\xi^{2-\eta}.
\end{align}
Here, $\rho$ is the stiffness parameter (or the ``decay constant'' for the superfluid phonon). Note that it does 
not depend on $\xi$ near the high-density critical point, which is generally away from the superfluid phase 
transition characterized by the amplitude mode of the diquark condensate.

From Eqs.~(\ref{eq:csSF}), (\ref{eq:zdef}), and (\ref{eq:paraxi}), we find the dynamic critical 
exponent $z$ as
\begin{align}
\label{eq:z}
z = 2 - \frac{\eta}{2}\,.
\end{align}
This dynamic critical exponent is different from all the exponents reported in the conventional classification by Hohenberg and Halperin 
\cite{Hohenberg:1977ym}. In this sense, we find that the high-density QCD critical point belongs to a new 
dynamic universality class.

\section{Conclusion and discussion}
\label{sec:conclusionHDCP}
So far, we have constructed the low-energy effective theory near the high-density QCD critical point and studied its static and dynamic critical phenomena. We have first shown that the static universality class of the system is the same as that of the high-temperature critical point, in Sec.~\ref{sec:StaticsHDCP}. From the speed of the superfluid phonon, $c_{\rm s}$, obtained in Eq.~(\ref{eq:csSF}), we found the critical slowing down of $c_{\rm s}$ in the vicinity of the critical point. Furthermore, we have calculated the dynamic critical exponent of the system, $z$, in  Eq.~(\ref{eq:z}), and found that the high-density QCD critical point belongs to a new dynamic universality class beyond the conventional Hohenberg and Halperin's classification \cite{Hohenberg:1977ym}. 

Let us discuss nonlinear effects on the dynamic critical phenomena. In general, there are nonlinear corrections to kinetic coefficients when a critical point is approached. Nevertheless, we can argue that the expression for the dynamic critical exponent (\ref{eq:z}) is an exact relation, which does not receive any renormalization effects due to $\U(1)_{\rm B}$ symmetry of the system. In fact, when we rewrite (\ref{eq:ntheta}) with some coupling constant $g_0$
\begin{align}
\label{eq:thetancano2}
[\theta(\bm{r}),n(\bm{r}')]&=g_0 \delta(\bm{r}-\bm{r}'),
\end{align}
the speed of the superfluid phonon, Eq.~(\ref{eq:csSF}) becomes
\begin{align}
\label{eq:csSF2}
c_{\rm s}\equiv \displaystyle\sqrt{\frac{g_0^2\rho}{\chi_{\rm B}}}.
\end{align}
Since Eq.~(\ref{eq:thetancano2}) is related to the $\U(1)_{\rm B}$ symmetry,  $g_0$ is not affected by any renormalization effects. Thus, the dynamic critical exponent obtained by using $c_{\rm s}$ is an exact relation. As is also shown in Appendix \ref{chap:EM}, the speed of the superfluid phonon, Eq.~(\ref{eq:cs2}), does not include any kinetic coefficients when the energy and momentum densities are taken into account.

Why the high-density QCD critical point belongs to a new dynamic universality class beyond the conventional classification? First, the presence of the superfluid phonon in the high-density critical point leads to a different dynamic universality class from that of the high-temperature QCD critical point. % regarding of the low-energy gapless modes which are generally essential  as we mentioned in Sec.~\ref{sec:DUCgeneral}. 
We next compare with the superfluid $\lambda$ transition of $^4$He, where the superfluid phonon also exists. Nevertheless, the interplay between the superfluid phonon and the chiral order parameter cannot be found in this condensed matter system. In fact, the second-order phase transition of the $\lambda$ transition is characterized by the massless superfluid gap.%
\footnote{Quantitatively, one can see a difference in the correlation length dependence of the stiffness parameter $\rho$, Eq.~(\ref{eq:paraxi}). Unlike the high-density QCD critical point, the phase transition is characterized by the superfluid gap in the helium system, so that the stiffness parameter depends on $\xi$ as $\rho\sim\xi^{-1}$ \cite{Hohenberg:1977ym}.}

Our findings suggest that the static quantities cannot distinguish two possible critical points in the QCD phase diagram, whereas the dynamic critical phenomena can distinguish them. Moreover, since the uniqueness of the dynamic critical phenomena of the high-density critical point is due to the presence of the superfluid phonon, observation of the critical slowing down of the superfluid phonon in the future heavy-ion collisions would provide indirect evidence of superfluidity of high-density QCD matter.

Other than the high-density QCD critical point, systems near critical points associated with $\mathcal G_1=\Z_2$ symmetry braking under $\mathcal G_2=\U(1)$ spontaneously symmetry breaking can also belong to this new dynamic universality class with appropriate background fields. The generalization of $\mathcal G_{1,2}$  enables us to classify the dynamic critical phenomena involving a general interplay between order parameters and Nambu-Goldstone modes.

%Since our analysis is based only on the symmetries and those  breaking patterns of the system, we can expect the same dynamic universality class in any $\U(1)$ broken systems near a critical point associated with $\Z_2$ symmetry breaking. 

\chapter{Dynamic critical phenomena induced by the chiral magnetic effect in QCD}
\chaptermark{Dynamic critical phenomena induced by CME}
\label{chap:DRGCME}
In this chapter, we study the interplay between the dynamic critical phenomena and the CME in QCD. In Sec.~\ref{sec:SetupCME}, we first explain our setup and summarize the symmetries and the hydrodynamic modes of the system. In Sec.~\ref{sec:FormCME}, we construct the Ginzburg-Landau theory and the nonlinear Langevin equation to describe the static and dynamic critical phenomena, respectively. We also give the MSRJD action corresponding to the Langevin theory and summarize its Feynman rules. In Sec.~\ref{eq:dRGCME}, we study the Ginzburg-Landau free energy and the MSRJD action by using the static and dynamic RG, respectively. We also show the results in Sec.~\ref{eq:dRGCME}. We conclude with Sec.~\ref{sec:CD}.

\section{Setup}
\label{sec:SetupCME}
\subsection{Symmetries}
We consider two-flavor QCD with massless up and down quarks at finite temperature $T$ and isospin chemical potential $\mu_{\rm I}$ in an external magnetic field ${\bm B}$. In this setup, there are two additional terms in the quark sector of massless two-flavor QCD Lagrangian. One is the finite isospin density term (the first term of Eq.~(\ref{eq:L'quark0})); the other is the coupling  to the background electromagnetic gauge field $A^\mu$ (the second term of Eq.~(\ref{eq:L'quark0})):
 \begin{align}
 \label{eq:L'quark0}
  \mathcal L' _{\rm quark} &= \mathcal L_{\rm quark}+ \mu_{\rm I} \bar q  \gamma^0 \tau^3 q + A_\mu \bar q  \gamma^\mu Q_{\rm e} q \\
  \label{eq:L'quark}
&= \mathcal L_{\rm quark}+ \frac{\mu_{\rm I}}{2} ( q^\dagger_{\rm u}    q_{\rm u}-q^\dagger_{\rm d}  q_{\rm d})+\frac{2}{3}A_\mu \bar q_{\rm u} \gamma^\mu q_{\rm u}-\frac{1}{3}A_\mu \bar q_{\rm u} \gamma^\mu q_{\rm u}.
 \end{align}
Here, $\mathcal L_{\rm quark}$ is the same Lagrangian given in Eq.~(\ref{eq:Lquark}) except for $m_{\rm u}=m_{\rm d}=0$ in the present case; $n_{\rm I} = \bar q \gamma^0 \tau^3 q$ and $\mu_{\rm I}$ are the isospin%
\footnote{The flavor space at $N_{\rm f}=2$ is specially called the isospin space in an analogy to the spin $1/2$. }
conserved-charge density and the isospin chemical potential, respectively; the third component of the generators for the $\SU(2)$ algebra, $\tau^3$ acts on the vector in the flavor space. In Eq.~(\ref{eq:L'quark}), we write the Dirac fields for the up and down quarks as $q_{\rm u}$ and $q_{\rm d}$, by using the vector notation of the flavor space:
\begin{align}
q=\left(
\begin{array}{c}
q_{\rm u}\\
q_{\rm d}
\end{array}
\right).
\end{align}
The electric charge matrix $Q_{\rm e}$ in Eq.~(\ref{eq:L'quark0}) are 
\begin{align}
Q_{\rm e}=\left(
\begin{array}{cc}
\dis \frac{2}{3} &0\\
0&\dis -\frac{1}{3} 
\end{array}
\right).
\end{align}
Note here that the up and down quarks possess the electric charge $2/3$ and $-1/3$ in the unit of the elementary charge. %From the expression of the term proportional to $\mu_{\rm I}$ in (\ref{eq:L'quark}),  $\mu_{\rm I}$ characterizes the chemical potential difference between up and down quarks.

In the presence of the electromagnetic gauge field (and/or the isospin chemical potential), the chiral symmetry Eq.~(\ref{eq:chiral}) is explicitly broken into its subgroup \cite{Shushpanov:1997sf}
\begin{align}
\label{eq:EB}
\SU(2)_{\rm L}\times \SU(2)_{\rm R} \ra \text{U(1)}^{\tau 3}_{\text{L}}\times\text{U(1)}^{\tau 3}_{\text{R}}.
\end{align}
Here we call $\text{U(1)}^{\tau 3}_{\text{L}}\times\text{U(1)}^{\tau 3}_{\text{R}}$ the partial chiral symmetry, which corresponds to the invariance under the following particular chiral transformation (corresponding to $\tau^3$),\footnote{
This transformation can be also written as
\begin{equation}
\label{eq:U(1)tau3A}
  q \rightarrow 
  q' =
  e^{-\i\alpha_{\text{V}} \tau^3} e^{-\i\alpha_{\text{A}} \tau^3\gamma^5} q ,
\end{equation}
where $\alpha_{\text{V}}$ and $\alpha_{\text{A}}$ denote the phase-rotating angles associated with $\text{U(1)}^{\tau^3}_{\text{V}}$ and $\text{U(1)}^{\tau^3}_{\text{A}}$ symmetries, respectively.}
\begin{align}
q_{\rm L}\rightarrow V^{\tau^3}_{\rm L} q_{\rm L},\quad 
q_{\rm L}\rightarrow V^{\tau 3}_{\rm R} q_{\rm R},\quad V^{\tau 3}_{\rm L,R}\equiv \e^{-\i \theta_{\rm L,R} \tau^3} q_{\rm L,R},
\end{align}
with $\theta_{\rm L,R} $ being the independent parameters.

Some remarks on our setup are in order here. First, we assume massless quarks so that possible quark-mass corrections to the CME can be ignored. Second, we consider finite $\mu_{\rm I}$ instead of baryon chemical potential $\mu_{\rm B}$, because both $n_{\rm I}$ and $n_{{\rm I}5}$ are conserved in massless QCD, so that $\mu_{\rm I}$ and $\mu_{{\rm I}5}$ are well-defined. On the other hand, the conservation of the axial charge  is violated by the axial anomaly, even in massless QCD, Eq.~(\ref{eq:QCDanomaly}).
%, and then the axial baryon chemical potential $\mu_{{\rm B}5}$ is not generically well-defined.%\footnote{One can readily see this fact in the wave equation of the CMW (\ref{eq:waveeqCMW}) for the baryon charge. That equation should receives correction from the gluon effects due to the axial anomaly. Unlike the baryon charge case, such an issue does not occur for the isospin charge.} %Therefore, in our setup, whether the CME affects the dynamic critical phenomena in QCD becomes a theoretically well-posed question.  

\subsection{Hydrodynamic variables}
\label{sec:HVs}
We here order the hydrodynamic variables of the system. First one is the two-component order parameter field characterizing the symmetry breaking of the partial chiral symmetry $\text{U(1)}^{\tau 3}_{\text{L}}\times\text{U(1)}^{\tau 3}_{\text{R}}$,
$\phi_\alpha$ ($\alpha=1,2$), i.e.,
\begin{align}
\label{eq:phispace}
\phi=\left(
\begin{array}{cc}
\sigma\\ \pi^3
\end{array}
\right),
\end{align}
with $\sigma = \bar q \tau^0 q$ and $\pi^3 = \bar q \i \gamma^{5} \tau^{3} q$ being the chiral condensate and the neutral pion, respectively ($\tau^0 = 1/2$).
We can show that the other variables in the chiral order parameter $\Phi$ in Eq.~(\ref{eq:chiralOP}) are not hydrodynamic modes as follows. At $N_{\rm f}=2$, the chiral order parameter $\Phi$ can be decomposed
 \begin{align}
\Phi= \sigma\tau^0 + \i \eta\tau^0 + \delta^a \tau^a + \i \pi^a \tau^a,
 \end{align}
where $\sigma = \bar q \tau^0 q$, $\eta = \bar q \i\gamma_5 \tau^0 q$, $\delta^a = \bar q \tau^{a} q$, and $\pi^a = \bar q \i \gamma^{5} \tau^{a} q$; we take the sum over repeated indices. In the absence of the magnetic field, $\sigma$ and $\pi^a$ become massless near the second-order chiral phase transition, while $\eta$ and $\delta^a$ acquire finite masses due to the $\text{U(1)}_\text A$ anomaly \cite{Pisarski:1983ms}.\footnote{Actually, in the presence of the $\text{U(1)}_\text A$ anomaly, the Ginzburg-Landau potential for the chiral order parameter $\Phi$ includes the so-called  Kobayashi-Maskawa-'t\,Hooft interaction \cite{tHooft:1976rip,Kobayashi:1970ji},
\begin{align}
-\frac{c}{2}(\det \Phi +\det \Phi^\dagger )= -\frac{c}{2}[\sigma^2+(\pi^a)^2] +\frac{c}{2}[\eta^2 +(\delta^a)^2],
\end{align} 
with $c$ being some positive parameter. By taking into account the mass terms 
\begin{align}
\frac{a}{2}{\rm Tr} \Phi^\dagger \Phi =\frac{a}{2}[\sigma^2+(\pi^a)^2+\eta^2 +(\delta^a)^2],
\end{align}
only $\sigma$ and $\pi^a$ can be massless at the phase transition point.
}
When we switch on the  magnetic field, the charged pions $\pi^{1,2}$ also acquire a mass proportional to $\sqrt{ |\Bv|}$ due to the explicit chiral symmetry breaking (\ref{eq:EB}). We can integrate out such massive degrees of freedom.
% according to the general idea to study dynamic critical phennomena in Sec.~\ref{sec:DUCgeneral}.

The second hydrodynamic variable is the conserved charge densities. We here only take into account the conserved charge densities associated with the symmetry $\text{U(1)}^{\tau 3}_{\text{L}}\times\text{U(1)}^{\tau 3}_{\text{R}}$ which are coupled to $\phi_\alpha$, i.e., $n_{\rm I}$ and the axial isospin density $n_{{\rm I}5} = \bar q \gamma^0 \gamma_5 \tau^3 q$. Although the energy and momentum densities can be also coupled to these hydrodynamic variables, we only focus on $\phi_\alpha$, $n_{\rm I}$, and $n_{\rm I5}$.

From now on, we omit the subscripts of $n_{\rm I}$, $n_{{\rm I}5}$, $\mu_{\rm I}$, and $\mu_{{\rm I}5}$ and rewrite these variables into $n$, $n_5$, and $\mu$ for notational simplicity.

\section{Formulation}
\label{sec:FormCME}
In this section, we give the formulation to study the static and the dynamic universality classes of the system. In Sec.~\ref{sec:LangevinCME}, we first derive the nonlinear Langevin equations following general arguments in Sec.~\ref{sec:general Langevin theory}. We also construct the Ginzburg-Landau theory in Sec.~\ref{sec:LangevinCME}. In Sec.~\ref{sec:DPTCME}, we show the MSRJD field theory equivalent to the derived Langevin equations with the help of  Sec.~\ref{sec:DPTgeneral}.

\subsection{Langevin theory}
\label{sec:LangevinCME}
Applying the general formation of the Langevin equation (\ref{eq:Nonlinear Langevin}) to this system, we obtain the  nonlinear Langevin equations for the hydrodynamic variables $\phi_\alpha$, $n$, and $n_5$ as follows:
\begin{align}
 \label{eq:Langephi}
 \frac{\partial \phi_\alpha}{\partial t}
 &= -\Gamma\frac{\delta F}{\delta \phi_\alpha} 
 - g\int_V \left[\phi_\alpha,n_5\right] 
 \frac{\delta F}{\delta n_5 } 
 + \xi_\alpha,\\
 \label{eq:Langen}
 \frac{\partial n}{\partial t}
 &= \lambda \na^2\frac{\delta F}{\delta n}
 -\int_V \left[n, n_5 \right] \frac{\delta F}{\delta n_5 }
 + \zeta,\\
 \label{eq:Langen5}
 \frac{\partial n_5}{\partial t}
 &= \lambda_5\na^2\frac{\delta F}{\delta n_5}
  - g\int _V \left[n_5 , \phi_\alpha \right] \frac{\delta F}{\delta \phi_\alpha}  - \int _V \left[n_5, n\right] \frac{\delta F}{\delta n}
 + \zeta_5,
\end{align}
We here use the simplified notation of the integrals (\ref{eq:simple}). The Ginzburg-Landau free energy $F$ is given by%\footnote{Note that $F$ is a functional of charge densities $n,n_5$ at finite fixed $\mu$. One can also write down the free energy $F'$ which is a functional of $\mu$ and is connected to $F$ by the Legendre transformation: $F=F'-\int \dd \rv \mu \bar n$ with $\bar n\equiv \delta F'/\delta \mu$. While $F'$ can involve linear terms of $\mu$, such as $-\int \dd \rv \mu \varepsilon_{\alpha\beta}\Bv \cdot(\na \phi_\alpha)\phi_\beta$, $F$ does not contain such terms, because of the cancellation due to the Legendre transformation.}
\begin{align}
 \label{eq:GL}
 F&=\int \dd \bm{r} \left[ \frac{r}{2}(\phi_\alpha)^2+
 \frac{1}{2}({\bm \nabla}\phi_\alpha)^{2}+u(\phi_\alpha)^2(\phi_\beta)^2+\frac{1}{2\chi}n^2+\frac{1}{2\chi_5}n_5^2+\gamma n\phi_\alpha^2\right],
\end{align}
%where its derivation is briefly reviewed around Eq.~(\ref{eq:Fgeneral}) and 
where we take the summations over repeated indices; $r,u$ and $\gamma$ are some expansion parameters; the isospin and axial isospin susceptibilities, $\chi$ and $\chi_5$, are defined as 
\begin{align}
 \label{chi}
 \chi \equiv \frac{\partial n}{\partial \mu}, \qquad 
 \chi_5 \equiv \frac{\partial n_5}{\partial \mu_5}.
\end{align}
The last term of Eq.~(\ref{eq:GL}), $\gamma n \phi_\alpha^2$, is forbidden at $\mu = 0$ by the charge conjugation symmetry, whereas it can appear at $\mu \neq 0$. Returning to the Langevin equations~(\ref{eq:Langephi})--(\ref{eq:Langen5}), $\Gamma,\lambda$, and $\lambda_5$ are the kinetic coefficients obtained by the derivative expansion (\ref{eq:gammaexpansion}), and $g$ is the coupling constant between $\phi_\alpha$ and  $n_5$. The noise terms $\xi_\alpha,\, \zeta$, and $\zeta_5$ satisfy the fluctuation-dissipation relations:
\begin{align}
\label{eq:FDROP}
\langle \xi_\alpha(\rv,t) \xi_\beta(\rv',t') \rangle&=2\Gamma\delta_{\alpha\beta}\delta(t-t')\delta^d(\rv-\rv'),\\
\label{eq:FDRn}
\langle \zeta (\rv,t) \zeta (\rv',t') \rangle&=-2\lambda\na^2\delta(t-t')\delta^d(\rv-\rv'),\\
\label{eq:FDRn5}
\langle \zeta_5 (\rv,t) \zeta_5 (\rv',t') \rangle&=-2\lambda_5\na^2\delta(t-t')\delta^d(\rv-\rv'),
\end{align}
and $\langle \xi_\alpha \zeta \rangle = \langle \xi_\alpha \zeta_5 \rangle= \langle \zeta \zeta_5 \rangle=0$.
We here postulate the following Poisson brackets from the symmetry algebra:
\begin{align}
 \label{eq:PBn5phi}
 \left [n_5 (\rv,t),\phi_\alpha(\rv',t) \right]
 &= \varepsilon_{\alpha\beta}\phi_\beta\delta(t-t')\delta^d(\rv-\rv'),\\
 \label{eq:PBn5n}
 \left [n (\rv,t),n_5(\rv',t) \right]
 &= C \Bv\cdot\na\delta(t-t')\delta^d(\rv-\rv').
\end{align}
Here, $\varepsilon_{\alpha\beta}$ denotes the anti-symmetric tensor in the order parameter space~(\ref{eq:phispace}), and $C$ is usually related to the anomaly coefficient (\ref{eq:C}) or the CME coefficient. However, in our analysis near the second-order phase transition, we regard $C$ as a free parameter, which will be determined by the RG equation of the system. Since nonlinear fluctuations of massless $\sigma$ can potentially renormalize $C$, it is a nontrivial question whether $C$ is exactly fixed, related to the nonrenormalization away from the second-order phase transition mentioned under Eq.~(\ref{eq:C}). Nevertheless, in Sec.~\ref{sec:DynamicsCME}, we will show that this anomaly coefficient or the CME coefficient does not receive the renormalization at the one-loop level. %Eq.~\eqref{eq:PBn5phi} can be interpreted as the classical limit of the corresponding quantum commutations. 

\subsection{Dynamic perturbation theory}
\label{sec:DPTCME}
We here summarize the MSRJD formulation of our Langevin theory. The field theoretical MSRJD action corresponding to the Langeinv theory in Sec.~\ref{sec:LangevinCME} is
\begin{align}
\label{eq:action}
S= \int \dd t \int \dd \rv \ (\mathcal L _ \phi +\mathcal L _n + \mathcal L _{\phi n}),
\end{align}
whose general form is given by Eq.~(\ref{eq:MSR}). We will explain each of the Lagrangian in Eq.~(\ref{eq:action}) and its brief derivation by assuming Eq.~(\ref{eq:MSR}) (see Sec.~\ref{sec:DUCgeneral} for the derivation of Eq.~(\ref{eq:MSR}) itself).

%, it is useful to determine $\mathcal F_{\rm I}[\psi]$ in Eq.~(\ref{eq:MSR}) (the definition of $\mathcal F_{I}[\psi] $ itself is given in Eq.~(\ref{eq:generalLangevin})). In particular, $\mathcal L_\phi $ corresponds to

First, $\mathcal L _ \phi $ represents the kinetic term of the order parameters $\phi_\alpha$ and the 4-point interaction term:
\begin{align}
\mathcal L_\phi  &= \tilde{\phi}_{\alpha}\left(\frac{\d }{\d t}+\Gamma(r-\na^{2})\right)\phi_{\alpha}-\Gamma\tilde{\phi}_{\alpha}^2+ 4\Gamma u \tilde{\phi}_\alpha \phi_\alpha \phi_\beta^2\,,
\end{align}
where $\tilde \phi _ \alpha $ denotes the responsible field for $\phi_\alpha$. Almost all of the terms come from $\mathcal F_{\phi_\alpha}[\psi]$ corresponding to the right-hand side of Eq.~(\ref{eq:Langephi}) (see Eq.~(\ref{eq:generalLangevin}) for the definition of $\mathcal F_{I}[\psi]$ with $\psi_{\rm I}=\phi_\alpha,n,n_5$), but the bilinear of the response field, $-\Gamma\tilde{\phi}_{\alpha}^2$, comes from the last term of Eq.~(\ref{eq:MSR}). The form of $\gamma_{IJ}(\na)$ in Eq.~(\ref{eq:MSR}) can be obtained by comparing the original definition of $\gamma_{IJ}(\na)$, Eq.~(\ref{eq:generalFDR}) and Eq.~(\ref{eq:FDROP}).
%Note also that the interaction terms among $\phi_\alpha$ and $\tilde \phi _ \alpha $ given by the last term will not be important in the following analysis. 

Next, $\mathcal L _n$ represents the bilinear part of the conserved charge densities $n$ and $n_5$, which are coupled to each other in the presence of the CME:
\begin{align}
\label{eq:Ln}
\mathcal L _n &=\frac{1}{2}(\tilde n, n, \tilde n_5, n_5) \mathcal M
\left(
\begin{array}{c}
\tilde n\\
n\\
\tilde n _5\\
n_5
\end {array}
\right),
\end{align}
where 
\begin{align}
\mathcal M \equiv 
 \left( \begin{array}{cccc}
2\lambda \na^2   
&  \dis \frac{\d }{\d t}-\frac{\lambda}{\chi} \na^2
&0
&\dis \frac{C}{\chi}_5 \Bv \cdot\na \\
%%%
-\dis \frac{\d }{\d t}-\frac{\lambda}{\chi} \na^2
&0
& \dis - \frac{C}{\chi}\Bv\cdot\na
&0 \\
%%%
0
& \dis \frac{C}{\chi} \Bv \cdot \na
&2 \lambda_5 \na^2
& \dis \frac{\d }{\d t}-\frac{\lambda_5}{\chi_5} \na^2\\
%%%
- \dis \frac{C}{\chi}_5 \Bv \cdot\na  
& 0
& -\dis \frac{\d }{\d t}-\frac{\lambda_5}{\chi_5} \na^2& 0
\end{array}
\right)
\end{align}
and $\tilde n$ and $\tilde n _ 5$ are the response fields for $n$ and $n_5$, respectively. Almost all terms come from $\mathcal F_{n}[\psi]$ and $\mathcal F_{n_5}[\psi]$ corresponding to the linear terms of the right-hand sides of Eqs.~(\ref{eq:Langen}) and (\ref{eq:Langen5}), respectively. Meanwhile, the bilinear terms of $\tilde n$ and $\tilde n_5$ originate from the last term of Eq.~(\ref{eq:MSR}). Here, Eqs.~(\ref{eq:FDRn}) and (\ref{eq:FDRn5}) give $\gamma_{nn}=\lambda \na^2$ and $\gamma_{n_5n_5}=\lambda_5 \na^2$, respectively.

The last term of Eq.~(\ref{eq:action}), $\mathcal L _{\phi n}$ represents the 3-point interaction between the order parameters $\phi_\alpha$ and the conserved charge densities $n$ and $n_5$:
\begin{align}
\label{eq:LOPCD}
\mathcal L _{\phi n}&=-\frac{g\varepsilon_{\alpha\beta}}{\chi_5}
\left(\tilde{\phi}_\alpha \phi_\beta n_5 +\chi_5 \tilde n_5 (\na^2 \phi_\alpha)\phi_\beta \right)
\notag\\
&\hspace{80pt}
+2\gamma\Gamma \tilde{\phi}_\alpha \phi_\alpha n
-\gamma\lambda\tilde n (\na^2\phi_\alpha^2)
+\gamma C\tilde n _5 \Bv\cdot(\na \phi^2_\alpha).
\end{align}
We obtain these terms from $\mathcal F_{I}[\psi]$ corresponding to the nonlinear terms among $\phi_\alpha$, $n$, and $n_5$ in the Langevin equations~(\ref{eq:Langephi})--(\ref{eq:Langen5}).

We make some remarks on the interaction term $\mathcal L _{\phi n}$. There are two types of interactions: the first line of Eq.~(\ref{eq:LOPCD}) ($\propto g$) originates from the Poisson bracket~(\ref{eq:PBn5phi}), the second line of Eq.~(\ref{eq:LOPCD}) ($\propto \gamma$) originates from the non-Gaussian term $\gamma n \phi_\alpha^2$ in the Ginzburg-Landau free energy (\ref{eq:GL}). The former ($\propto g$) exists even at $\mu=0$ and gives the couplings between different components of the order parameters field reflecting $\varepsilon_{\alpha\beta}$ in the interaction vortex; the latter ($\propto \gamma$) exists as long as $\mu\neq0$ and gives the couplings between the same order parameter components. Note also that the last term of Eq.~(\ref{eq:LOPCD}) may potentially generate the nonlinear corrections to the anomaly coefficient $C$. Nevertheless, we will show that this is not the case from the explicit computation.

\subsubsection{Feynman rules}
We here summarize the Feynman rules for the action~(\ref{eq:action}) (see Ref.~\cite{Tauber} to obtain the Feynman rules from the MSRJD action. See also standard textbooks of quantum field theory, e.g., Ref.~\cite{Peskin:1995ev} for getting the Feynman rules from a field theoretical action)

The bare propagator of the order parameter, $G ^{0}_{\alpha\beta}$, is obtained by calculating the 2-point correlation $\langle \phi_\alpha \tilde \phi_\beta \rangle$ from the Gaussian part of $\mathcal L _\phi$ in momentum space:
\begin{align}
\label{eq:G0op}
G^{0}_{\alpha\beta} (\bm{k},\omega) = G^{0} (\bm{k},\omega) \delta _{\alpha\beta}&\equiv\frac{\delta _{\alpha\beta}}{-\i\omega+\Gamma(r+\bm{k}^2)},
\end{align}
which is diagonal with respect to $\alpha$ and $\beta$. The bare propagator of the conserved fields, $D^{0}_{ij}$, is obtained by calculating the 2-point correlation $\langle n_i \tilde n_j \rangle$ from $\mathcal  L _n$, where $n_i =n,n_5$. In particular, the inverse matrix of $D^0_{ij}$ has the following expression:
\begin{align}
\label{eq:matrix}
\mathcal [D^0 (\kv,\omega)] ^{-1}=
\left(
\begin{array}{cc}
-\i\omega+\dis \frac{\lambda}{\chi}\kv^2
&\i \dis \frac{C}{\chi_5} \Bv\cdot\kv
\\
\i \dis \frac{C}{\chi}  \Bv\cdot\kv&-\i\omega + \dis \frac{\lambda_5}{\chi_5}\kv^2
\end{array}
\right)\,,
\end{align}
which has off-diagonal components with respect to $i$ and $j$ because of the CME. 

As we have briefly mentioned in Sec.~\ref{sec:dynamics}, there is a 2-point noise vortex corresponding to the bilinear of the response fields in the action. In particular, from the bilinear term of the response field of the order parameter $\tilde \psi_\alpha$ in $\mathcal L _\phi$, we obtain the bare noise vertex of the order parameter as $2\Gamma \delta_{\alpha\beta}$. Similarly, we get the bare noise vertex of the conserved charge densities as
\begin{align}
L^{0}(\kv)=
\left(
\begin{array}{cc}
2 \lambda \kv^2&0
\\0&2\lambda_5\kv^2
\end{array}
\right), 
\end{align}
which is related to the bare correlation function of the conserved charge densities, $B^0_{ij}$, through the following equation: 
\begin{align}
B^{0}_{ij}(\kv,\omega)&
=D_{il}^{0}(\kv,\omega)L_{lk}^{0}(\kv)[D^{0}(-\kv,-\omega)]^{T}_{kj}=D_{il}^{0}(\kv,\omega)L_{lk}^{0}(\kv)[D^{0}(\kv,\omega)]^{\dagger}_{kj}.
\end{align}
Here, $B^0_{ij}$ can be calculated explicitly by the 2-point correlation $\langle n_i n_j \rangle$ from $\mathcal L _n$ as follows:
\begin{align}
B^{0}_{11}(\kv,\omega)&=\frac{ 2\lambda \kv^2 (\omega^2+\lambda_5^2\kv^4/\chi_5^2)+2\lambda_5\kv^2(C\Bv\cdot\kv/\chi_5)^2}{\left|\det[ D^0(\kv,\omega)]^{-1}\right|^2},\\
B^{0}_{12}(\kv,\omega)&=B^{0}_{21}(\kv,\omega)=\frac{2\left(\lambda/\chi+\lambda_5/\chi_5\right) \kv^2 C(\Bv\cdot\kv)\omega}{\left|\det[ D^0(\kv,\omega)]^{-1}\right|^2},\\
B^{0}_{22}(\kv,\omega)&=\frac{2\lambda_5\kv^2(\omega^2+\lambda^2 \kv^4/\chi^2)+2\lambda \kv^2(C \Bv\cdot\kv/\chi)^2 }{\left|\det[ D^0(\kv,\omega)]^{-1}\right|^2}.
\end{align}

\begin{figure}[t]
\begin{center}
\subfigure[$U^0_{\alpha ;\beta \gamma \delta}$]{ \includegraphics[bb=0 0 127 112,height=3.5cm]{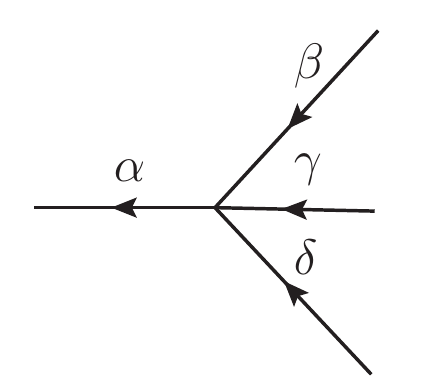}\label{Fig:phi}}~
\subfigure[$V^0_{\alpha;\beta i}$]{ \includegraphics[bb=0 0 115 98,height=3.5cm]{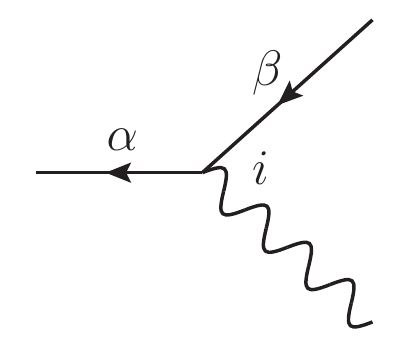}\label{Fig:phin}}~
\subfigure[$V^0_{i;\alpha \beta}$]{ \includegraphics[bb=0 0 140 108,height=3.5cm]{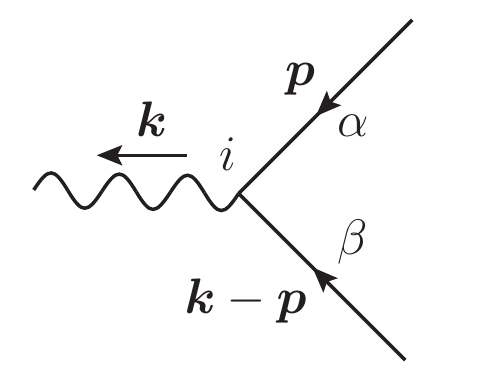}\label{Fig:phin2}}
\caption{Interaction vertices}
\label{fig:IV}
\end{center}
\end{figure}

The 4-point interaction vertex is obtained from $\mathcal L _{\phi}$ as 
\begin{align}
U^0_{\alpha ; \beta \gamma \delta}=-4u \Gamma \delta_{\alpha\beta} \delta_{\gamma \delta},
\end{align}
where the indices $\alpha ;\beta \gamma \delta$ are the shorthand notation of the fields $\tilde \phi _\alpha \phi _\beta \phi _\gamma \phi _\delta$. 
%From now on, we write the indices of the response field on the left side of a semicolon and the indices for the hydrodynamic fields on the right side of a semicolon. 
The 3-point interaction vertices are obtained as
\begin{align}
\label{eq:bareV}
V^0_{ \alpha; \beta i}
&=\left(
\begin{array}{c}
-2\gamma\Gamma\delta_{\alpha \beta}\\
g\varepsilon_{\alpha\beta}/\chi_5
\end{array}
\right),\\
V^0_{i; \alpha \beta }(\kv,\pv)
&=\left(
\begin{array}{c}
-2\gamma\lambda \kv^2 \delta_{\alpha \beta}\\
g[(\kv-\pv)^2-\pv^2]\varepsilon_{\alpha\beta}-2\i\gamma C \Bv \cdot \kv \delta_{\alpha \beta}
\end{array}
\right),
\end{align}
where we use the vector notation for the label $i$ which classifies $n_i=n,n_5$. The indices $\alpha; \beta i$ and $i; \alpha \beta $ are the shorthand notation of $\tilde  \phi _\alpha \phi_\beta n_i$ and  $\tilde n_i  \phi _\alpha \phi_\beta $, respectively. The vertex $V^0_{i; \alpha \beta }(\kv,\pv)$ is the function of the outgoing momentum $\kv$ of $n_i$ and the ingoing momentum $\pv$ of $\phi_\alpha$ (see also Fig.~\ref{Fig:phin2} for the configuration of the external momentum). 

\begin{figure}[t]
\begin{center}
\subfigure[$2\Gamma \delta_{\alpha\beta}$]{ \includegraphics[bb=0 0 88 111,height=3.5cm]{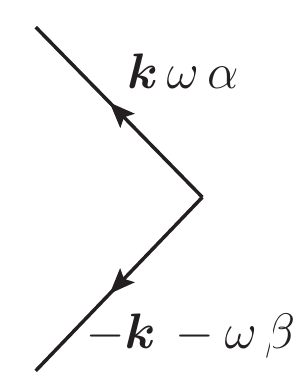}\label{Fig:a}}
\subfigure[$L^0_{ij}$]{ \includegraphics[bb=0 0 102 112,height=3.5cm]{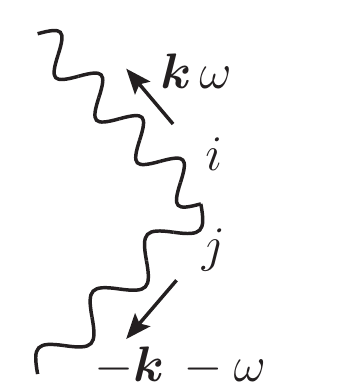}\label{Fig:b}}
\caption{Noise vertices}
\label{fig:NV}
\end{center}
\end{figure}

We next summarize the diagrammatic representations of the above propagators, the noise- and interaction- vortices. First, we depict $G^0_{\alpha\beta}$ by the plane line, and $D^{0}_{ij}$ by the wavy line with the outgoing and ingoing components $i$ and $j$. Since $G^0_{\alpha\beta}$ is diagonal with respect to $\alpha,\beta$ as one can confirm in Eq.~(\ref{eq:G0op}), we omit writing both of the outgoing and ingoing indices $\alpha$ and $\beta$ in the diagram for $G^0_{\alpha\beta}$. Instead, we shall write $\alpha$ alone at the center of plane lines. Each noise vertex of the order parameters and the conserved charge densities can be understood as the diagrams with two outgoing lines as represented in Fig.~\ref{fig:NV}. Note that the number of ingoing and outgoing lines in one of the diagrams corresponds to the power of the hydrodynamic variable $\psi=\phi_\alpha,n_i$ and the response field $\tilde \psi=\tilde \phi_\alpha,\tilde n_i$ in the action, respectively \cite{Tauber}. 
As is shown in Fig~\ref{fig:IV}, each of the interaction vertices has one outgoing and three or two ingoing lines. 

Ordinary Feynman rules are applied to obtain full $n$-point correlations. Among others, we obtain the full propagators $G_{\alpha\beta}$ and $D_{ij}$ by using the self-energies of the order parameter, $\Sigma_{\alpha\beta}$, and those of the conserved charge densities, $\Pi_{ij}$, as
\begin{align}
\label{eq:DysonG}
G^{-1}_{\alpha\beta}(\bm{k},\omega)&=[G^0_{\alpha\beta}(\bm{k},\omega)]^{-1}-\Sigma_{\alpha\beta}(\bm{k},\omega),\\
\label{eq:DysonD}
D^{-1}_{ij}(\bm{k},\omega)&= [D^0_{ij}(\bm{k},\omega)]^{-1} -\Pi_{ij}(\bm{k},\omega).
\end{align}
We also get the three-point vertex function $V_{\alpha;\beta i}(\kv_1,\kv_2,\omega_1,\omega_2)$  by computing one-particle irreducible diagrams with outgoing $\tilde \phi _ \alpha$ and ingoing $\phi_\beta,\, n_i$. Here, $\kv_1$ and $\omega_1$ denote the ingoing momentum and frequency of $\phi_\beta$, and $\kv_2$ and $\omega_2$ denote the ingoing momentum and frequency of $n_i$. From the energy and momentum conservation laws, $\tilde \phi_\beta$ has the outgoing momentum $\kv_3 = \kv_1+\kv_2$ and the frequency $\omega_3 = \omega_1+\omega_2$. For later purposes, it is convenient to divide $V_{\alpha;\beta i}$ into its bare contribution $V^0_{\alpha;\beta i}$ and the correction term $ \mathcal V _{\alpha;\beta i}$ as follows:
\begin{align}
\label{eq:mathcalVdef}
V_{\alpha;\beta i}(\kv_1,\kv_2,\omega_1,\omega_2)=V^0_{\alpha;\beta i}+\mathcal V _{\alpha;\beta i}(\kv_1,\kv_2,\omega_1,\omega_2).
\end{align}

\section{Renormalization-group analysis}
\label{eq:dRGCME}
In this section, we apply the RG to the effective theories derived in Sec.~\ref{sec:FormCME}. In Sec.~\ref{sec:statics}, we apply the static RG to the Ginzburg-Landau free energy~(\ref{eq:GL}) within the $\epsilon$ expansion. In Sec.~\ref{sec:DynamicsCME}, we study the MSRJD action (\ref{eq:action}) by using the dynamic RG with the help of Appendix~\ref{sec:calculation}. We remark those results in Sec.~\ref{sec:DUCCME}.

\subsection{Statics}
\label{sec:statics}
In a single RG transformation, we integrate out the degrees of freedom in the momentum shell $\Lambda/b<q<\Lambda$ (with $\Lambda$ being the momentum cutoff and $b>1$, $q\equiv|\qv|$), and we rescale the coordinate and the fields in the following way (For the details about the static RG, see Sec.~\ref{sec:StaticRGgeneral}):
\begin{align}
\label{eq:scalingcord}
\rv&\rightarrow \rv'=b^{-1}\rv,\\
\label{eq:scalingphi}
\phi_\alpha(\bm{r})&\rightarrow \phi'_\alpha(\bm{r}')=b^a \phi_\alpha(\bm{r}),\\n(\bm{r})&\rightarrow n'(\bm{r}')=b^c n(\bm{r}),\\
\label{eq:scalingn5}
n_5(\bm{r})&\rightarrow n'_5(\bm{r}')=b^{c_5} n_5 (\bm{r}).
\end{align} 
Here $a$,  $c$, and $c_5$ are some constants computed in the calculation below. Hereafter, we work with the spatial dimension $d\equiv4-\epsilon$ with small $\epsilon$ and perform the calculation to leading orders in the expansion of $\epsilon$ as we remarked around Eq.~(\ref{eq:epsilonexpansion}). 

Let us write the static parameters at the $l$th stage of the renormalization procedure, $r_l,\, u_l,\, \chi_l$ and $\gamma_l$. Then, these parameters satisfy the same recursion relation as that of the so-called model C for the system with a single (two-component) order parameter and the conserved charge densities~\cite{HHSI}. In particular, we quote Eqs.~(4.5)--(4.8) in  Ref.~\cite{HHSI}. The recursion relations for the static parameters in the leading-order of $\epsilon$ are given as
\begin{align}
r_{l+1}&=b^{d-2a}\{ r_l + 8\bar u _l [\Lambda^2(1-b^{-2})-2r_l \ln b]\}, \\
\label{eq:RRu}
\bar u_{l+1}&=b^{d-4a} \bar u_l \left( 1-40\bar u _l \ln b \right), \\
\label{eq:RRchi}
\chi_{l+1}^{-1}&=b^{d-2c}\chi_l^{-1}(1-4v_l \ln b),\\
\label{eq:RRchi5}
(\chi_5)_{l+1}^{-1}&=b^{d-2c_5}(\chi_5)_l^{-1},\\
\label{eq:RRgamma}
\gamma_{l+1}&=b^{2-c}\gamma_l \left[1- \left( 16 \bar u _l +4v_l \right) \ln b \right].
\end{align}
Here, the left-hand sides are the parameters at the $(l+1)$-th step of the RG, and we introduced the following quantities,
\begin{align}
\label{eq:Parastas}
v\equiv\frac{\gamma^2 \chi \Lambda^{-\epsilon}}{8\pi^2},\quad \bar u\equiv  \frac{u \Lambda^{-\epsilon}}{8\pi^2}.
\end{align}
Note that one may not find the equation corresponding to Eq.~(\ref{eq:RRchi5}) in Ref.~\cite{HHSI} which we quoted. Nevertheless, Eq.~(\ref{eq:RRchi5}) can be readily introduced because there is only Gaussian term for $n_5$ in the Ginzburg-Landau free energy~(\ref{eq:GL}) and $(\chi_5)_l$ is affected only by the trivial scale transformation (\ref{eq:scalingn5}). From the condition that $(\chi_5)_l$ remains finite at the fixed point, we find the value of $c$  from Eq.~(\ref{eq:RRchi5}) as
\begin{align}
\label{eq:c5}
c_5=\frac{d}{2}.
\end{align}
It is also worth writing that the recursion relations (\ref{eq:RRu})--(\ref{eq:RRgamma}) are the extension of Eqs.~(\ref{eq:r're}) and (\ref{eq:u're}) to include the effects of the conserved charge density $n$ and $n_5$. In particular, $v$ characterizes the nonlinear coupling between the order parameter and the conserved-charge density $n$.

Let us now compute $a$ and $c$. Because the anomalous dimension $\eta$ is zero to the order of $\epsilon$ \cite{Chaikin}, the exponent $a$ is solely determined by Eq.~(\ref{eq:zeta}) as
\begin{align}
\label{eq:a}
a=\frac{d-2}{2}\,.
\end{align}
To determine $c$, we evaluate the interplay of the RG flow between $v$ and $\bar u$. Combining Eqs.~(\ref{eq:RRchi}) and (\ref{eq:RRgamma}), we obtain the recursion relation for $v_l$,
\begin{align}
\label{eq:RRv}
v_{l+1}&=b^\epsilon v_l \left[1- \left( 32 \bar u_l  + 4v_l  \right) \ln b \right].
\end{align}
The RG equations corresponding to Eqs.~(\ref{eq:RRu}) and (\ref{eq:RRv}) become%
\footnote{
The derived recursion relations typically have the following form:
\begin{align}
A_{l+1} = b^{c_A} A_l (1+ B_l \ln b). \nonumber
\end{align}
with $A_l$ and $B_l$ being the variables and $c_A$ being some constant. One can derive the RG equation for $A_l$ by setting $b = e^l$ and taking the limit $l \rightarrow 0$ as 
\begin{align}
\frac{{\rm d}A_l}{{\rm d}l} = (c_A + B_l) A_l \nonumber
\end{align}}
\begin{align}
\label{eq:REu}
\frac{{\rm d}\bar u_l}{{\rm d}l} = (d - 4 a - 40 \bar u_l) \bar u_l. 
\\
\label{eq:REv}
\frac{{\rm d}v_l}{{\rm d}l} = (\epsilon - 32 \bar u_l  - 4v_l) v_l,
\end{align}
where Eq.~(\ref{eq:REu}) is the two-component order parameter version of Eq.~(\ref{eq:ubarRG}), and  Eq.~(\ref{eq:REv}) is the RG equation that  evaluates the coupling between the order parameter $\phi_\alpha$ and the conserved charge density $n$. 

Equations~\eqref{eq:REu}-\eqref{eq:REv} tell us the fixed-point values of $\bar u_l$ and $v_l$ as \cite{HHSI}
\begin{align}
\bar u_\infty= \frac{\epsilon}{40}, \quad
\label{eq:FVv}
\dis v_\infty = \frac{\epsilon}{20}\,.
\end{align}
Returning to Eq.~(\ref{eq:RRchi}), we arrive at
\begin{align}
\label{eq:c}
\dis c=\frac{3d-2}{5}\,.
\end{align}

\subsubsection{Static critical exponents}
The physical parameters near the transition temperature $T_c$ are solely given by the loop-correction terms proportional to $\ln b$ in the static recursion relations \cite{Onuki:PTD}. In particular, we obtain the renormalized isospin charge susceptibility $\chi(T)$  by taking into account the correction calculated in Eq.~(\ref{eq:RRchi}) to the bare value $\chi_0$ at the cutoff scale $\Lambda$,
\begin{align}
\label{eq:chiT}
\chi(T)=\chi_0 [1+4v_\infty \ln (\Lambda\xi)]\sim \xi^{\epsilon/5},
\end{align}
where we use the relation $1+x \ln \Lambda \xi +O(x^2)=(\Lambda \xi)^x$ for $x \ll 1$, by regarding $x=4 v_\infty$ as a small parameter when $\epsilon\ll 1$. 
Defining the critical exponents $\nu$ and $\alpha$ in the usual manner,
\begin{align}
\label{eq:eps}
\xi\sim\tau^{-\nu},\quad\chi\sim\tau^{-\alpha}, 
\end{align}
with $\dis\tau\equiv{(T-T_c)}/{T_c}$ being the reduced temperature, we obtain 
\begin{equation}
\label{eq:alpha/nu}
\dis \frac{\alpha}{\nu}=\frac{\epsilon}{5}\,. 
\end{equation}

\subsection{Dynamics}
\label{sec:DynamicsCME}
As we mentioned it around Eq.~(\ref{eq:dynamicrescalegeneral}), we also need to rescale the response fields $\tilde \psi = \tilde \phi_\alpha,\tilde n_i$ in addition to those for the hydrodynamic variables $\psi=\phi_\alpha,n_i$, (\ref{eq:scalingphi})--(\ref{eq:scalingn5}) in the dynamic RG:
\begin{align}
\label{eq:rescalephitilde}
\tilde{\phi}_\alpha(\bm{r})&\rightarrow \tilde{\phi}'_\alpha(\bm{r}')=b^{\tilde{a}} \tilde{\phi}_\alpha(\bm{r}),\\
\tilde{n}(\bm{r})&\rightarrow \tilde{n}'(\bm{r}')=b^{\tilde{c}} \tilde{n}(\bm{r}),\\
\label{eq:rescalen5tilde}
\tilde{n}_5 (\bm{r})&\rightarrow \tilde{n}'_5(\bm{r}')=b^{\tilde{c}_5} \tilde{n}_5(\bm{r}).
\end{align}
We will determine $\tilde a$, $\tilde c$, and $\tilde c_5$ in the following calculations. 

We compute the full inverse propagators at the ($l+1$)th renormalization step. Then, we get the recursion relations of the dynamic parameters $\Gamma_l,\, \lambda_l,\, \lambda_5,$ and $C_l$, in an analogy to those of the static quantities, (\ref{eq:RRu})--(\ref{eq:RRgamma}).  In particular, we here demonstrate the derivation of the recursion relation for $\Gamma_l$ as an example. First, we calculate the full inverse propagator for the order parameter as
\begin{align}
\label{eq:G0-1expand}
[G_{\alpha\alpha}(\kv',\omega')]^{-1}_{l+1}
&=[G_{\alpha\alpha}(\kv,\omega)]^{-1}_l b^{-\tilde a -a}\notag\\
&=\left [-\i\omega \left(1-\i\left.   \frac{\d \Sigma_{\alpha\alpha}(\bm{0},\omega)}{\d \omega}   \right|_{\omega\rightarrow0}\right)+\Gamma_l r_l-\Sigma_{\alpha\alpha}(\zev,0) \right.\notag\\
&\qquad\left.
+\left(  \Gamma_l
-\frac{1}{2}\left.   \frac{\d^2 \Sigma_{\alpha\alpha}(\bm{k},0)}{\d \bm{k}^2} \right|_{\bm{k}\rightarrow \bm{0}} \right)   \kv^2 +\cdots
\right]b^{-\tilde a -a}.
\end{align}
Here, we use Eq.~(\ref{eq:DysonG}) and expand $\Sigma_{\alpha\alpha}$ with respect to the frequency $\omega$ and wave number $\kv$.
By regarding the term proportional to $\kv^2$ on the right-hand side as $\Gamma_{l+1}\kv'^2$ and including the overall factor $b^{z+d}$ coming from rescaling the measure of the action, we obtain the recursion relation for $\Gamma_l$,
\begin{align}
\label{eq:RRGamma}
\Gamma_{l+1}&=\Gamma_{l}
\left(
1-
\frac{1}{2\Gamma_l} \left.   \frac{\d^2 \Sigma_{\alpha\alpha}(\bm{k},0)}{\d \bm{k}^2}    \right|_{\bm{k}\rightarrow \bm{0}}
\right)b^{d+z-\tilde a -a-2}.
\end{align}
Furthermore, we normalize the term proportional to $-\i\omega$ including the overall factor, 
\begin{align}
\label{eq:atildea}
1&=\left(  1-\i \left.   \frac{\d \Sigma_{\alpha\alpha}(\bm{0},\omega)}{\d \omega}   \right|_{\omega\rightarrow0}  \right)b^{d-\tilde a -a}.
\end{align}

Similarly, we can derive the following recursion relations by computing the inverse propagator for conserved charge densities, $[D_{ij}(\kv',\omega')]^{-1}_{l+1}$, and three-point vertex $[V_{\alpha;\beta i} (\kv_1,\kv_2,\omega_1,\omega_2)]_{l+1}$ instead of $
[G_{\alpha\alpha}(\kv',\omega')]^{-1}_{l+1}$ as we used in Eq.~(\ref{eq:G0-1expand}):
\begin{align}
\label{eq:RRlambdachi}
\frac{\lambda_{l+1}}{\chi_{l+1}}&=\frac{\lambda_l}{\chi_{l}}\left(1-\frac{\chi_l}{2\lambda_l}  \left.    \frac{\d^2 \Pi_{11} (\bm{k},0)}{\d \bm{k}^2}    \right|_{\bm{k}\rightarrow \bm{0}}  \right)b^{d+z-\tilde{c}-c-2},\\
\label{eq:RRlambda5chi5}
\frac{(\lambda_5)_{l+1}}{(\chi_5)_{l+1}}&=\frac{(\lambda_5)_l}{(\chi_5)_l}\left(1-\frac{(\chi_5)_l}{2(\lambda_5)_l}  \left.    \frac{\d^2 \Pi_{22} (\bm{k},0)}{\d \bm{k}^2}    \right|_{\bm{k}\rightarrow \bm{0}}  \right)b^{d+z-\tilde{c}_5-c_5-2},\\
\label{eq:RRCchi}
\frac{C_{l+1}}{\chi_{l+1}}\Bv &=\left(\frac{C_l}{\chi_l}\Bv +\left.   \i\frac{\d \Pi_{21}(\bm{k},0)}{\d \bm{k}}    \right|_{\bm{k}\rightarrow \bm{0}}\right)b^{d+z-\tilde{c}_5-c-1},\\
\label{eq:RRCchi5}
\frac{C_{l+1}}{(\chi_5)_{l+1}}\Bv &=\left(\frac{C_l}{(\chi_5)_l}\Bv +\left.   \i\frac{\d \Pi_{12}(\bm{k},0)}{\d \bm{k}}    \right|_{\bm{k}\rightarrow \bm{0}}\right)b^{d+z-\tilde{c}-c_5-1},\\
\label{eq:RRgchi5}
\frac{g_{l+1}}{(\chi_5)_{l+1}}\varepsilon_{\alpha\beta}&=\frac{g_l}{(\chi_5)_l} \left(  \varepsilon_{\alpha\beta}  + \frac{(\chi_5)_l}{g_l} 
\left.  \mathcal  V_{\alpha;\beta 2} (\kv_1,\kv_2,\omega_1,\omega_2)     \right|_{\kv_{1,2}\rightarrow \zev, \ \omega_{1,2}\rightarrow 0}\right)b^{d+z-\tilde a -a -c_5},
\end{align}
with the following constraints,
\begin{align}
\label{eq:ctildec}
1&=\left(  1-\i \left.   \frac{\d \Pi_{11} (\bm{0},\omega)}{\d \omega}   \right|_{\omega\rightarrow0}  \right)b^{d-\tilde c - c}\,,\\
\label{eq:ctilde5c5}
1&=\left(  1-\i \left.   \frac{\d \Pi_{22} (\bm{0},\omega)}{\d \omega}   \right|_{\omega\rightarrow0}  \right)b^{d-\tilde c _5 - c _5}\,.
\end{align}
Therefore, once we can evaluate the self-energies $\Sigma_{\alpha\alpha}(\kv,\omega)$, $\Pi_{ij}(\kv,\omega)$ and vertex function corrections $V_{\alpha;\beta2}(\kv_1,\kv_2,\omega_1,\omega_2)$, we obtain the recursion relations for the dynamic parameters in the MSRJD action. We put the details of the calculations on $\Sigma_{\alpha \alpha}$, $\Pi_{ij}$, and $\mathcal V_{\alpha;\beta 2}$ in Appendix~\ref{sec:calculation}. 

In the calculations of Appendix~\ref{sec:calculation}, the following parameters are introduced:
\begin{gather}
\label{eq:Paras}
f \equiv\frac{g ^2 \Lambda^{-\epsilon}}{8\pi^2 \lambda_5 \Gamma},\quad w\equiv\frac{\Gamma \chi}{\lambda},\quad w_5 \equiv\frac{\Gamma \chi_5 }{\lambda_5},\quad h\equiv\frac{C B}{\sqrt{\lambda\lambda_5}\Lambda},\\
\label{eq:Paras2}
X\equiv\frac{2}{\sqrt{(1+w)(1+w_5)+h^2}+\sqrt{(1+w)(1+w_5)}}\sqrt{\frac{1+w}{1+w_5}},\quad X'\equiv \frac{1+w_5}{1+w}X,
\end{gather}
where $B \equiv |{\bm B}|$. By using these new parameters and based on the detailed analysis in appendix~\ref{sec:calculation}, we can obtain the recursion relations for the dynamic parameters $\Gamma_l,\, \lambda_l,\, \lambda_5,\, C_l$ and $g_l$ at the one-loop level:
\begin{align}
\label{eq:RRGamma2}
\Gamma_{l+1}&=b^{z-2}\Gamma_{l}
\left[1- (4v_l w_l X'_l-f_l X_l) \ln b \right],\\
\label{eq:RRlambda2}
\lambda_{l+1}&=b^{z-d+2c-2}\lambda_l,\\
\label{eq:RRlambda52}
(\lambda_5)_{l+1}&=b^{z-d+2c_5-2}(\lambda_5)_l \left(  1+\frac{f_l}{2} \ln b \right),\\
\label{eq:RRC}
C_{l+1} &=b^{z+c+c_5-d-1}C_l, \\
\label{eq:RRg}
g_{l+1}&=b^{z - d + c _5}g_l.
\end{align}
Here, Eqs.~(\ref{eq:RRGamma2}),~(\ref{eq:RRlambda2})--(\ref{eq:RRC}),~(\ref{eq:RRg}) are derived in Appendix~\ref{sec:SigmaCME},~\ref{sec:PiCME},~\ref{sec:VCME}, respectively. Among these recursion relations, Eq.~(\ref{eq:RRC}) shows that the CME coefficient $C$ is not renormalized by the critical fluctuations of the order parameter in this order.
This may be viewed as an extension of the nonrenormalization theorem for the CME coefficient at the second-order chiral phase transition, where $\sigma$ becomes massless.

To obtain the fixed point solutions of the recursion relations, it is useful to obtain the recursion relations for the parameters defined in Eq.~\eqref{eq:Paras}. By using the recursion relations for the dynamic parameters (\ref{eq:RRGamma2})--(\ref{eq:RRg}) and the static parameters (\ref{eq:RRchi})--(\ref{eq:RRgamma}), we reach
\begin{align}
\label{eq:RRf}
f_{l+1}&=b^{\epsilon}f_l \left[1+ \left(4v_l w_l X'_l-f_l X_l -\frac{1}{2}f_l \right) \ln b \right],\\
\label{eq:RRw5}
(w_5)_{l+1}&=(w_5)_l \left[1 - \left(4v_l w_l X'_l-f_l X_l + \frac{1}{2}f_l \right) \ln b \right],\\
\label{eq:RRw}
w_{l+1}&=w_l \left[1 - \left(4v_l w_l X'_l-f_l X_l - 4 v_l \right) \ln b \right],\\
\label{eq:RRh}
h_{l+1}&=b h_l \left(1- \frac{f_l}{4} \ln b \right).
\end{align}
The dynamic RG equations corresponding to Eqs.~(\ref{eq:RRf})--(\ref{eq:RRh}) can be derived in a way similar to 
Eqs.~\eqref{eq:REu}-\eqref{eq:REv} as (see also the footnote under Eq.~(\ref{eq:REu}) for the derivation of the RG equation from the recursion relation) 
\begin{align}
\label{eq:REf}
\frac{{\rm d}f_l}{{\rm d}l}&= \left(\epsilon+ 4v_l w_l X'_l-f_l X_l -\frac{1}{2}f_l \right) f_l\,,\\
\label{eq:REw5}
\frac{{\rm d}(w_5)_l}{{\rm d}l}&= \left(-4v_l w_l X'_l + f_l X_l - \frac{1}{2}f_l \right) (w_5)_l\,,\\
\label{eq:REw}
\frac{{\rm d}w_l}{{\rm d}l}&=\left(-4v_l w_l X'_l + f_l X_l + 4 v_l \right)w_l\,,\\
\label{eq:REh}
\frac{{\rm d}h_l}{{\rm d}l}&= \left(1- \frac{f_l}{4} \right)h_l\,,
\end{align}
from which we find four possible nontrivial fixed-point values of $f,\, w_5,\,w,\,h$:%
\footnote{Besides, there is a trivial fixed point, $f_\infty=(w_5)_\infty=w_\infty=h_\infty=0$, which is stable only for $\epsilon<0$, and is not considered here.}
\begin{align}
\label{eq:FPVs}
({\rm i})&\quad f_\infty=\epsilon,\quad (w_5)_\infty=1,\quad w_\infty=h_\infty=0;\\
\label{eq:FPVs2}
({\rm ii})&\quad f_\infty=\frac{2}{3}\epsilon,\quad (w_5)_\infty= w_\infty= h_\infty=0;\\
\label{eq:FPVs3}
({\rm iii})&\quad f_\infty=\epsilon,\quad (w_5)_\infty=\frac{3}{7},\quad w_\infty= h_\infty=\infty;\\
\label{eq:FPVs4}
({\rm iv})&\quad f_\infty=2\epsilon,\quad (w_5)_\infty=0,\quad w_\infty= h_\infty=\infty.
\end{align}

Some remarks on the fixed points above are in order here. 
Since the magnetic field is external ($B\neq 0$), the fixed points (i) and (ii) with $h_\infty=0$ should be interpreted as $C=0$.
We should note that the RG equations (\ref{eq:REf})--(\ref{eq:REw}) are nonuniform in the limits $ w\rightarrow \infty$ and $h^2 \rightarrow \infty$. If one takes $w\rightarrow \infty$ first by fixing $h^2$ to some particular value, the fixed point (iii) is obtained; if one takes $h^2 \rightarrow \infty$ first by fixing $w$ to some particular value, the fixed point (iv) is obtained. 
In other words, the fixed point (iii) corresponds to the case $w_\infty \gg h_\infty^2 \gg1$, and the fixed point (iv) corresponds to the case $h_\infty^2  \gg w_\infty \gg1$.
The competition between $w\rightarrow\infty$ and $h^2 \rightarrow \infty$ in Eq.~(\ref{eq:Paras2}) are characterized by the strength of the following parameter,
\begin{align}
\label{eq:h2/w}
 \frac{h^2}{w}=\frac{C^2B^2}{\lambda_5\Gamma\chi\Lambda^2}.
\end{align}
From the expression of Eq.~(\ref{eq:h2/w}), we can see which parameters among $C,\lambda$ and $\lambda_5$ are dominant near the fixed points, for a finite kinetic coefficient of the order parameter, $\Gamma$, and finite static susceptibilities, $\chi$ and $\chi_5$.\footnote
{One can confirm the finiteness of $\Gamma, \chi,$ and $\chi_5$  by putting back the fixed-point values of $v, f, w_5, w$, and $h$ to the recursion relations (\ref{eq:RRchi}), (\ref{eq:RRchi5}), and (\ref{eq:RRGamma2}) with the help of  Eqs.~(\ref{eq:c5}) and (\ref{eq:c}).}
By looking at $f_\infty$, $(w_5)_\infty$, and the fixed-point value of \eqref{eq:h2/w}, one can see the fixed point (iii) corresponds to $C\rightarrow0$ and $\lambda\rightarrow0$ with finite $\lambda_5$, where the CME can be neglected compared to the diffusion effect; the fixed point (iv) corresponds to $C\rightarrow\infty$, $\lambda\rightarrow 0$, and $\lambda_5\rightarrow\infty$ with $C^2/\lambda_5\rightarrow\infty$, where the diffusion effect can be neglected compared to the CME. In short, we can regard the competition between the two limits $w \to \infty$ and $h^2 \to \infty$  as the competition between the CME and the diffusion of the axial isospin density $n_5$.

\subsubsection{Stability of fixed points}
\label{sec:stability}
We here summarize the stability analysis of the fixed points (i)--(iv). In order to investigate the influence of the CME, we study the stability of the fixed points at $C=0$, namely the fixed points (i) and (ii). For this purpose, we consider the linear perturbations around these fixed points, 
\begin{align}
f_l=f_\infty+\delta f,\quad (w_5)_l=(w_5)_\infty+\delta w_5,\quad w_l=\delta w,\quad h_l=\delta h.
\end{align}
Substituting these expressions into Eqs.~(\ref{eq:REf})--(\ref{eq:REh}) and setting $v_l=v_\infty = \epsilon/20$ from Eq.~\eqref{eq:FVv},%
\footnote {Here we can ignore the fluctuation of $v_l$, because all of the fluctuation of $v_l$ will be multiplied by $O(\delta w)$ in Eqs.~(\ref{eq:REf})--(\ref{eq:REh}) if we try to substitute $v_l=v_{\infty}+\delta v$.}  
the linearized equations with respect to $\delta f$, $\delta w_5$, $\delta w_5$, and $\delta h$ read
\begin{align}
\label{eq:linearized}
\frac{\dd}{\dd l}
\left(\begin{array}{c}
\delta f\\
\delta w_5\\
\delta w\\
\delta h
\end{array}\right)
=\mathcal M'
\left(\begin{array}{c}
\delta f\\
\delta w_5\\
\delta w\\
\delta h
\end{array}\right),
\end{align}
where we define 
\begin{align}
\mathcal M' \equiv\left(\begin{array}{cccc}
-f_\infty\left(\theta_\infty+\dis \frac{1}{2}\right)&f_\infty^2 \theta_\infty^2&4 v_\infty f_\infty&0\\
(w_5)_\infty\left(\theta_\infty-\dis \frac{1}{2}\right)&\ f_\infty\left[\theta_\infty -\dis \frac{1}{2}-(w_5)_\infty\theta_\infty^2\right]&-4v_\infty(w_5)_\infty&0\\
0&0&f_\infty\theta_\infty+4v_\infty&0\\
0&0&0&1-\dis \frac{f_\infty}{4}
\end{array}\right)
\end{align}
and
\begin{align}
\label{eq:FPvs1and2}
\theta_\infty\equiv \frac{1}{1+(w_5)_\infty}=
 \left\{ 
 \begin{array}{c}
  \dis{\frac{1}{2}} \quad  {\rm for \ the \ case} \ ({\rm i}), 
   \vspace{5pt} \\
   1 \quad  \ {\rm for \ the \ case} \ ({\rm ii}).
 \end{array}
  \right. 
\end{align}
We can test the stability of the fixed points (i) and (ii) by substituting each of the fixed point values into $\mathcal M'$.  Because of $(w_5)_\infty\left(\theta_\infty-1/2\right)=0$ 
for both cases (i) and (ii), the matrix $\mathcal M'$ defined in Eq.~(\ref{eq:linearized}) 
is reduced to an upper triangular matrix. Thus, the eigenvalues of $\mathcal M'$ are just given by its diagonal components for each fixed point:
\begin{align}
\label{eq:eigenvalues}
({\rm i})&\quad\left(-\epsilon,\,-\frac{\epsilon}{4},\,\frac{7}{10}\epsilon,\,1-\frac{\epsilon}{4} \right)\quad{\rm and }\quad({\rm ii})\quad\left(-\epsilon,\,\frac{\epsilon}{3},\,\frac{13}{15}\epsilon,\,1-\frac{\epsilon}{6} \right).
\end{align}
\begin{figure}[htb]
 \centering
 \includegraphics[bb= 0 0 810 768,  width=7.7cm]{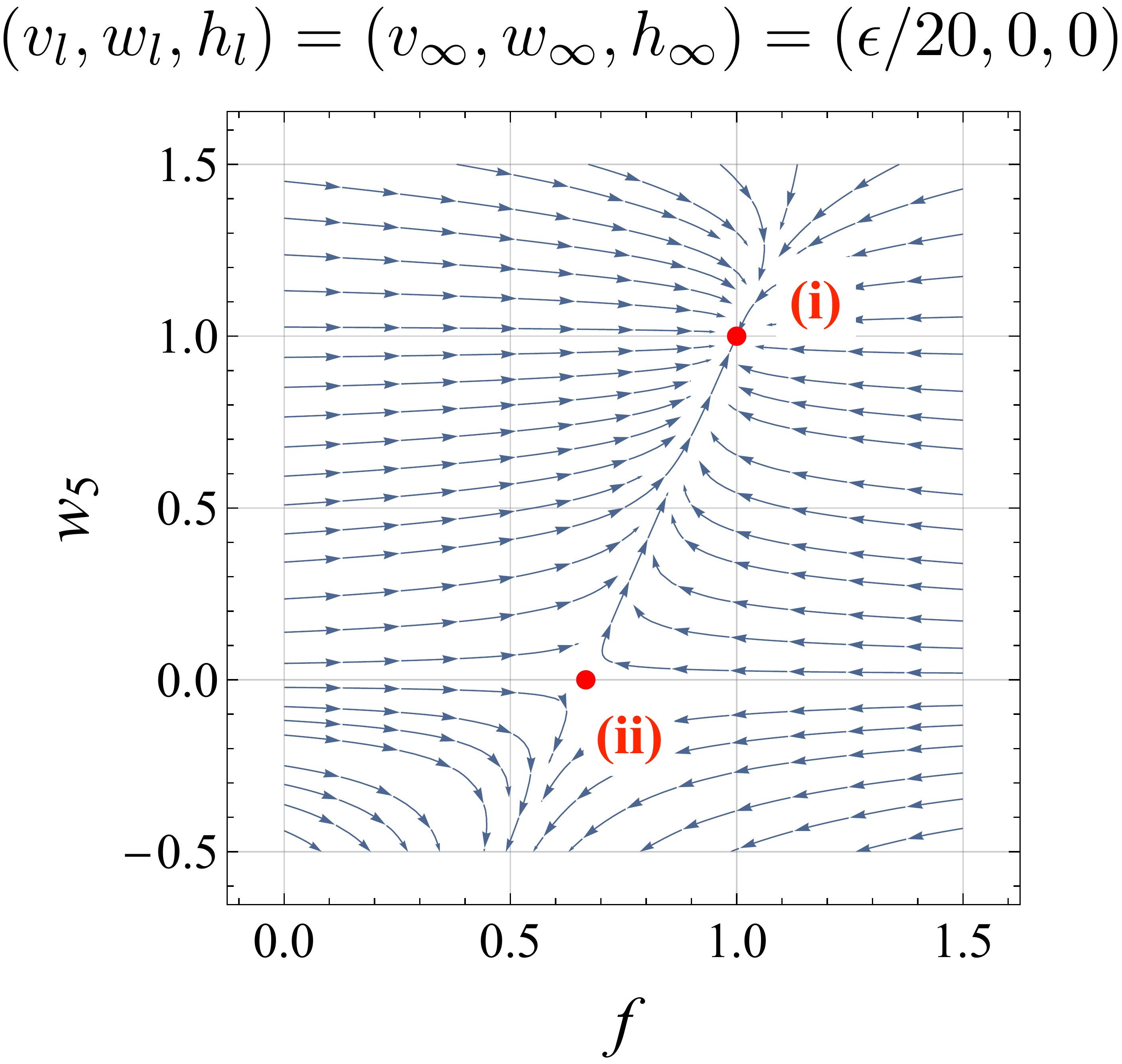}\\
 \caption{RG flow of the parameters $f$ and $w_5$ for fixed $w$ and $h$ ($\epsilon = 1$), which shows the fixed points (i) and (ii). }
 \label{fig:flow1}
\end{figure}%
From this result, we find that the fixed point (ii) is unstable in the $w_5$ direction and that the RG flow runs to the fixed point (i) 
(see also Fig.~\ref{fig:flow1}
showing the RG flows in the $(f,w_5)$ plane at $w=h=0$). 
We also find that the fixed points (i) and (ii) are unstable in the directions of $w$ and $h$, showing that $\lambda$ and $C$ are relevant. It follows that small but nonzero values of $w$ and $h$ grow around the fixed point (i). 

\newpage

\begin{figure}[H]
\includegraphics[bb= 0 0 700 768, width=1\linewidth]{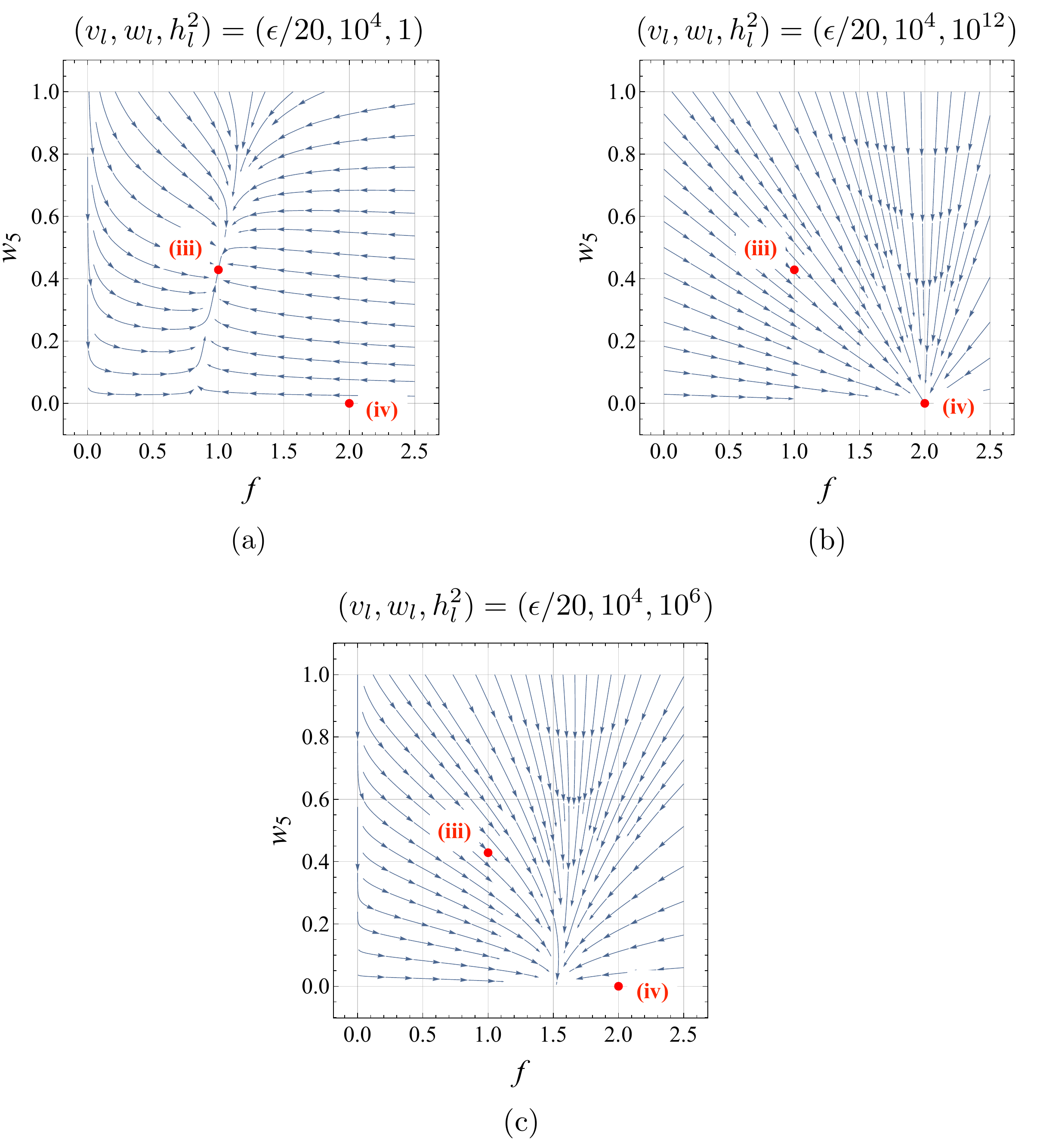}
\caption{RG flows of $f$ and $w_5$ ($\epsilon = 1$). We fix the values of $w$ and $h^2 $ in several cases: (a) $w\gg h^2 \gg1$, (b) $h^2 \gg w\gg1$, and (c) $w\sim h^2 $. These figures show the existence of the fixed points (iii) and (iv), and the flow to their intermediate values.} \label{fig:flow2}
\end{figure}

\begin{figure}[t]
 \centering
 \includegraphics[bb=0 0 330 330, 
 width=6.8cm]{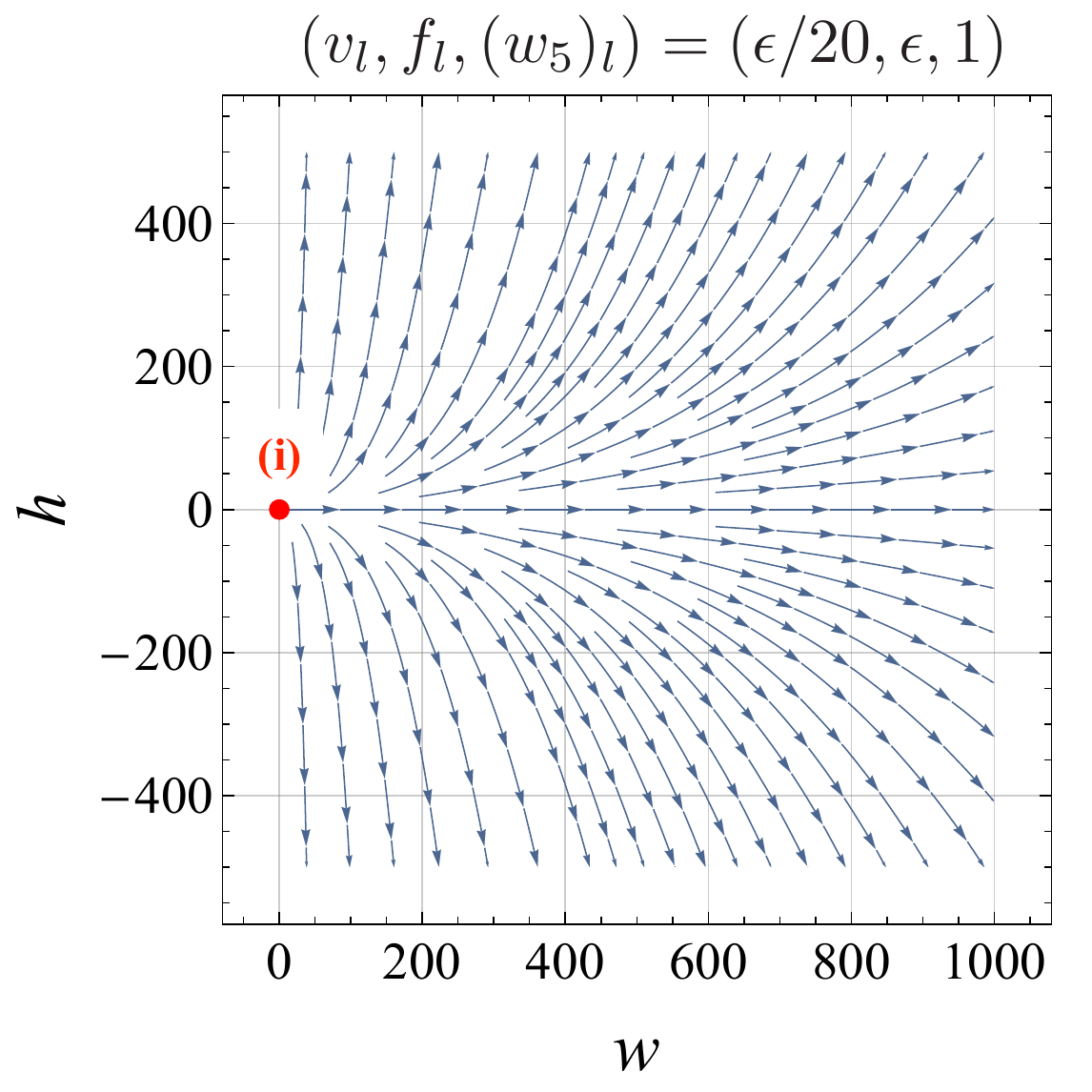}
 \caption{ RG flow of $w$ and $h$ with fixing  
 $f$ and $w_5$ to those values at the fixed point (i) ($\epsilon = 1$).
 }
 \label{fig:flow3}
\end{figure}

There are two possibilities of the final destination of the flow from the fixed point (i): the fixed point (iii) and the fixed point (iv). Let us first qualitatively understand the conditions for obtaining each of the fixed points (iii) and (iv) by using RG flow diagrams. For this purpose, we first forcibly {\it fix} $w$ and $h$ to some finite values and investigate the RG flows in the $(f,w_5)$ plane. As one can see in the RG flows in Figs.~ \ref{fig:flow2}(a) and \ref{fig:flow2}(b), $f$ and $w_5$ flow to the fixed point (iii) when $w\gg h^2 \gg1$, while they flow to the fixed point (iv) when $h^2 \gg w\gg1$, related to the properties of the fixed points (iii) and (iv) noted under Eq.~(\ref{eq:FPVs4}). On the other hand, Fig.~\ref{fig:flow2}(c) shows that  $(f,w_5)$ flow to the intermediate values between the fixed-point values of (iii) and (iv) when $w \sim h^2$. Next, we vary $w$ and $h$ around those values of the fixed point (i) with fixed $f$ and $w_5$, and study which one is more relevant (iii) or (iv) around the fixed points (i). As is shown in Fig.~\ref{fig:flow3}, the points in the $(w,h)$ plane flow in the direction along the $h$ axis unless $w \gg h$. Therefore, the system eventually flows to the fixed point (iv) for most of the parameter region around the fixed point (i).

Next, we consider the RG flows in all the parameter space $(f,w_5,w,h)$ without fixing the parameters. Here, we first set the initial parameters near the fixed point (i) and consider the flow equations at a finite flow time. As is shown within the linear-stability analysis in Appendix \ref{sec:crossover}, the initial parameter region that flows to the fixed point (iv) is much broader than the region that flows to the fixed point (iii) in the $(w,h)$ plane. Therefore, in the almost whole region of the $(w, h)$ plane near the fixed point (i), the fixed point (iv) is favorable rather than the fixed point (iii).
%Similarly to the RG flows in the previous paragraph, when the initial values of $w$ and $h$ are varied, all the parameters move between the fixed-point values of (iii) and (iv).
%(where a crossover between the dynamic universality classes corresponding to the fixed points (iii) and (iv) is also discussed)
We can also study the RG flow from the initial values near the fixed point (iii). Expect for the case of the RG evolution, starting from the parameters exactly at the fixed point (iii), all the parameters will eventually take the fixed-point values of (iv). This is because $h$ grows much more rapidly than $w$ due to the additional scaling factor $b$ in the recursion relation~(\ref{eq:RRh}) for $h$, compared to the relation~(\ref{eq:RRw}) for $w$. From the above discussion, it follows that the fixed point (iv) is stable in the almost whole region at finite $w$ and $h$, while generally at a finite flow time there is a small parameter region that flows to the fixed point (iii).

\newpage

\newpage

\subsection{Physical consequences}
\subsubsection{Dynamic universality class}
\label{sec:DUCCME}
From the fixed point values (i)--(iv) obtained in Eqs.~(\ref{eq:FPVs})--(\ref{eq:FPVs4}), we can evaluate the dynamic critical exponent which characterizes the dynamic universality class of the system. Practically, we substitute the fixed point values into the recursion relation~(\ref{eq:RRGamma2}) for $\Gamma_l$, in each case of the fixed-point values (i)--(iv). 

 The fixed points (i) and (iii) have the dynamic critical exponent of model E, $z=d/2$. As is also summarized in Tab.~\ref{tab:DUC} the dynamic universality class of model E is generally determined only by two-component order parameter (mapped into $\U(1)$) and {\it one} conserved density that are coupled through  the Poisson brackets. 
In our case, the order parameter field $\phi_\alpha$ and the axial isospin density $n_5$ are essential, whereas the isospin density $n$ does not affect the dynamic universality class.\footnote{The difference between $n$ and $n_5$ can be seen in the Poisson brackets: there is a nonzero Poisson bracket among $n_5$ and $\phi_\alpha$ in Eq.~(\ref{eq:PBn5phi}), whereas there are no nontrivial Poisson brackets among $n$ and $\phi_\alpha$.} 

On the other hand, the fixed point (iv) has the dynamic critical exponent $z=2$.  Up to $O(\epsilon)$, this exponent is the same as that of model A, which is the simplest class determined only by nonconserved order parameters (see also  Tab.~\ref{tab:DUC}). In the system near the fixed point (iv), the internal-momentum loop dominated by the CMW (the wavy lines in Fig.~\ref{fig:Sigma}) is suppressed, so that not only $n$ but also $n_5$ do not affect the dynamic universality class. One can confirm that the factors $X$ and $wX'$ stemming from Fig.~\ref{fig:Sigma} vanish. 

Now we remind the stability analysis in Sec.~\ref{sec:stability}: the small fluctuation of the fixed point (i) in the absence of the CME can lead to the fixed point (iv). We find that the inclusion of the CME can change the dynamic universality class from model E into model A, corresponding to the stable fixed point (i) and (iv), respectively. Strictly speaking, there is a small parameter region that leads to model E, even $\Bv \neq {\bm 0}$ and $C\neq 0 $ as we also noted in \ref{sec:stability}. Nevertheless, such a region is small compared to the region that leads to model A. %We note that the dynamic universality class in table \ref{tab:DUC} remains unchanged even when the isospin chemical potential $\mu$ is absent.%\footnote{One can easily confirm the dynamic universality class at $\mu = 0$ from the recursion relations which are obtained by setting $v=0$ in Eqs.~(\ref{eq:REf})--(\ref{eq:REh}).}

\subsubsection{Critical attenuation of the CMW}
As a result of the static critical behavior, Eq.~(\ref{eq:eps}), we find the critical attenuation of the CMW: in the vicinity of the  second-order chiral phase transition, the speed of the CMW tends to zero as
\begin{align}
\label{eq:CMWcs}
v_{\rm CMW}^2\equiv \frac{C^2 B^2}{\chi\chi_5} \sim \xi^{-\frac{\alpha}{\nu}},
\end{align}
where $v_{\rm CMW}$ is the speed of the CMW \cite{Kharzeev:2010gd} which have been already seen in its wave equation (\ref{eq:waveeqCMW}). We have already obtained the ratio $\alpha/\nu$ in Eq.~(\ref{eq:alpha/nu}). This phenomenon is analogous to the critical attenuation of the speed of sound near the critical point associated with the liquid-gas phase transition \cite{Onuki:1997}.
% and the superfluid transition of liquid $^4$He \cite{Pankert:1986}.

\section{Conclusion and discussion}
\label{sec:CD}
In this chapter, we have studied the critical dynamics near the second-order chiral phase transition in massless two-flavor QCD under an external magnetic field and investigated the influence of the CME on the dynamic critical phenomena in QCD. We found that the inclusion of the CME and the resulting CMW can change the dynamic universality class of the system from the model E into model A. We also found the critical attenuation of the CMW analogous to that of the sound wave in the liquid-gas phase transition. 

We now discuss the analogy of the critical attenuation between the CMW and the sound wave of the compressive fluids near the liquid-gas critical point. Let us first recall the critical attenuation of sounds near the critical point associated with the liquid-gas phase transition, where the order parameter $\psi_{\rm LG}$ is a linear combination of the energy density $\varepsilon$ and the mass density $\rho_m$. In this case, the speed of sound, $c_s$, is attenuated with the correlation-length dependence \cite{Onuki:1997},
\begin{align}
\label{eq:cs}
c_s^2\equiv \left( \frac{\partial P}{\partial \rho_m} \right)_{\! S}
= \frac{T\left( \dis \frac{\partial P}{\partial T} \right)^{\! 2}_{\! \rho_m}}{\rho_m^2 C_V\dis\left(1- \frac{C_V}{C_P}  \right)} \sim \xi^{-\frac{\alpha}{\nu}},
\end{align} 
where $P$ and $S$ are the pressure and total entropy per unit mass of the fluids, respectively. To obtain the last expression of Eq.~(\ref{eq:cs}), we use some thermodynamic relations and the fact that the specific heat with constant volume $C_V\equiv T (\partial S/ \partial T)_\rho$, and that with constant pressure $C_P\equiv T (\partial S/ \partial T)_P$, diverge near the critical point as $C_V\sim\xi^{\frac{\alpha}{\nu}}$ and $C_P\sim\xi^{\frac{\gamma}{\nu}}$, respectively. Here, the critical exponents $\nu$, $\alpha$, and $\gamma$ defined by Eq.~(\ref{eq:eps}) and $\psi_{\rm LG}\sim\tau^\gamma$ are determined by the static universality class of the 3D Ising model, $\alpha \approx 0.1,\,\nu \approx 0.6,\,\gamma \approx 1.2$. We also use the approximation ${C_V}/{C_P}\ll 1$ near the critical point. Remarkably, Eq.~(\ref{eq:cs}) takes exactly the same form as that of the CMW which we obtained in Eq.~(\ref{eq:CMWcs}), although the values of $\alpha$ and $\nu$ themselves are different due to the difference of the static universality classes.

When the reduced temperature $\tau$ is sufficiently larger than $\bar m_{\rm q}\equiv m_{\rm q}/T_{\rm c}$, quark mass effects on the critical phenomena can be negligible. 
%, $\tau \gg \bar m_{q}^{-\beta \delta }$, with $\sigma\sim \tau^\beta$ and $ \sigma \sim \bar m_{\rm q}^{1/\delta} $ at $\tau=0$ (see, e.g., Ref \cite{Goldenfeld:1992qy} for the crossover theory). 
Our analysis in the chiral limit would be relevant to such a parameter region around the high-temperature QCD critical point under an external magnetic field. %In particular, the model A dynamics is demonant  at $h^{1/(\nu h_h)}\gg \tau \gg \bar m_{\rm q}^{-\beta \delta }$.

\chapter{Summary and outlook}
\label{chap:Summary}

In this thesis, we have studied the novel dynamic critical phenomena induced by superfluidity and the CME in QCD. 

In Sec.~\ref{chap:DCPHD} we have first elucidated the influence of the superfluidity on the critical phenomena near the high-density QCD critical point. In particular, we have found that the static universality class is the same as that of the high-temperature critical point, independently of the existence of the superfluid phonon. On the other hand, we have found that the 
superfluid phonon exhibits the critical slowing down when the critical point is approached. Furthermore, we have found that the dynamic universality class of the high-density critical point is not only different from that of the high-temperature critical point but also all of the conventional classes studied by Hohenberg and Halperin  \cite{Hohenberg:1977ym}. %This new dynamic universality class of the high-density critical point stems from the interplay specific to QCD between the chiral order parameter and the superfluid phonon. 

%the observation of the dynamic critical phenomena near the high-density critical point  is essential to distinguish the possible two QCD critical points. For this purpose, 
Experimental signatures related to the dynamic critical phenomenon in the heavy-ion collisions can distinguish the possible two QCD critical points. Though the {\it vanishing} speed of the superfluid phonon characterizes the high-density critical point, little research has considered the superfluid phonon itself in the context of the heavy-ion collisions. It would be essential to investigate the role of the superfluid phonon on the evolution of the hot and dense medium created in the heavy-ion collisions. 
%It would be interesting to study the signature of the high-density critical point (also the case of the high-temperature critical point) in the dynamical evolution of the heavy-ion systems.

In Sec.~\ref{chap:DRGCME} we have studied the interplay between the dynamic critical phenomena and the CME in QCD. For this purpose, we considered the dynamic critical phenomena of the second-order chiral phase transition under an external magnetic field. Then, we have found that the inclusion of the CME and the resulting CMW can change the dynamic universality class of the system from the model E into model A within the conventional classification. We have also found that the speed of the CMW tends to vanish near the phase transition point and that the speed can be characterized by the same static critical exponents as those of the sound wave in the vicinity of the liquid-gas critical point.

While we have limited our analysis in Sec.~\ref{chap:DRGCME} to the background magnetic field, dynamical electromagnetic fields may affect the critical dynamics in QCD. In massless QCD coupled to dynamical electromagnetic fields in an external magnetic field, there appears a nonrelativistic photon with a quadratic dispersion relation due to the quantum anomaly \cite{Sogabe:2019gif}. At finite temperature, such a novel gapless mode will be crucial for dynamic critical phenomena near the second-order chiral phase transition. This study will be reported in detail elsewhere.

%Second one is to consider another type of chiral transport phenomena, such as the chiral vortical effect in the rotation. 

%From our findings in Sec.~\ref{chap:DRGCME}, we can learn the potential of the CME to change the universality classes far from equilibrium as well as the dynamic universality class near equilibrium. In particular, it would be interesting to study the effects of the chiral transport to the Karder-Parisi-Zhang (KPZ) universality class for the randomly surface growing phenomena, such as the propagation of flame front \cite{Kardar:1986xt}. When the universal scaling behavior emerges in the system which possesses a chiral charge, the chiral transport can qualitatively change the characteristic exponents of the nonequilibrum scaling dynamics in a similar manner as the dynamic critical exponent as we have shown in Sec.~\ref{chap:DRGCME}. 

\appendix
\makeatletter

\chapter{Static RG of the scalar field theory}
\label{chap:staticRG}
In this appendix, we show the detailed RG analysis on the (one component) scaler field theory (\ref{eq:GLW}). In particular, we derive the RG equations~(\ref{eq:rbarRG}) and (\ref{eq:ubarRG}), and the Wilson-Fisher fixed-point solution~(\ref{eq:WFbody}).

\section{Perturbative RG equation}
\begin{figure}[t]
\begin{tabular}{cc}
\begin{minipage}{.7\textwidth}
\subfigure[1-loop]{ \includegraphics[bb=0 0 142 99,height=2cm]{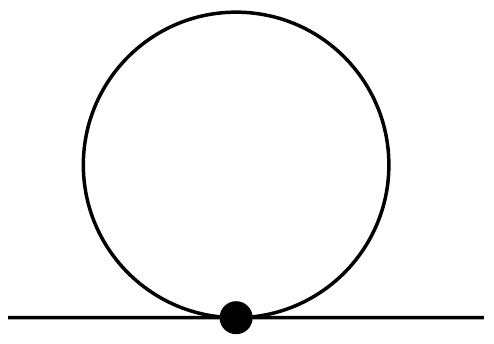}\label{fig:Sigma1}}~
\subfigure[2-loop]{ \includegraphics[bb=0 0 312 227,height=4cm]{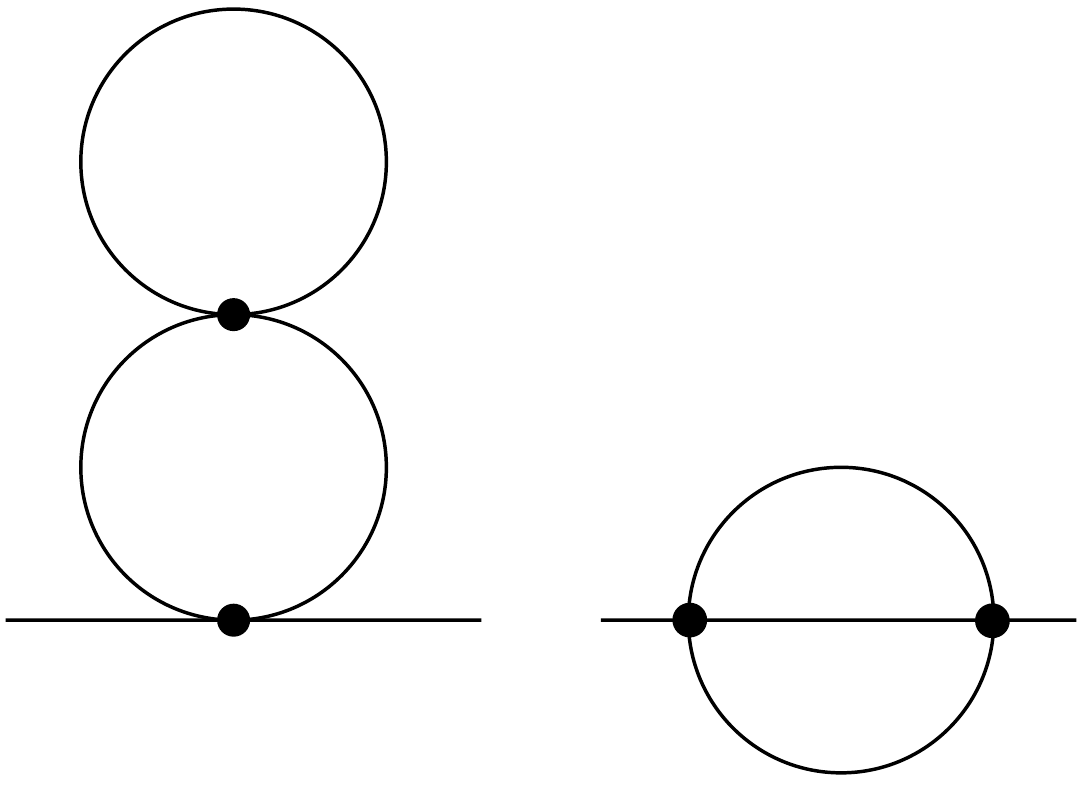}\label{fig:NV2}}
\caption{$\Sigma$}
\label{fig:Sigmaapp}
\end{minipage}~
\begin{minipage}{.3\textwidth}
 \includegraphics[bb=0 0 198 141,height=2.5cm]{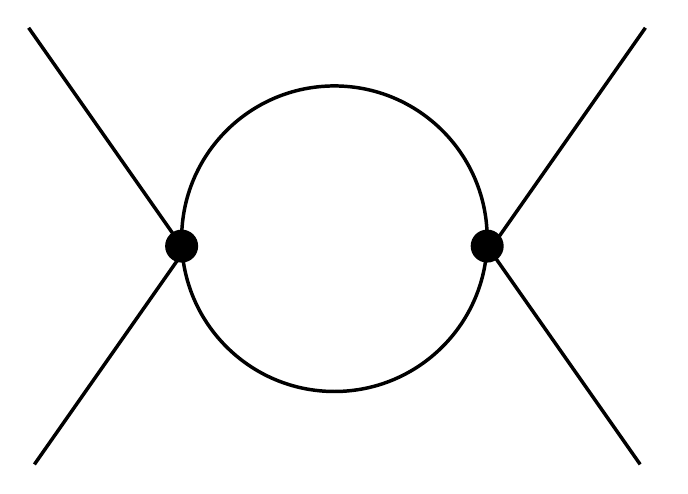}\label{fig:V1}
\caption{$V$}
\label{fig:Vapp}
\end{minipage}
\end{tabular}
\end{figure}

To derive the RG equations of the Ginzburg-Landau free energy (\ref{eq:GLW}), we first construct its perturbation theory. We start by decomposing the free energy into the following form:
\begin{align}
\label{eq:decompapp}
\beta F_\Lambda[\psi] =
\beta F_{0\Lambda}[\psi] +
\beta F'_\Lambda[\psi] + \int \dd \rv h\psi,
\end{align}
where the free part and the perturbative part are given by\begin{align}
\beta F_{0\Lambda}[\psi]&=\int \dd\rv  \left[ \frac{r}{2} \psi^{2}(\rv )+\frac{1}{2}(\na\psi(\rv ))^{2}\right]\notag\\
&=\int_{\qv}  \frac{1}{2}(r+\q^{2}) \psi(\q)\psi(-\q),\\
\beta F_{\Lambda}'[\psi]&=u \int \dd\rv  \psi^{4}(\rv )\notag\\
&=u \int_{\q_{1}}\int_{\q_{2}}\int_{\q_{3}}\int_{\q_{4}} \psi(\q_{1})\psi(\q_{2})\psi(\q_{3})\psi(\q_{4})\delta(\q_{1}+\q_{2}+\q_{3}+\q_{4}),
\end{align}
respectively. Here, we use the simple notation of the momentum integral,
\begin{align}
\int _{\qv}\equiv \int \frac{\dd^d  \qv}{(2\pi)^d}.
\end{align}
Using Eqs.~(\ref{eq:fielddecomp}) and (\ref{eq:decompapp}), we can write the partition function as
\begin{align}
Z&=\int \mathcal{D} \psi^{>} \mathcal{D} \psi^{<} \e^{-\beta F_{\Lambda}[\psi]}=\int \mathcal{D} \psi^{<} \e^{-\beta F_{\Lambda/b}[\psi^{<}]},
\end{align}
with $\beta F_{\Lambda/b}[\psi^<]$ being the effective free energy of the low-momentum field $\psi^<$:
\begin{align}
\label{eq:B6}
\e^{-\beta F_{\Lambda/b}[\psi^{<}]}&\equiv \int \mathcal{D} \psi^{>} \e^{
-\beta F_{0\Lambda}[\psi] -
\beta F'_\Lambda[\psi] -\beta \int \dd \rv h\psi}.
\end{align}
One can write this effective free energy by using the generating functional $\mathcal Z[h]$, \begin{align}
\mathcal Z[h] &\equiv \left<  \e^{-\beta \int \dd \rv h\psi^>} \right>,
\end{align}
with the expectation value defined in the following form;
\begin{align}
\ex{O[\psi^{<},\psi^{>}]}&\equiv (Z^>)^{-1}\int \mathcal{D} \psi^{>} 
\e^{
-\beta F_{0\Lambda}[\psi] -
\beta F'_\Lambda[\psi] }
O[\psi^{<},\psi^{>}],\\
Z^>&\equiv\int \mathcal{D} \psi^{>}\e^{
-\beta F_{0\Lambda}[\psi] -
\beta F'_\Lambda[\psi] }.
\end{align}
In fact, we rewrite Eq.~(\ref{eq:B6}) into
\begin{align}
\label{eq:effective}
\e^{-\beta F_{\Lambda/b}[\psi^{<}]+\beta\int \dd \rv h \psi^<}&=\int \mathcal{D} \psi^{>}  \e^{
-\beta F_{0\Lambda}[\psi] -
\beta F'_\Lambda[\psi] -\beta \int \dd \rv h\psi^>}
=\mathcal Z[h].
\end{align}
Then, we find that $\beta F_{\Lambda/b}[\psi^<]$ satisfies
\begin{align}
\beta F_{\Lambda/b}[\psi^<] = -\ln \mathcal Z[h] +\beta \int \dd \rv h\psi^<.
\end{align}
This relation means that the generating functional obtained from integrating over $\psi^>$ gives the effective free energy for $\psi^<$. Therefore, we can get the effective free energy by calculating the sum of each vortex function, 
\begin{flalign}
\label{eq:effectiveaction}
&\beta F_{\Lambda/b}[\psi^< ]\notag\\
&=\int_{\q} \frac{\delta (\beta F_{\Lambda/b}[\psi^< ])}{\delta \psi^< (\q) \delta \psi^< (-\q)}
%(r'+K'\q^{2})
 \psi^< (\q)\psi^< (-\q)\notag\\
&\quad+\int_{\q_{1}\cdots\q_{4}}
\frac{\delta (\beta F_{\Lambda/b}[\psi^< ])}{\delta \psi^< (\q_{1}) \delta \psi^< (\q_{2})\delta \psi^< (\q_{3}) \delta \psi^< (\q_{4})}
\psi^< (\q_{1})\psi^< (\q_{2})\psi^< (\q_{3})\psi^< (\q_{4})\notag\\
&\hspace{60pt}\times 
\delta(\q_{1}+\cdots+\q_{4})+\cdots,
\end{flalign}
where the 2- and 4- points vortex functions are given by
\begin{align}
\label{eq:Gamma2}
\displaystyle\frac{\delta (\beta F_{\Lambda/b}[\psi^< ])}{\delta \psi^< (\q) \delta \psi^< (-\q)}&=\frac{1}{2}(r+\q^{2})+\Sigma (\q),\\
\label{eq:Gamma4}
\frac{\delta (\beta F_{\Lambda/b}[\psi^< ])}{\delta \psi^< (\q_{1}) \delta \psi^< (\q_{2})\delta \psi^< (\q_{3}) \delta \psi^< (\q_{4})}&=u+V(\q),
\end{align}
respectively. Here, $\Sigma$ and $V$ are the 1-particle irreducible diagrams including the internal momentum loops of $\psi^>$. The lowest-loop contributions can be obtained as Figs.~\ref{fig:Sigmaapp} and \ref{fig:Vapp}.
The rules for calculating these diagrams are summarized as follows:

(i) Assign momentums for each of internal and external lines satisfying the momentum conservation laws;

(ii) Multiply the free propagator $\dis \frac{1}{r+\qv^2}$ %$\left <  \psi^> (\qv)  \psi^> (-\qv)  \right>_0 = \dis \frac{1}{r+\qv^2}$ 
for each of the internal lines;

(iii) Multiply the free 4-point vortex function $u$ for each of the vertices (black dots);

(iv) Integrate over the momentum of the internal lines; 
   
(v) Multiply some numerical factor.\footnote
{For the given diagram with $N_{\rm v }$ vertices, the numerical factor  is given by 
\begin{align}
\frac{N_{\rm v }!}{N_{\rm sym}} \times \dis \prod_{i=1}^{N_{\rm v }} \frac{4!}{N_{i>}! N_{i<}!},
\end{align}
where $N_{i>}$ and $N_{i<}$ are the number of the internal and the external lines connecting to the $i$-th vortex, and $N_{\rm sym}$ is the geometrical symmetry factor.}

Next we carry out the rescaling of the vortex functions (\ref{eq:Gamma2}) and (\ref{eq:Gamma4}) and complete the RG transformation. 
Writing the vortex functions after one RG step as
\begin{align}
\displaystyle\frac{\delta (\beta F_{\Lambda}[\psi'])}{\delta \psi'(\q') \delta \psi'(-\q')}&=\frac{1}{2}(r'+K'\q'^2)+O(\q'^{4}),\\
\frac{\delta (\beta F_{\Lambda}[\psi'])}{\delta \psi'(\q'_{1}) \delta \psi'(\q'_{2})\delta \psi'(\q'_{3}) \delta \psi'(\q'_{4})}&=u'+O(\q'^{2}),
\end{align}
we obtain the parameters one step after the RG, 
\begin{align}
\label{eq:r'}%%%
r'&=b^{2}\left[r
+6u\int^>_{\q}  \frac{1}{r+\qv^2}
+O(u^{2})\right],\\
\label{eq:K'}%%%
K'&=1+O(u^{2}),\\
\label{eq:u'}%%%
u'&=b^{\epsilon}\left[ u-36u^{2}\int^>_{\q} \frac{1}{(r+\qv^2)^2}\right].
\end{align}
Here, the overall factors $b^2$ and $b^\epsilon$ come from the rescaling. The short-hand notation 
\begin{align}
\label{eq:int>}
\int_\q^> \equiv \int^\Lambda_{\Lambda/b} \frac{\dd^d \qv}{(2\pi)^d}
\end{align}
denotes the integration over the high momentum loop. We can carry out the integrals in Eqs.~(\ref{eq:r'}) and (\ref{eq:u'}) as
\begin{align}
\label{eq:int1}
\int^>_{\q}  \frac{1}{r+\qv^2} &= \int_\q^>  \frac{1}{\qv^2}  \frac{1}{1+r/\qv^2}\notag\\
&\simeq \int_\q^>  \frac{1}{\qv^2} \left(  1- \frac{r}{\qv^2} \right)\notag\\
&= \int^\Lambda_{\Lambda/b} \frac{\dd^d \qv}{(2\pi)^d}  \frac{1}{\qv^2}  -r\int^\Lambda_{\Lambda/b} \frac{\dd^d \qv}{(2\pi)^d}  \frac{1}{\qv^4}\notag\\
&=\frac{\Lambda^{2-\epsilon}}{8 \pi^2}\left(1-\frac{1}{b^2}\right)-r\frac{\Lambda^{-\epsilon}}{8 \pi^2}\ln b,\\
%%%
\int^>_{\q} \frac{1}{(r+\qv^2)^2}& \simeq \int^>_{\q} \frac{1}{\qv^4}\notag\\
\label{eq:int2}
&=\frac{\Lambda^{-\epsilon}}{8 \pi^2}\ln b. 
\end{align}
Here, we consider the system sufficiently near the phase transition point and set $r\ll1$. We also use the formulas~(\ref{eq:integral1/q^2}) and (\ref{eq:integral1/q^4}). Then, we can write Eqs.~(\ref{eq:r'}) and (\ref{eq:u'}) into 
\begin{align}
r'&=b^{2}\left[r
+6uK_d\Lambda^{2-\epsilon} \left(  1-\frac{1}{b^2}-\frac{r}{\Lambda^2} \ln b
\right)\right],\\
u'&=b^{\epsilon}\left( u-36u^{2} K_d \Lambda^{-\epsilon}\ln b \right).
\end{align}
By defining the dimension less parameters,
\begin{align}
\bar r \equiv \frac{r}{\Lambda^2},\quad\bar u \equiv \frac{K_d u}{\Lambda^\epsilon},
\end{align}
with 
\begin{align}
K_d\equiv \frac{1}{8\pi^2},
\end{align}
we can rewrite these into 
\begin{align}
\label{eq:r'2}
\bar r'&=b^{2}\left[\bar r
+6\bar u  \left(  1-\frac{1}{b^2}-\bar r \ln b
\right)\right],\\
\label{eq:u'2}
\bar u'&=b^{\epsilon}\left( \bar u-36\bar u^{2} \ln b \right).
\end{align}
These are the recursion relations which relate the parameters before and after the RG procedure (at a single step). 

We will next derive the differential equations equivalent to the recursion relation. In order to do this, we consider the thin momentum shell with $0<l\ll1$,
\begin{align}
b\equiv \e^l\simeq 1+l.
\end{align}
Then, Eqs.~(\ref{eq:r'2}) and (\ref{eq:u'2}) reduce to
\begin{align}
\bar r'&=(1+2l)\left(\bar r
+24\bar ul -6 \bar u \bar r l\right)=\bar r
+24\bar ul -6 \bar u \bar r l+ 2 \bar r l +O(l^2),\\
\bar u'&=(1+\epsilon l)\left( \bar u-36\bar u^{2} l \right)=\bar u-36\bar u^{2} l +\epsilon \bar u l.
\end{align}
By writing 
\begin{align}
\frac{\dd \bar r}{\dd l}&\equiv \lim_{l\ra 0}\frac{\bar r'-\bar r}{l},\quad\frac{\dd \bar u}{\dd l}\equiv \lim_{l\ra 0}\frac{\bar u'-\bar u}{l},
\end{align}
we obtain the RG equations,
\begin{align}
\label{eq:RGEqr}
\frac{\dd \bar r}{\dd l}&= 2\bar r +12 \bar u -12 \bar r\bar u,\\
\label{eq:RGEqu}
\frac{\dd \bar u}{\dd l}&=\epsilon \bar u-36\bar u^{2},
\end{align}
which give Eqs.~(\ref{eq:rbarRG}) and (\ref{eq:ubarRG}).

\section{Wilson-Fisher fixed point}
\label{sec:WFappen}
We first assume $r$ and $u$ are order of $\epsilon$ near the critical point. Then we can neglect the last term in the right-hand side of Eq.~(\ref{eq:RGEqr}) and any other higher-order terms coming from the perturbation expansion of $u$. Since the typical length scale disappear at the critical point, the RG invariance emerges in the system. Therefore, we can set
\begin{align}
\frac{\dd \bar r}{\dd l} = \frac{\dd \bar u}{\dd l}=0,
\end{align}
which yields two fixed-point solutions: one is the Gaussian fixed point:
\begin{align}
\bar r =  \bar u =  0,
\end{align}
and the other is the so-called Wilson-Fisher fixed point (we get Eq.~(\ref{eq:WFbody})):
\begin{align}
\bar r =- \frac{\epsilon}{6}\quad \bar u = \frac{\epsilon}{36}.
\end{align}
This solution is consistent with the assumption at the beginning of this subsection. From the liner stability analysis, the Gaussian fixed point is unstable with respect to $\bar u$, whereas the Wilson-Fisher fixed point is stable \cite{Chaikin}.

\section{Useful intergrals}

\begin{align}
\label{eq:integral1/q^2}
 \int^\Lambda_{\Lambda/b} \frac{\dd^d \qv}{(2\pi)^d} 
 \frac{1}{\qv^2} &= \frac{1}{(2\pi)^d} \int \dd \theta \int \dd \varphi_1 \int \dd \varphi_2 ... \int^\Lambda_{\Lambda/b} \dd q q^{d-3} \sin^{d-2} \theta \sin^{d-3} \varphi_1 ...\notag\\  
&= \frac{\Lambda^{d-2}}{(2\pi)^d} \int \dd \theta \int \dd \varphi_1 \int \dd \varphi_2 ... \int^1_{1/b} \dd x x^{d-3} \sin^{d-2} \theta \sin^{d-3} \varphi_1 ...\notag\\  
&\simeq  \frac{\Lambda^{2-\epsilon}}{(2\pi)^4} \int^\pi_0 \dd \theta \sin^2 \theta \int^\pi_0 \dd \varphi_1 \sin \varphi_1 \int^{2\pi}_0 \dd \varphi_2 \int^1_{1/b} \dd x x \notag\\  
&=\frac{\Lambda^{2-\epsilon}}{8 \pi^2}\left(1-\frac{1}{b^2}\right),
\end{align}

\begin{align}
\label{eq:integral1/q^4}
 \int^\Lambda_{\Lambda/b} \frac{\dd^d \qv}{(2\pi)^d} 
 \frac{1}{\qv^4} &= \frac{1}{(2\pi)^d} \int \dd \theta \int \dd \varphi_1 \int \dd \varphi_2 ... \int^\Lambda_{\Lambda/b} \dd q q^{d-5} \sin^{d-2} \theta \sin^{d-3} \varphi_1 ...\notag\\  
&= \frac{\Lambda^{d-4}}{(2\pi)^d} \int \dd \theta \int \dd \varphi_1 \int \dd \varphi_2 ... \int^1_{1/b} \dd x x^{d-3} \sin^{d-2} \theta \sin^{d-3} \varphi_1 ...\notag\\  
&\simeq  \frac{\Lambda^{-\epsilon}}{(2\pi)^4} \int^\pi_0 \dd \theta \sin^2 \theta \int^\pi_0 \dd \varphi_1 \sin \varphi_1 \int^{2\pi}_0 \dd \varphi_2 \int^1_{1/b} \dd x \frac{1}{x} \notag\\  
&=\frac{\Lambda^{-\epsilon}}{8 \pi^2}\ln b.
\end{align}

\newpage \thispagestyle{empty}

\chapter{Coupling to energy-momentum densities}
\label{chap:EM}

In Secs.~\ref{sec:StaticsHDCP} and \ref{sec:dynamicsHDCP}, we have ignored the contributions of the energy and momentum densities, $\varepsilon$ 
and $\bm{\pi}$. In this appendix, we show that these contributions do not affect the static and 
dynamic universality classes of the high-density QCD critical point studied in Secs.~\ref{sec:StaticsHDCP} and \ref{sec:dynamicsHDCP}. In particular, we obtain the expression of the superfluid phonon in the presence of $\varepsilon$ 
and $\bm{\pi}$, Eq.~(\ref{eq:cs2}), which will be mentioned in the discussion in Sec.~\ref{sec:conclusionHDCP}.

\section{Statics}\label{sec:EMsta}
Let us first consider the Ginzburg-Landau functional  in terms of $\sigma$, $n_{\rm B}$, 
$\varepsilon$, $\bm{\pi}$, and $\theta$, up to the second-order expansion. Similarly to the 
argument in Sec.~\ref{sec:StaticsHDCP}, the time reversal symmetry $\mathcal T$ prohibits the mixing between 
$x_{i}\equiv\sigma,n_{\rm B}, \varepsilon$ and $\bm{\pi},\theta$:
\begin{align}
F[\sigma,n_{\rm B},\varepsilon,\bm{\pi},\theta]&=F[\sigma,n_{\rm B},\varepsilon]+F[\bm{\pi},\theta].
\end{align}
The presence of $m_{\rm q}$ and $\mu_{\rm B}$ generally allows the mixing among $\sigma, n_{\rm B},$ and 
$\varepsilon$ in the $\mathcal{T}$-even sector as
\begin{gather}
\label{F_even}
F[\displaystyle \sigma,n_{\rm B},\varepsilon]=\frac{1}{2}\int \dd \bm{r} x_{i}\beta_{ij}(\bm{\nabla})x_{j}, \\
\beta_{ij}(\bm{\nabla})=V_{ij}-v_{ij}\bm{\nabla}^{2}.
\end{gather}
Here, the subscripts $i, j$ are the shorthand notations for $x_{i}, x_{j}$, respectively. The $\mathcal{T}$-odd 
sector is given by
\begin{gather}
\label{F_odd}
F[\displaystyle \bm{\pi},\theta]=\frac{1}{2}\int \dd \bm{r}\left[V_{\pi\pi}\bm{\pi}^{2}+2V_{\pi\theta}\bm{\pi}\cdot\bm{\nabla}\theta+V_{\theta\theta}(\bm{\nabla}\theta)^{2}\right].
\end{gather}
We call $V_{ij}$,\ $v_{ij}$, $V_{\pi\pi}$, $V_{\pi\theta}$, and $V_{\theta\theta}$ the 
Ginzburg-Landau parameters. 

By completing the square, Eq.~(\ref{F_odd}) can be written as 
\begin{align}
F[\displaystyle \bm{\pi},\theta]=\frac{1}{2}\int \dd \bm{r}\left[V_{\pi\pi}({\bm \pi}')^{2}+V_{\theta\theta}'(\bm{\nabla}\theta)^{2}\right],
\end{align}
where ${\bm \pi}' \equiv \bm{\pi}+V_{\pi\theta} \bm{\nabla}\theta/V_{\pi\pi}$ and 
$V_{\theta\theta}' \equiv V_{\theta\theta}-V_{\pi\theta}^{2}/V_{\pi\pi}$. We assume
$V_{\theta\theta}'>0$ and redefine ${\bm \pi}'$ and $V_{\theta\theta}'$ as ${\bm \pi}$ and 
$V_{\theta\theta}$ for simplicity in the following caluculation. 

In a similar manner to Eq.~(\ref{eq:corr}) when the $\varepsilon$ and $\bm{\pi}$ are absent, the correlation length $\xi$ is 
defined from the correlation function of $\sigma$. The form of the correlation function~(\ref{eq:Delta}) can be generalized into  
\begin{align}
\xi\sim(\det V)^{-\frac{1}{2}}\,
\end{align}
in the present case with $V$ being the $3 \times 3$ matrix in $x_{i}\equiv\sigma,n_{\rm B}, \varepsilon$ space. Thus, the critical point is characterized by the condition, $\det V=0$, analogously to the condition 
$\Delta = 0$ in Eq.~(\ref{eq:Delta}). At the critical point, only one of the linear combinations of 
$\sigma$, $n_{\rm B}$, and $\varepsilon$ becomes massless. This number of the gapless mode is the same as that in the absence of the $\varepsilon$ and $\pi$ as we discussed around Eq.~(\ref{eq:FnB}), the static universality class remains the same as that of Sec.~\ref{sec:StaticsHDCP}.

We define the generalized susceptibilities, 
\begin{align}
\displaystyle \chi_{ij} \equiv \left. \frac{\delta\langle x_{i}\rangle_{X_j}}{\delta X_{j}} \right|_{X_{j}=0} \,,
\end{align}
where $\langle x_{i}\rangle_{X_j}$ is the the expectation value of $x_{i}$ in the same definition of Eq.~(\ref{eq:expect}) except for the replacement $\beta F \rightarrow \beta F+\displaystyle \int \dd \bm{r}x_{j}X_{j}$ 
(where the summation over the index $j$ is {\it not} implied). Here, $X_{\varepsilon} \equiv -\beta$, 
$X_{n} \equiv \beta \mu_{\rm B}$, and $X_{\sigma} \equiv \beta m_{q}$ with $\beta$ denotes the inverse temperature. Here and from now on, we omit the subscript of $n_{\rm B}$ when $n_{\rm B}$ is in the subscript of some character, e.g., $X_{n}$ rather than $X_{n_{\rm B}}$.
One can show the following relation between $\displaystyle \chi_{ij}$ and $V_{ij}$
\begin{align}
\label{eq:chiepep}\chi_{\varepsilon\varepsilon}&=\frac{1}{\mathcal{V}}\langle {\varepsilon}^{2}\rangle_{\bm{q}\rightarrow {\bm 0}}=T(V^{-1})_{\varepsilon\varepsilon}\,,\\
\chi_{\varepsilon n}&=\frac{1}{\mathcal{V}}\langle \varepsilon n_{\rm B} \rangle_{\bm{q}\rightarrow {\bm 0}}=T(V^{-1})_{\varepsilon n}\,,\\
\label{eq:chinn}
\chi_{nn}&=\frac{1}{\mathcal{V}}\langle n_{\rm B}^{2}\rangle_{\bm{q}\rightarrow {\bm 0}}=T(V^{-1})_{nn}\,,
\end{align} 
where $(V^{-1})_{ij}$ denotes the $(i,j)$ component of the inverse matrix of $V$. Since 
$(V^{-1})_{ij}\propto(\det V)^{-1}$, one finds
\begin{align}
\label{eq:chixis}
\chi_{\varepsilon\varepsilon}\sim\chi_{\varepsilon n}\sim\chi_{nn}\sim\xi^{2-\eta},
\end{align}
which can be regard as a generalization of Eq.~(\ref{eq:suscptibilities}).

\section{Dynamics}
\label{sec:fulldynamics}
\subsection{Full Langevin equations}
We next construct the Langevin equations 
of the full hydrodynamic variables $x_{i}=\sigma,n_{\rm B}, \varepsilon$ and $\bm{\pi},\theta$. The 
Langevin equations read 
\begin{align}
\label{eq:langex}
\frac{\d x_i}{\d t}&=-\gamma_{ij}(\na)\frac{\delta F}{\delta x_{j}}
-\int_{ V} \left[\tilde{x}_{i},\bm{\pi} \right]\cdot\frac{\delta F}{\delta\bm{\pi}} -\displaystyle \int_{V} \left[\tilde{x}_{i},\theta \right]\frac{\delta F}{\delta\theta}+\xi_{i},\\
%%%
\label{eq:langepi}
\frac{\d \bm{\pi}}{\d t}
&=\Gamma_{\pi\pi}\bm{\nabla}\bm{\nabla}\cdot\frac{\delta F}{\delta\bm{\pi}}+\Gamma_{\pi\pi}'\bm{\nabla}^{2}
\frac{\delta F}{\delta\bm{\pi}}-\Gamma_{\pi\theta}\bm{\nabla}\frac{\delta F}{\delta\theta}-\displaystyle \int_V \left[\bm{\pi},\tilde{x}_{i}\right]\frac{\delta F}{\delta x_{i}}+\bm{\xi}_{\pi},\\
%%%
\label{eq:langetheta2}
\frac{\d \theta}{\d t}
&=-\Gamma_{\pi\theta}\bm{\nabla}\cdot\frac{\delta F}{\delta\bm{\pi}}-\Gamma_{\theta\theta}\frac{\delta F}{\delta\theta}-\displaystyle \int_V \left[\theta,\tilde{x}_{i}\right]\frac{\delta F}{\delta x_{i}}+\xi_{\theta},
\end{align}
where we use the shorthand notation of the integrals, Eq.~ (\ref{eq:simple}). Summation over repeated indices $i, j$ is also understood. Here 
\begin{align}
\gamma_{ij}(\na)\equiv\left(\begin{array}{ccc}
\Gamma_{\sigma\sigma} &-\Gamma_{\sigma n}\bm{\nabla}^{2} &-\Gamma_{\sigma \varepsilon}\bm{\nabla}^{2}\\
-\Gamma_{\sigma n}\bm{\nabla}^{2}& -\Gamma_{n n}\bm{\nabla}^{2} &-\Gamma_{n \varepsilon}\bm{\nabla}^{2}\\
-\Gamma_{\sigma \varepsilon}\bm{\nabla}^{2}& -\Gamma_{n \varepsilon}\bm{\nabla}^{2}& -\Gamma_{\varepsilon \varepsilon}\bm{\nabla}^{2}
\end{array}\right)\,,
\end{align}
$\Gamma_{ab}$ ($a,b=x_{i},\pi,\theta$), and $\Gamma_{\pi\pi}'$ are the kinetic 
coefficients. We also define the net variables $\tilde \sigma \equiv \bar q q$, $\tilde n_{\rm B}\equiv \bar q \gamma^0 q$, and $\tilde{\varepsilon} \equiv T^{00}$. On the other hand, we note again that $\sigma=\tilde \sigma -\sigma_{\rm eq}$, $n_{\rm B}=\tilde n_{\rm B}  -(n_{\rm B})_{\rm eq}$, $\varepsilon \equiv \tilde \varepsilon -\varepsilon_{\rm eq}$ are the fluctuations around the equilibrium values $\sigma_{\rm eq}$, $(n_{\rm B})_{\rm eq}$, and $\varepsilon_{\rm eq}$. The noise terms $\xi_{i},\ \bm{\xi}_{{\pi}},$ and 
$\xi_{\theta}$ above are not important in the following calculation. In order to write down the above Langevin equations, we use the momentum conservation law and the Onsager's principle, 
similarly to the derivation in Sec.~\ref{sec:dynamics}.

We postulate the following Poisson brackets, in addition to Eqs.~(\ref{eq:ntheta}),
\begin{align}
\label{eq:brasig}
[\bm{\pi}(\bm{r}),\tilde{\sigma}(\bm{r}')]&=\tilde{\sigma}(\bm{r}')\bm{\nabla}\delta(\bm{r}-\bm{r}'),\\
\label{eq:brapin}[\bm{\pi}(\bm{r}),\tilde{n}_{\rm B}(\bm{r}')]&=\tilde{n}_{\rm B}(\bm{r}')\bm{\nabla}\delta(\bm{r}-\bm{r}'), \\
[\bm{\pi}(\bm{r}),\tilde{s}(\bm{r}')]&=\tilde{s}(\bm{r}')\bm{\nabla}\delta(\bm{r}-\bm{r}'),\\
\label{eq:brathetas}
[\theta(\bm{r}),\tilde{s}(\bm{r}')]&=0,
\end{align}
where $\tilde{s}(\bm{r})$ denotes the entropy density. The Poisson brackets concerning 
$\tilde{\varepsilon}(\bm{r})$ can be derived as
\begin{align}
\label{eq:brapiepsilon}
[\bm{\pi}(\bm{r}),\tilde{\varepsilon}(\bm{r}')]&=\left(T\tilde{s}(\bm{r})+\mu_{\rm B} \tilde{n}_{\rm B}(\bm{r})\right) \bm{\nabla}\delta(\bm{r}-\bm{r}'),\\
\label{eq:brathetaepsilon}
[\theta(\bm{r}),\tilde{\varepsilon}(\bm{r}')]&=\mu_{\rm B}\delta(\bm{r}-\bm{r}').
\end{align}
Here, we used the thermodynamic relation $\dd \varepsilon=T \dd s+\mu_{\rm B} \dd n_{\rm B}$ and the 
definition of the Poisson brackets 
$\displaystyle \left[\bm{\pi}(\bm{r}),\tilde{y}_{i}(\bm{r}')\right] \equiv \delta \tilde{y}_{i}(\bm{r}')/\delta \bm{u}(\bm{r})$ 
for $\tilde{y}_{i}\equiv \tilde{n}_{\rm B},\tilde{s},\tilde{\varepsilon}$, with $\bm{u}(\bm{r})$ being the infinitesimal 
translation of the coordinate, $\bm{r}\rightarrow\bm{r}+\bm{u}(\bm{r})$. 
(See, e.g., Ref.~\cite{Volovick} for the details of the Poisson brackets.)

We consider the small fluctuations of variables around the equilibrium values, 
$\sigma(\bm{r}) = \tilde{\sigma}(\bm{r})-\sigma_{\text{eq}}$, $n_{\rm B}(\bm{r}) = \tilde{n}_{\rm B}(\bm{r}) - (n_{\rm B})_{\text{eq}}$, 
and $s(\bm{r}) = \tilde{s}(\bm{r}) - s_{\text{eq}}$. Then, the linearized Langevin equations neglecting the higher order fluctuations become:
\begin{align}
\label{eq:linearsig}
\frac{\d \sigma}{\d t} &= -\left(\Gamma_{\sigma \sigma} V_{\sigma i} - \Gamma_{\sigma j} v_{j i} \bm{\nabla}^{2}\right) x_i 
- \displaystyle \sigma_{\text{eq}}V_{\pi\pi}\bm{\nabla}\cdot\bm{\pi},\\
%%%%%%
\frac{\d n_{\rm B}}{\d t}&= \Gamma_{n j}v_{j i}\bm{\nabla}^{2}x_i 
-\displaystyle  (n_{\rm B})_{\text{eq}}V_{\pi\pi}\bm{\nabla}\cdot\bm{\pi}-\displaystyle V_{\theta\theta}\bm{\nabla}^{2}\theta,\\
%%%%%%
\frac{\d \varepsilon}{\d t} &= \Gamma_{\varepsilon j}v_{j i}\bm{\nabla}^{2}x_i 
-w_{\text{eq}}V_{\theta\theta}\bm{\nabla}\cdot\bm{\pi}-\mu V_{\theta\theta}\bm{\nabla}^{2}\theta,\\
%%%%%
\frac{\d \bm{\pi}}{\d t} &= -\left[\sigma_{\mathrm{e}\mathrm{q}}V_{\sigma i}+ (n_{\rm B})_{\text{eq}}V_{ ni}+w_{\text{eq}}V_{\varepsilon i}\right]\bm{\nabla}x_i+\Gamma_{\pi\pi}V_{\pi\pi}\bm{\nabla}\bm{\nabla}\cdot\bm{\pi}
+\Gamma_{\pi\pi}'V_{\pi\pi}\bm{\nabla}^{2}\bm{\pi}, \\
%%%%%
\label{eq:lineartheta}
\frac{\d \theta}{\d t} & = -\left[\left(V_{n i}+\mu_{\rm B} V_{\varepsilon i}\right)-\left(v_n+\mu_{\rm B} v_{\varepsilon i}\right)\bm{\nabla}^{2}\right]x_i -\Gamma_{\pi\theta}V_{\pi\pi}\bm{\nabla}\cdot\bm{\pi}
+\Gamma_{\theta\theta} V_{\theta\theta}\bm{\nabla}^{2}\theta,
\end{align}
where $w_{\text{eq}} = Ts_{\text{eq}}+\mu_{\rm B} (n_{\rm B})_{\text{eq}}$. 
% (Note again that the subscripts $i,j$ are shorthand notation for $x_i, x_j$.)

\subsection{Decomposition of momentum density}
In this section, we decompose the momentum density $\pi^{i}$ 
into the longitudinal and transverse parts with respect to momentum ${\bm q}$ in $(t, {\bm q})$ space, 
\begin{align}
\pi^{i}=\pi_{\rm L}^{i}+\pi_{\rm T}^{i}, \quad \pi^i_{\rm L} = (P_{\rm L})^{ij} \pi^j, 
\quad \pi^i_{\rm T} = (P_{\rm T})^{ij} \pi^j,
\end{align}
where $P_{\rm L,T}$ are the longitudinal and transverse projection operators defined by
\begin{align}
(P_{\rm L})^{ij} \equiv \frac{q^i q^j}{|{\bm q}|^2}\,, \quad  
(P_{\rm T})^{ij} \equiv \delta_{ij} - \frac{q^i q^j}{|{\bm q}|^2}\,.
\end{align}

In the following calculation, we show that the dynamics of $\pi_{\rm T}^{i}$ is decoupled from that of the other hydrodynamic variables. We first write the linearized Langevin equations (\ref{eq:linearsig})--(\ref{eq:lineartheta}) to 
the leading order of ${\bm q}$, formally as
\begin{align}
\label{eq:sigpi}
\frac{\d x_k}{\d t} &= A_{x_k}(\bm{q}\cdot\bm{\pi}) + f(x_k,\theta)\,, \\
\label{eq:pipi}
\frac{\d \bm \pi}{\d t} &= A_{\pi} {\bm q}({\bm q}\cdot{\bm \pi}) 
+ B_{\pi} |{\bm q}|^2 {\bm \pi} + {\bm q} g(x_k)\,,\\
\label{eq:thetapi}
\frac{ \d \theta}{\d t}&=A_{\theta}(\bm{q}\cdot\bm{\pi})+h(x_{k},\theta)\,,
\end{align}
where $f(x_k,\theta)$, $g(x_k)$ and $h(x_k,\theta)$ are some functions which may involve $x_k$ and 
$\theta$, but not involve $\bm{\pi}$. The coefficients $A_{x_k}$, $A_{\pi}$, $A_{\theta}$, and $B_{\pi}$ are the parameters which depend on the Ginzburg-Landau parameters $V_{ij},\ v_{ij}$, kinetic 
coefficients, and equilibrium values of the thermodynamic quantities. The explicit forms of these functions ($f$, $g$, and $h$) and the 
coefficients ($A_{x_k}$, $A_{\pi}$, $A_{\theta}$, and $B_{\pi}$) are not important for our purpose. By using ${\bm \pi}_{\rm L,T}$, we can write 
Eqs.~(\ref{eq:sigpi})--(\ref{eq:thetapi}) into
\begin{align}
\frac{ \d x_k}{\d t} &= A_{x_k}(\bm{q}\cdot\bm{\pi}_{\rm L}) + f(x_k,\theta)\,, \\
\frac{\d {\bm \pi}_{\rm L}}{\d t } &=  (A_{\pi} + B_{\pi}) |{\bm q}|^2 
{\bm \pi}_{\rm L} + {\bm q} g(x_k)\,, \\
\frac{\d {\bm \pi}_{\rm T}}{\d t} &= B_{\pi} |{\bm q}|^2 {\bm \pi}_{\rm T}\,,\\
\frac{\d \theta}{\d t}&=A_{\theta}(\bm{q}\cdot\bm{\pi}_{\text{L}})+h(x_{k},\theta)\,.
\end{align}
One can confirm that the dynamics of $\pi^{i}_{\rm L}$ and $\pi_{\rm T}^{i}$ are decoupled from each other at the 
mean-field level.

\subsection{Hydrodynamic modes}
Thanks to the decomposition in the previous section, we can readily obtain the hydrodynamic mode for $\bm{\pi}_{T}$,
\begin{align}
\label{eq:pitdynamics}
\left(i\omega-B_{\pi}\bm{q}^{2}\right)\bm{\pi}_{\mathrm{T}}=0,\quad B_{\pi}=\Gamma_{\pi\pi}'V_{\pi\pi},
\end{align}
which is a diffusion mode.

The other Langevin equations which involve ${\bm \pi}_{\rm L}$ (but not $\bm{\pi}_{\rm T}$) can be expressed in the form of the matrix equation,
\newpage
\begin{align}
\label{eq:langeem}
\mathcal M_{\rm EM}
\left(\begin{array}{c}
\sigma\\
 n_{\rm B}\\
\varepsilon\\
\bm{\pi}_{\mathrm{L}}\\
\theta
\end{array}\right)=0\,,
\end{align}
where
\begin{align}
&\mathcal M_{\rm EM} \notag\\
&\equiv \left(\begin{array}{rrrrr}
\i\omega-A_{\sigma\sigma}-a_{\sigma\sigma}\bm{q}^{2}& -A_{\sigma n}-a_{\sigma n}\bm{q}^{2} &-A_{\sigma\varepsilon}-a_{\sigma\varepsilon}\bm{q}^{2}& -\i a_{\sigma\pi}\bm{q}& 0\\
%%%
-a_{n\sigma}\bm{q}^{2} & \i\omega-a_{nn}\bm{q}^{2}&  -a_{n\varepsilon}\bm{q}^{2}  &-\i a_{n\pi}\bm{q}  &-a_{n\theta}\bm{q}^{2}\\
%%%
-a_{\varepsilon\sigma}\bm{q}^{2} & -a_{\varepsilon n}\bm{q}^{2} & \i\omega-a_{\varepsilon\varepsilon}\bm{q}^{2} & -\i a_{\varepsilon\pi}\bm{q}&  -a_{\varepsilon\theta}\bm{q}^{2}\\
%%%
-\i a_{\pi\sigma}\bm{q} & -\i a_{\pi n}\bm{q} & -\i a_{\pi\varepsilon}\bm{q} & \i\omega-a_{\pi\pi}\bm{q}^{2} & 0\\
%%%
-A_{\theta\sigma}-a_{\theta\sigma}\bm{q}^{2} &-A_{\theta n}-a_{\theta n}\bm{q}^{2} &-A_{\theta\varepsilon}-a_{\theta\varepsilon}\bm{q}^{2}& -\i a_{\theta\pi}\bm{q} &\i\omega-a_{\theta\theta}\bm{q}^{2}
\end{array}\right).
\end{align}
Here $A_{ab}$ and $a_{ab}$ are the parameters depending on $V_{ij},\ v_{ij}$, 
kinetic coefficients, and thermodynamic quantities. Note here that $A_{ab},\ a_{ab}$ 
are not symmetric with respect to $a$ and $b$. 
The eigenfrequencies of Eq.~(\ref{eq:langeem}) can be found from $\det\mathcal{M}=0$,
which yields
\begin{align}
\label{eq:proper}
\omega^{5}&+\i\left(A_{\sigma\sigma}+g_{4}\bm{q}^{2}\right)\omega^{4}
-\left[g_{3}\bm{q}^{2}+O(\bm{q}^{4})\right]\omega^{3}-\i\left[g_{2}\bm{q}^{2}+O(\bm{q}^{4})\right]\omega^{2}\notag\\
&\qquad\qquad\quad\qquad\qquad+\left[g_{1}\bm{q}^{4}+O(\bm{q}^{6})
\right]\omega+\i\left[g_{0} \bm{q}^{4}+O(\bm{q}^{6})\right]=0,
\end{align}
where $g_0,..., g_{4}$ are some functions of $A_{ab}$ and $a_{ab}$. 
Among others, we give the explicit 
expressions only for $A_{\sigma\sigma}$, $g_{2}$, and $g_{0}$ which will be used in the following arguments,
\begin{align}
A_{\sigma\sigma}&\equiv \Gamma_{\sigma\sigma},\\
%%%
\label{eq:zpara}
g_{2}&\equiv
\Gamma_{\sigma\sigma}
\left[(n_{\rm B})_{\rm eq}^2V_{\pi\pi}+V_{\theta\theta}\right]\left|\begin{array}{cc}
V_{\sigma\sigma} &V_{\sigma n}\notag\\
V_{\sigma n} &V_{nn}
\end{array}\right|\\
&\quad+2\Gamma_{\sigma\sigma} \left[(n_{\rm B})_{\rm eq}w_{\text{eq}}V_{\pi\pi}+\mu_{\rm B} V_{\theta\theta}\right]\left|\begin{array}{cc}
V_{\sigma\sigma}&V_{\sigma n}\\
V_{\sigma\varepsilon}&V_{n\varepsilon}
\end{array}\right|\notag \\
&\quad+\Gamma_{\sigma\sigma} (w_{\text{eq}}^{2}V_{\pi\pi}+\mu_{\rm B}^{2} V_{\theta\theta})\left|\begin{array}{cc}
V_{\sigma\sigma}&V_{\sigma\varepsilon}\\
V_{\sigma\varepsilon}&V_{\varepsilon\varepsilon}
\end{array}\right|,\\
%%%
\label{eq:psi}g_{0}
&\equiv \displaystyle \Gamma_{\sigma\sigma}V_{\pi\pi}V_{\theta\theta}T^{2} s_{\text{eq}}^{2} \det V.
\end{align}

Equation~(\ref{eq:proper}) can be factorized as
\begin{align}
&\left[\omega+\i A_{\sigma\sigma}+\i\left(g_{4}-\frac{g_{3}}{A_{\sigma\sigma}}+\frac{g_{2}}{A_{\sigma\sigma}^{2}}\right)\bm{q}^{2}\right]\notag \\
&\times\left[\omega-s_{+}|\bm{q}|+\i\frac{t_{+}}{2}\bm{q}^{2}\right]\left[\omega+s_{+}|\bm{q}|+\i\frac{t_{+}}{2}\bm{q}^{2}\right]\notag \\
\label{eq:proper2}&\times\left[\omega-s_{-}|\bm{q}|+\i\frac{t_{-}}{2}\bm{q}^{2}\right]\left[\omega+s_{-}|\bm{q}|+\i\frac{t_{-}}{2}\bm{q}^{2}\right]=0,
\end{align}
where $s_{\pm}$ and $t_{\pm}$ satisfy
\begin{align}
s_{+}^{2}+s_{-}^{2}&=\displaystyle \frac{g_{2}}{A_{\sigma\sigma}},\\
s_{+}^{2}s_{-}^{2}&=\displaystyle \frac{g_{0}}{A_{\sigma\sigma}},\\
t_{+}+t_{-}&=\displaystyle \frac{g_{3}}{A_{\sigma\sigma}}-\frac{g_{2}}{A_{\sigma\sigma}^{2}},\\
s_{+}^{2}t_{-}+s_{-}^{2}t_{+}&=\displaystyle \frac{g_{1}}{A_{\sigma\sigma}}-\frac{g_{0}}{A_{\sigma\sigma}^{2}}.
\end{align}

From Eq.~(\ref{eq:proper2}), we finds the hydrodynamic modes of the system other than the diffusion 
mode described by Eq.~(\ref{eq:pitdynamics}): one relaxation mode and two pairs of phonons. 
The speeds of phonons $s_{\pm}$ can be obtained from the solution of
\begin{align} 
s^{4}-\frac{g_{2}}{A_{\sigma\sigma}}s^{2}+\frac{g_{0}}{A_{\sigma\sigma}}=0.
\end{align}
Note here that Eq.~(\ref{eq:psi}) shows that $g_{0} \rightarrow 0$ when the critical point
is approached $\det V\rightarrow 0$. Therefore, near the critical point, we obtain 
\begin{align}
s_{+}^{2}=\displaystyle \frac{g_{2}}{A_{\sigma\sigma}}-\frac{g_{0}}{g_{2}}\,,\quad s_{-}^{2}=\frac{g_{0}}{g_{2}}\,.
\end{align}

\subsection{Dynamic critical exponent}
From the results above, we can show that one of the phonons with the speed $c_{\rm s}\equiv s_{-}$ 
exhibits the critical slowing down. By using Eqs.~(\ref{eq:zpara}) and (\ref{eq:psi}), 
together with Eqs.~(\ref{eq:chiepep})--(\ref{eq:chinn}), we have
\begin{align}
\label{eq:cs2}
c_{\rm s}^{2}&=\displaystyle \frac{V_{\pi\pi}V_{\theta\theta}T^{3}s_{\mathrm{e}\mathrm{q}}^{2}}{\kappa_{nn}\chi_{nn}+2\kappa_{n\varepsilon}\chi_{n\varepsilon}+\kappa_{\varepsilon\varepsilon}
\chi_{\varepsilon\varepsilon}}\,,
\end{align}
where
\begin{align}
\kappa_{nn}&\equiv (n_{\rm B})_{\rm eq}^{2}V_{\pi\pi}+V_{\theta\theta},\\
\kappa_{n\varepsilon}&\equiv (n_{\rm B})_{\rm eq} w_{\mathrm{e}\mathrm{q}}V_{\pi\pi}+\mu_{\rm B} V_{\theta\theta},\\
\kappa_{\varepsilon\varepsilon}&\equiv w_{\mathrm{e}\mathrm{q}}^{2}V_{\pi\pi}+\mu_{\rm B}^{2}V_{\theta\theta}\,.
\end{align}
Using Eq.~(\ref{eq:chixis}), $V_{\pi\pi}\sim \xi^0$, and $V_{\theta\theta}\sim \xi^0$ 
near the critical point, we obtain
\begin{align}
\label{eq:cs22}
c_{\rm s}^2 \sim \xi^{-2 + \eta}.
\end{align} 
Comparing Eq.~(\ref{eq:cs22}) with Eq.~(\ref{eq:zdef}), we obtain the same dynamic critical exponent $z$ as that 
given by Eq.~(\ref{eq:z}). Therefore, the dynamic universality class remains the same as that without $\varepsilon$ and ${\bm \pi}$ in Sec.~\ref{sec:dynamics}.

\newpage \thispagestyle{empty}

\chapter{Calculation of the self-energies and the vertex function}
\chaptermark{Calculation on $\Sigma$, $\Pi$, and $\mathcal V$}
\label{sec:calculation}
As we mentioned under Eq.~(\ref{eq:ctilde5c5}), in this appendix, we evaluate the self-energies of the order parameter, $\Sigma_{\alpha\beta}(\kv,\omega)$, and those of the conserved charge densities, $\Pi_{ij}(\kv,\omega)$, to the first order of $\epsilon$, namely at the one-loop level. From these expressions, 
%we can compute their derivatives with respect to $\omega$, $\kv$ or ${\bm k}^2$ that are used in Eqs.~(\ref{eq:RRGamma})--(\ref{eq:RRCchi5}), (\ref{eq:ctildec}), and (\ref{eq:ctilde5c5}). 
we can derive the recursion relations for the dynamic parameters, Eqs.~(\ref{eq:RRGamma2})--(\ref{eq:RRC}). In particular, we calculate $\Pi_{ij}$ and derive Eqs.~(\ref{eq:RRlambda2})--(\ref{eq:RRC}) in Sec.~\ref{sec:PiCME}; we calculate $\Sigma_{\alpha\alpha}$ and derive Eq.~(\ref{eq:RRGamma2}) in Sec.~\ref{sec:SigmaCME}. We also evaluate the three-point vertex function used in Eq.~(\ref{eq:RRgchi5}) at the one-loop level, and derive Eq.~(\ref{eq:RRg}).

\section{Self-energy $\Pi$}
\label{sec:PiCME}
\begin{figure}[t]
\begin{center}
\includegraphics[bb=0 0 215 137,height=4cm]{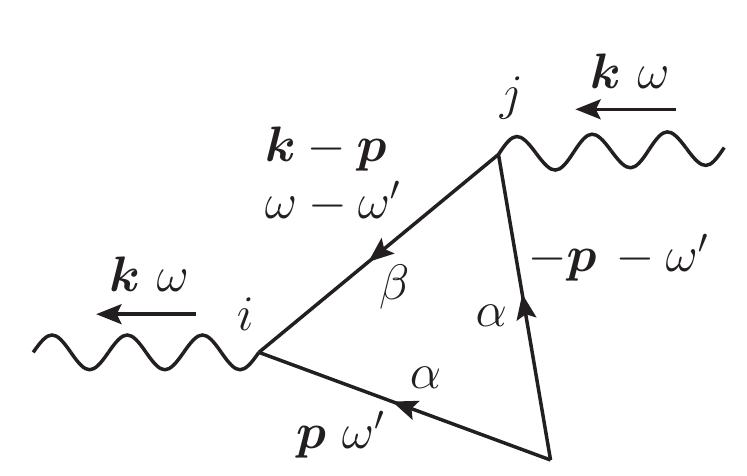}
\caption{Diagram for $\Pi_{ij}$ at one-loop level} \label{fig:Pi}  
\end{center}
\end{figure}

Let us begin with the self-energy for the conserved-charge densities, $\Pi_{ij}(\kv,\omega)$, which corresponds to the diagrams with outgoing $\tilde n_i$ and ingoing $n_j$. The leading diagram is given by Fig.~\ref{fig:Pi}. According to the Feynman rules on the analytic translation of the diagrams of the (plane and wavy) lines and the noise vortices (Fig.~\ref{fig:NV}) and the interaction vortices (Fig.~\ref{fig:IV}), summarized in Sec.~\ref{sec:DPTCME}, we can calculate Fig.~\ref{fig:Pi} as
\begin{align}
\label{eq:Pi0CME}
\Pi_{ij}(\kv,\omega)
&=
\int \frac{\dd \omega'}{2\pi}\int \frac{\dd^d p}{(2\pi)^d}
V^0_{i; \alpha \beta}(\kv,\pv) G^0(\kv-\pv,\omega-\omega') |G^0(\pv,\omega')|^2 2\Gamma V^0_{\beta;\alpha j}
\notag\\
&=
\int \frac{\dd^d p}{(2\pi)^d}
\frac{ F _{ij}}
{\{-\i\omega+\Gamma[r+(\kv-\pv)^2]+\Gamma(r+\pv^2)\}(r+\pv^2)}\,,
\end{align}
where we defined $F_{ij}$ as the components of the following matrix:
\begin{align}
F\equiv
\left(
\begin{array}{cc}
8\gamma^2\Gamma\lambda \kv^2&0\\
8\i\gamma^2 C \Gamma \Bv\cdot\kv&-2g^2[(\kv-\pv)^2-\pv^2]/\chi_5
\end{array}
\right).
\end{align}
From this expression of the self energy $\Pi_{ij}$, we can calculate its derivatives which are required in the recursion relations~(\ref{eq:RRlambdachi})--(\ref{eq:RRCchi5}) and the normalization conditions~(\ref{eq:ctildec}) and (\ref{eq:ctilde5c5}).

It is easy to calculate the latter conditions. 
Because $\Pi_{ij}(\zev,\omega)=0$, the normalization conditions~(\ref{eq:ctildec}) and (\ref{eq:ctilde5c5}) reduce to
\begin{align}
\label{eq:ctilde2}
\tilde{c}&=d-c,\quad
\tilde{c}_5=d-c_5.
\end{align} 

Let us calculate the derivatives needed in the recursion relations~(\ref{eq:RRlambdachi})--(\ref{eq:RRCchi5}),
\begin{align}
\left.  \i \frac{\d \Pi_{21}(\bm{k},0)}{\d \bm{k}}    \right|_{\bm{k}\rightarrow \bm{0}}&=-4\gamma^2 C\Bv
\int \frac{\dd^d p}{(2\pi)^d}\left(
\frac{1}{\pv^4}+O(\epsilon)\right)=-\frac{\gamma^2 C\Bv \Lambda^{-\epsilon}}{2\pi^2} \ln b,
\label{eq:dPidk21}\\
%%%%
\frac{1}{2}\left.  \frac{\d^2 \Pi_{11} (\bm{k},0)}{\d \bm{k}^2}    \right|_{\bm{k}\rightarrow \bm{0}}
&=4\gamma^2\lambda
\int \frac{\dd^d p}{(2\pi)^d}\left(
\frac{1}{\pv^4}+O(\epsilon)\right)
\label{eq:dPidk11}
=\frac{\gamma^2\lambda\Lambda^{-\epsilon}}{2\pi^2} \ln b,
\end{align}
where we carried out the integral over ${\bm p}$ in the shell $\Lambda/b<|\pv|<\Lambda$ using the standard formula in the dimensional regularization:
\begin{align}
\int \frac{\dd^d p}{(2\pi)^d}\frac{1}{\pv^4}=\frac{\Lambda^{-\epsilon}}{8\pi^2}\ln b.
\end{align}
We also compute
%%%%
\begin{align}
\frac{1}{2}\left.  \frac{\d^2 \Pi_{22} (\bm{k},0)}{\d \bm{k}^2}    \right|_{\bm{k}\rightarrow \bm{0}}
&=\frac{g^2}{\chi_5 \Gamma}
\int \frac{\dd^d p}{(2\pi)^d}\left(
\frac{\cos 2\theta}{\pv^4}+O(\epsilon)\right)
\label{eq:dPidk22}
=-\frac{g^2 \Lambda^{-\epsilon}}{16\pi^2 \chi_5 \Gamma} \ln b,
\end{align}
where we parameterize $\pv$ in the limit of $d\rightarrow 4$ by
\begin{align}
\label{eq:ppara}
(p_1,p_2,p_3,p_4) = p(\cos\theta,\,\sin\theta\cos\phi,\,\sin\theta\sin\phi\cos\varphi,\,\sin\theta\sin\phi\sin\varphi)
\end{align}
with $\kv=(1,0,0,0)$ and use
\begin{align}
\int \frac{\dd^d p}{(2\pi)^d}\frac{\cos 2 \theta}{\pv^4}=-\frac{\Lambda^{-\epsilon}}{16\pi^2} \ln b\,.
\end{align}

Now it ie ready to rewrite the recursion relations~(\ref{eq:RRlambdachi})--(\ref{eq:RRCchi5}). By using Eqs.~(\ref{eq:ctilde2})--(\ref{eq:dPidk11}), and (\ref{eq:dPidk22}), we obtain
\begin{align}
\label{eq:RRlambdachi2}
\frac{\lambda_{l+1}}{\chi_{l+1}}&=b^{z-2}\frac{\lambda_l}{\chi_{l}}\left(1- 4v_l \ln b
  \right),\\
\label{eq:RRlambda5chi52}
\frac{(\lambda_5)_{l+1}}{(\chi_5)_{l+1}}&=b^{z-2}
\frac{(\lambda_5)_l}{(\chi_5)_l}
\left(1+\frac{f_l}{2}\ln b \right),\\
\label{eq:RRCchi2}
\frac{C_{l+1} }{\chi_{l+1}}&=b^{z+c_5-c-1}\frac{C_l }{\chi_l}\left(1-4v_l \ln b \right),\\
\frac{C_{l+1} }{(\chi_5)_{l+1}}&=b^{z+c-c_5-1}\frac{C_l }{(\chi_5)_l}.
\end{align}
Then, we can derive Eqs.~(\ref{eq:RRlambda2})--(\ref{eq:RRC}) by using Eqs.~(\ref{eq:RRchi}), (\ref{eq:RRchi5}), and (\ref{eq:RRlambdachi2})--(\ref{eq:RRCchi2}).

\section{Self-energy $\Sigma$}
\label{sec:SigmaCME}
\begin{figure}[t]
\begin{center}
\subfigure[]{ \includegraphics[bb=0 0 215 122,height=4cm]{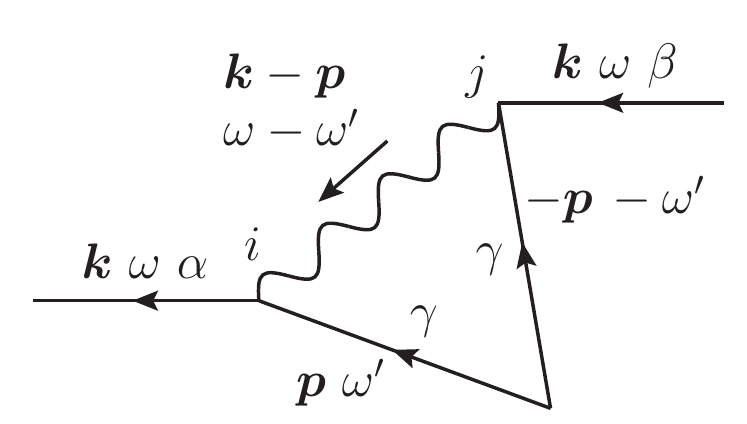}\label{fig:Sigma(a)}}~
\subfigure[]{ \includegraphics[bb=0 0 218 133,height=4cm]{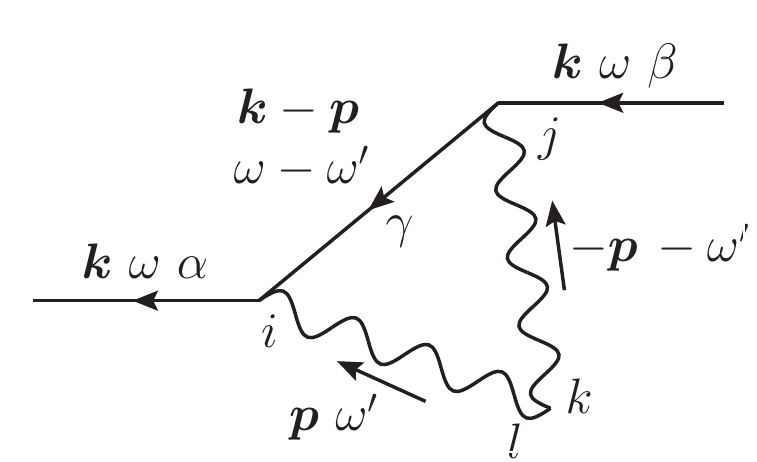}\label{fig:Sigma(b)}}
\caption{Diagrams for $\Sigma_{\alpha\beta}$ at one-loop level}\label{fig:Sigma}
\end{center}
\end{figure}
Next, let us evaluate the self-energy of the order parameter field $\phi_\alpha$, $\Sigma_{\alpha\beta}(\kv,\omega)$, 
whose diagrams are 
shown in Fig.~\ref{fig:Sigma},
\begin{align}
\Sigma_{\alpha\beta}(\kv,\omega)&=\Sigma^{(a)}_{\alpha\beta}(\kv,\omega)+\Sigma^{(b)}_{\alpha\beta}(\kv,\omega),
\end{align} 
where $\Sigma^{(a)}$ and $\Sigma^{(b)}$ correspond to figures~\ref{fig:Sigma(a)} and \ref{fig:Sigma(b)}, respectively. Similarly to the expression for Eq.~(\ref{eq:Pi0CME}), the  diagonal ($\alpha\alpha$) components used in the recursion relation~(\ref{eq:RRGamma}) and the normalization condition~(\ref{eq:atildea}) are given by
\begin{align}
&\Sigma_{\alpha\alpha}^{(a)}(\kv,\omega)\notag\\
&=\int \frac{\dd \omega'}{2\pi}\int \frac{\dd^d p}{(2\pi)^d}
V^0_{\alpha; \gamma i} D^0_{ij}(\kv-\pv,\omega-\omega') |G^0(\pv,\omega')|^2 2\Gamma V_{j;\gamma\alpha} (\kv-\pv,-\pv)\notag\\
&=
4\gamma^2\Gamma\int \frac{\dd^d p}{(2\pi)^d}
\frac{
 \lambda(\kv-\pv)^2[-\i \omega +\Gamma(r+\pv)^2 +\lambda_5(\kv-\pv)^2/\chi_5 ]+  C^2 [\Bv\cdot(\kv-\pv)]^2/\chi_5  }
{(r+\pv^2)\det [D ^0(\kv-\pv,\omega+\i\Gamma(r+\pv^2))]^{-1}}\notag\\
&\quad+\frac{g^2}{\chi_5}\int \frac{\dd^d p}{(2\pi)^d}
\frac{(\pv^2-\kv^2)[-\i\omega+ \Gamma(r+\pv)^2 +\lambda(\kv-\pv)^2/\chi]
}
{(r+\pv^2)\det [D ^0(\kv-\pv,\omega+\i\Gamma(r+\pv^2))]^{-1}}\,,
\label{eq:Sigma(a)2} 
\end{align}

\begin{align}
&\Sigma_{\alpha\alpha}^{(b)}(\kv,\omega)\notag\\
&=\int \frac{\dd \omega'}{2\pi}\int \frac{\dd^d p}{(2\pi)^d}
V^0_{\alpha; \gamma i}G^0(\kv-\pv,\omega-\omega') B^0 _{ij}(\pv,\omega') V_{\gamma;\alpha j}\notag\\
&=8\gamma^2\Gamma^2
\int \frac{\dd \omega'}{2\pi}\int \frac{\dd^d p}{(2\pi)^d}
\frac{\lambda\pv^2(\omega'^2+\kappa_5\pv^2)+\lambda_5 (C\Bv\cdot\pv)^2/\chi_5^2 }{|\det [D^0(\pv,\omega')]^{-1}|^2\{-\i(\omega-\omega')+\Gamma[r+(\kv-\pv)^2]\}}\notag\\
&\quad - \frac{2g^2}{\chi_5^2}\int \frac{\dd \omega'}{2\pi}\int \frac{\dd^d p}{(2\pi)^d}
\frac{ \lambda_5 \pv^2(\omega'^2+\kappa\pv^2)+\lambda (C\Bv\cdot\pv)^2/\chi^2}{|\det [D^0(\pv,\omega')]^{-1}|^2\{-\i(\omega-\omega')+\Gamma[r+(\kv-\pv)^2]\}}\,.
\label{eq:Sigma(b)b}
\end{align}
Note that we do not take a sum over $\alpha$ here.

The integral over $\omega'$ in Eq.~(\ref{eq:Sigma(b)b}) can be performed by closing contour below and picking up the poles on a lower half plane,
\begin{align}
\det [D^0(\pv,\omega')]^{-1}
&=\left[-\i\omega'+\kappa_+ \pv^2+ \i\Omega(\pv)\right]\left[-\i\omega'+\kappa_+ \pv^2- \i\Omega(\pv)\right]=0,
\label{eq:detM-1}
\end{align}
where we defined
\begin{align}
\Omega(\pv)\equiv\sqrt{\frac{(C\Bv\cdot\pv)^2}{\chi\chi_5} - \kappa_-^2  \pv^4 },\quad\kappa_\pm\equiv\frac{1}{2}\left(\frac{\lambda}{\chi}\pm\frac{\lambda_5}{\chi_5}\right).
\end{align}
The result of the contour integral is given by
\begin{align}   
\Sigma_{\alpha\alpha}^{(b)}(\kv,\omega)&=-2\gamma^2\Gamma^2 \int \frac{\dd^d p}{(2\pi)^d} \left(
\frac{ - \lambda \kappa_- \pv^4 +  (C\Bv\cdot\pv)^2/\chi_5  }
{\i\Omega(\pv)}K_-(\pv,\kv)-\lambda\pv^2K_+(\pv,\kv)\right)\notag\\
&\quad
+\frac{ g^2}{2\chi_5^2}
 \int \frac{\dd^d p}{(2\pi)^d} \left(
\frac{ \lambda_5 \kappa_- \pv^4 + (C\Bv\cdot\pv)^2/\chi  }
{\i\Omega(\pv)}K_-(\pv,\kv)-\lambda_5 \pv^2K_+(\pv,\kv)
\right),
\label{eq:Sigma(b)2}
\end{align}
where we introduced
\begin{align}
\label{eq:K} 
K_{\pm}(\pv,\kv)&\equiv
\frac{1}{ [ \kappa_+ \pv^2 + \i\Omega(\pv)] [-\i\omega+\Delta_+(\pv,\kv)+ \i\Omega(\pv)]}
\notag\\
&\quad \pm
\frac{1}{ [ \kappa_+ \pv^2 - \i\Omega(\pv)] [-\i\omega+\Delta_+(\pv,\kv)- \i\Omega(\pv)]},\\
\Delta_+(\pv,\kv)&\equiv\Gamma[r+(\kv-\pv)^2]+\kappa_+\pv^2.
\end{align}
By using an identity,
\begin{align}
 \det [D^0 (\pv,\omega+\i\Gamma[r+(\kv-\pv)^2])]^{-1} = [-\i\omega+\Delta_+(\pv,\kv)]^2+ \Omega(\pv)^2,
\end{align}
together with some straightforward calculations, we obtain
\begin{align}
\Sigma_{\alpha\alpha}^{(b)}(\kv,\omega)
&=4\gamma^2\chi\Gamma^2 \int \frac{\dd^d p}{(2\pi)^d} 
\dis\frac{-\i\omega+\lambda_5\pv^2/\chi_5+\Gamma[r+(\kv-\pv)^2]}{  \det [D^0 (\pv,\omega+\i\Gamma[r+(\kv-\pv)^2])]^{-1}} \nonumber \\
& \quad -\frac{ g^2}{\chi_5} \int \frac{\dd^d p}{(2\pi)^d} 
\dis\frac{-\i\omega+\lambda \pv^2/\chi+\Gamma[r+(\kv-\pv)^2]}{  \det [D^0 (\pv,\omega+\i\Gamma[r+(\kv-\pv)^2])]^{-1}  }.
\label{eq:Sigma(b)4}
\end{align}

Combining Eqs.~(\ref{eq:Sigma(a)2}) and (\ref{eq:Sigma(b)4}), we find 
\begin{align}
\Sigma_{\alpha\alpha}(\kv,\omega)
&=
4\gamma^2\chi\Gamma\int \frac{\dd^d p}{(2\pi)^d}
\frac{
 \Theta(\pv,\kv,0)\Theta_5(\pv,\kv,\omega)+   [C \Bv\cdot(\kv-\pv)]^2/(\chi\chi_5)  }
{(r+\pv^2)[ \Theta(\pv,\kv,\omega)\Theta_5(\pv,\kv,\omega) + (C\Bv\cdot\pv)^2 /(\chi\chi_5)  ]}\notag\\
&\quad -\frac{g^2}{\chi_5}\int \frac{\dd^d p}{(2\pi)^d}
\frac{(r+\kv^2)\Theta(\pv,\kv,\omega)
}
{(r+\pv^2)[ \Theta(\pv,\kv,\omega)\Theta_5(\pv,\kv,\omega)+ (C\Bv\cdot\pv)^2 /(\chi\chi_5)  ]},
\label{eq:Sigma_alal}
\end{align}
where $\Theta(\pv,\kv,\omega)$ and $\Theta_5(\pv,\kv,\omega)$ are defined by
\begin{align}
\Theta(\pv,\kv,\omega)&\equiv-\i\omega+\Gamma(r+\pv^2)+\lambda(\kv-\pv)^2/\chi, \\
\Theta_5(\pv,\kv,\omega)&\equiv-\i\omega+\Gamma(r+\pv^2)+\lambda_5(\kv-\pv)^2/\chi_5 .
\end{align}
Note that $\Sigma_{11}=\Sigma_{22}$ and the first term on the right-hand side of Eq.~(\ref{eq:Sigma_alal}) is independent of $\kv$ when $\omega=0$. From the expression~(\ref{eq:Sigma_alal}), we obtain
\begin{align}
\label{eq:dSigdk2mu}
&\frac{1}{2} \left.   \frac{\d^2 \Sigma_{\alpha\alpha}(\bm{k},0)}{\d \bm{k}^2}    \right|_{\bm{k}\rightarrow \bm{0}}\notag\\
&=-\frac{g^2}{\chi_5} \int \frac{\dd^d p}{(2\pi)^d}  \left( \frac{ \Theta(\pv,\zev,0)}{(r+\pv^2)[\Theta(\pv,\zev,0)\Theta_5(\pv,\zev,0)+ (C\Bv\cdot\pv)^2 /(\chi\chi_5) ] }
 + O(\epsilon)  \right)
,
\end{align}

\begin{align}
&\i \left.   \frac{\d \Sigma_{\alpha\alpha}(\bm{0},\omega)}{\d \omega}   \right|_{\omega\rightarrow0}\notag\\
&=
-4\gamma^2\chi\Gamma \int \frac{\dd^d p}{(2\pi)^d}
\frac{\Theta_5(\pv,\zev,0)}{(r+\pv^2)\{\Theta(\pv,\zev,0)\Theta_5(\pv,\zev,0)+ (C\Bv\cdot\pv)^2 /(\chi\chi_5) \} }\notag\\
&\quad+\frac{g^2}{\chi_5}
\int \frac{\dd^d p }{(2\pi)^d}
\frac{r[ \Theta^2(\pv,\zev,0)-(C\Bv\cdot\pv)^2/(\chi\chi_5)^2 ]}
{(r+\pv^2) [\Theta(\pv,\zev,0)\Theta_5(\pv,\zev,0) +(C\Bv\cdot\pv)^2/(\chi\chi_5)^2   ]^2}.
\label{eq:Sigmaomega}
\end{align}
The $O(\epsilon)$ terms in Eq.~\eqref{eq:dSigdk2mu} proportional to $r$ are irrelevant for the following discussion. 
Let us carry out the integral over $\pv$ in the shell $\Lambda/b<|\pv|<\Lambda$. 
Setting $\Bv=(1,0,0,0)$ and using the parameterization~(\ref{eq:ppara}), we obtain
\begin{align}
&\frac{1}{2\Gamma} \left.   \frac{\d^2 \Sigma_{\alpha\alpha}(\bm{k},0)}{\d \bm{k}^2}    \right|_{\bm{k}\rightarrow \bm{0}}
\notag\\
&=-\frac{2f}{\sqrt{(1+w)(1+w_5)+h^2}+\sqrt{(1+w)(1+w_5)}}\sqrt{\frac{1+w}{1+w_5}}\ln b,
\label{eq:Sigmakk}
\end{align}
where $w,\,w_5,\,h,\,f$ are defined in Eq.~(\ref{eq:Paras}). In order to obtain Eq.~(\ref{eq:Sigmakk}), we carry out the integral over $\theta$ by using the 
following formula (with $a$ being a real constant):
\begin{align}
\label{eq:INTtheta}
\int_0^\pi \dd\theta
\frac{\sin^2\theta}{a^2+\cos^2\theta}
=\frac{\pi}{a(\sqrt{a^2+1}+a^2)}.
\end{align}
To the leading order of $\epsilon$, we can ignore the term proportional to $r$ on the right-hand side of (\ref{eq:Sigmaomega}). 
Then, by comparing Eqs.~(\ref{eq:dSigdk2mu}) and (\ref{eq:Sigmaomega}), we can rewrite Eq.~(\ref{eq:Sigmaomega}) as
\begin{align}
&\i \left.   \frac{\d \Sigma_{\alpha\alpha}(\bm{0},\omega)}{\d \omega}   \right|_{\omega\rightarrow0}\notag\\
&=-\frac{8vw}{\sqrt{(1+w)(1+w_5)+h^2}+\sqrt{(1+w)(1+w_5)}}\sqrt{\frac{1+w_5}{1+w}}\ln b,
\label{eq:Sigmaomega2}
\end{align}
where $v$ is defined in Eq.~(\ref{eq:Parastas}). Using Eqs.~(\ref{eq:Sigmakk}) and (\ref{eq:Sigmaomega2}), 
Eqs.~(\ref{eq:RRGamma}) and (\ref{eq:atildea}) become
\begin{gather}
\label{eq:recurionGamma2}
\Gamma_{l+1}=\Gamma_{l}
\left(1+f_l X_l \ln b
\right)b^{d+z-\tilde a -a-2}, \\
\label{eq:atilde2}
1=\left(  1+4 v_l w_l  X'_l \ln b \right)b^{d-\tilde a -a},
\end{gather}
respectively, where $X$ and $X'$ are defined in Eq.~(\ref{eq:Paras2}). 
By substituting Eq.~(\ref{eq:atilde2}) into Eq.~(\ref{eq:recurionGamma2}), we finally arrive at Eq.~(\ref{eq:RRGamma2}).

\section{Vertex function $\mathcal V$}
\label{sec:VCME}
\begin{figure}[t]
\centering
\begin{minipage}[c]{0.75\hsize}
\subfigure[]{\includegraphics[bb=0 0 215 122,height=3cm]{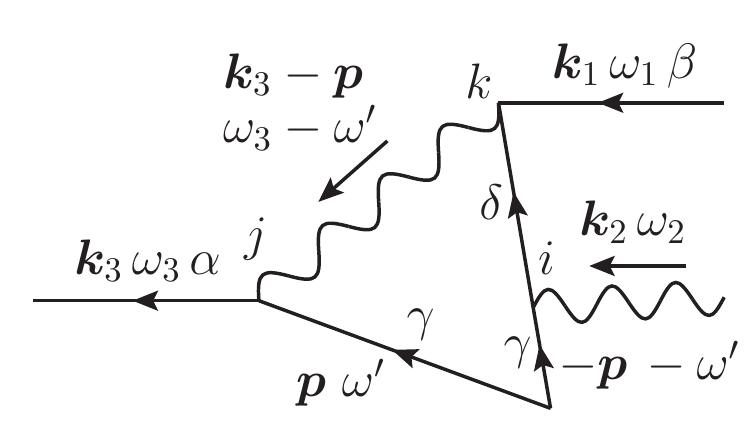}\label{Fig:phina}}~
\subfigure[]{ \includegraphics[bb=0 0 217 127,height=3cm]{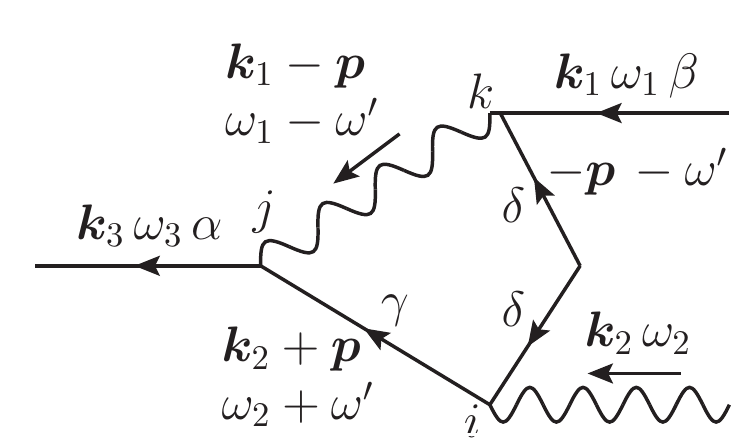}\label{Fig:phinb}}~
\end{minipage}\\
\begin{minipage}[c]{0.75\hsize}
\subfigure[]{\includegraphics[bb=0 0 222 125,height=3cm]{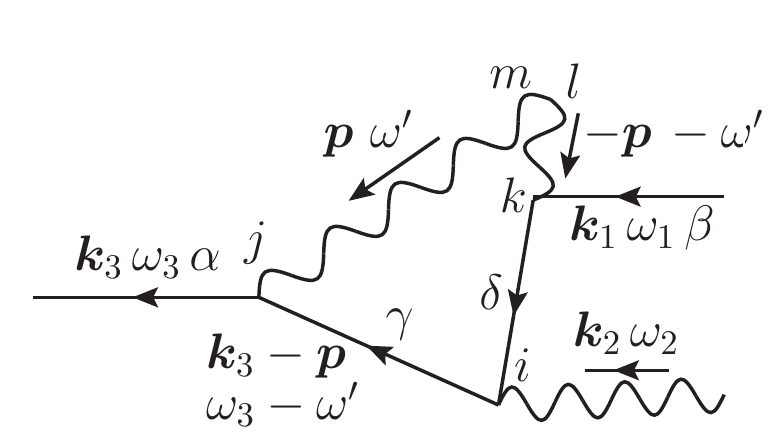}\label{Fig:phinc}}~
\subfigure[]{\includegraphics[bb=0 0 215 122,height=3cm]{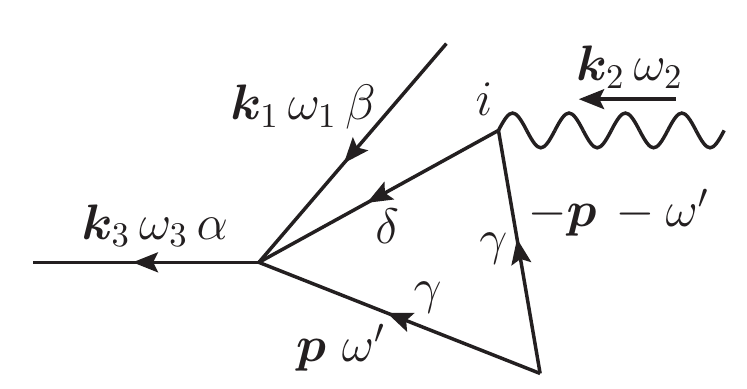}\label{Fig:phind}}
\end{minipage}
\caption{Diagrams for $V_{\alpha;\beta i}$ at one-loop level}\label{Fig:VC}
\end{figure}

The lowest-order diagrams for $\mathcal V_{\alpha;\beta i}$ are depicted in Fig.~\ref{Fig:VC}. We find that these diagrams with $\alpha\neq\beta$ and $i=2$ satisfy the following identity in the limit $\kv_1,\kv_2\rightarrow\zev$ and $\omega_1,\omega_2\rightarrow0$:
\begin{align}
\label{eq:V}
 \mathcal V_{\alpha;\beta2} (\zev,\zev,0,0)  = -\i \frac{g \varepsilon_{\alpha\beta}}{\chi_5}    \left.   \frac{\d \Sigma_{\gamma\gamma}(\bm{0},\omega)}{\d \omega}   \right|_{\omega\rightarrow0}.
\end{align}
Note that the contribution from Fig.~\ref{Fig:phind} vanishes as long as $\alpha\neq\beta$. By substituting Eqs.~(\ref{eq:V}), (\ref{eq:RRchi5}), and (\ref{eq:atildea}) into Eq.~(\ref{eq:RRgchi5}), we obtain Eq.~(\ref{eq:RRg}). In particular, the loop correction of the interaction vortex, Eq.~(\ref{eq:V}), is canceled by the response-field scaling factor which includes the loop correction originating from Eq.~(\ref{eq:atildea}). 

Instead of showing Eq.~(\ref{eq:V}) from the explicit calculation as we carried out in Secs.~\ref{sec:PiCME} and \ref{sec:SigmaCME}, we show Eq.~(\ref{eq:V}) as a consequence of the symmetry algebra. Our derivation is similar to that of Ref.~\cite{ONWT} (see its section~III.\,A), where the Ward-Takahashi identity for O($N$) symmetric systems is derived. We shall begin with the generating functional,
\begin{align}
Z[\tilde \jmath, j, \tilde \mu_5,\mu_5]
 \equiv 
 \left<  \exp \int \dd t \int \dd \rv \left(  \tilde \jmath _\alpha \tilde \phi_\alpha + j_\alpha\phi_\alpha + \tilde \mu_5 \tilde n_5 + \mu_5 n_5  \right)   \right>.
\end{align}
Here, $\tilde \jmath _\alpha,\, j_\alpha,\,\tilde \mu_5,$ and $\mu_5$ are the external fields of 
$\tilde \phi_\alpha, \,\phi_\alpha,\, \tilde n_5$ and 
$n_5$, 
respectively. We define
\begin{align}
\label{eq:Phi}
\tilde \Phi_\alpha \equiv \frac{\delta \ln Z}{\delta \tilde \jmath_\alpha},\quad \Phi_\alpha \equiv \frac{\delta \ln Z}{\delta j_\alpha},\quad
\tilde N_5 \equiv \frac{\delta \ln Z}{\delta \tilde \mu_5},\quad N_5 \equiv \frac{\delta \ln Z}{\delta \mu_5},
\end{align}
which reduce to the expectation values of $\tilde \phi_\alpha, \,\phi_\alpha,\, \tilde n_5$ and $n_5$, respectively, when the external fields are set to zero. 
We also define the effective actions by 
the Legendre transformations of Eq.~(\ref{eq:Phi}):
\begin{align}
\label{eq:W}
W[\tilde \Phi,\Phi,\tilde \mu_5,\mu_5]&\equiv-\ln Z[\tilde \jmath,j, \tilde \mu_5,\mu_5] +   \int \dd t \int \dd \rv \left(\tilde \jmath_\alpha \tilde \Phi_\alpha+j_\alpha\Phi_\alpha  \right),\\
\Gamma[\tilde \Phi,\Phi,\tilde N_5,N_5]&\equiv W[\tilde \Phi,\Phi,\tilde \mu_5,\mu_5]+ \int \dd t \int \dd \rv \left(\tilde \mu_5 \tilde N_5+\mu_5 N_5 \right).
\end{align}
One can show the following identities which will be used later (see, e.g., Sec.~4.\,4 of Ref.~\cite{Tauber} for the derivations):
\begin{gather}
\label{eq:id1}
\frac{\delta W}{\delta \Phi_\alpha}=
\frac{\delta \Gamma}{\delta \Phi_\alpha}=j_\alpha,\quad
\frac{\delta \Gamma}{\delta N_5}=\mu_5,\\
\label{eq:id2}
\frac{\delta^2 \Gamma}{\delta \tilde \Phi_\alpha (\kv,\omega) \delta \Phi_\beta (\kv,\omega)}=G^{-1}_{\alpha\beta}(-\kv,-\omega),\\
\label{eq:id3}
\frac{\delta^2 \Gamma}{\delta \Phi_\alpha  \delta  \Phi_\beta }=\frac{\delta^2 \Gamma}{\delta \Phi_\alpha \delta N_5}=0,\\
\label{eq:id4}
\frac{\delta^3 \Gamma}{\delta N_5\delta \tilde \Phi_\alpha \delta \Phi_\beta}=V_{\alpha;\beta 2}.
\end{gather}
Here, note that $V_{\alpha;\beta 2}$ in Eq.~(\ref{eq:id4}) satisfies Eq.~(\ref{eq:mathcalVdef}).

We use the condition that $W[\tilde \Phi,\Phi,\tilde \mu_5,\mu_5]$ is invariant under the $\text{U(1)}_{\text A}^{\tau^3}$ transformation defined in Eq.~(\ref{eq:U(1)tau3A}). Note that $W[\tilde \Phi,\Phi,\tilde \mu_5,\mu_5]$ is a functional of $\mu_5$, and changing $\mu_5$ corresponds to the $\text{U(1)}_{\text A}^{\tau^3}$ transformation, because $n_5$ is its generator. Let us consider the variation of $\mu_5$ at $t=0$, $\delta \mu_5=\vartheta\mu_5$ with $\vartheta$ being a small parameter. Then, the free energy~(\ref{eq:GL}) has an additional contribution,
\begin{align}
\delta F=\int \dd\rv\, n_5 \delta\mu_5.
\end{align}
By using the reversible term of Eq.~(\ref{eq:Langephi}), an infinitesimal $\text{U(1)}_{\text A}^{\tau^3}$ transformation can be written as
\begin{align}
\label{eq:U(1)Amu5} 
\delta  \Phi_\alpha = g\int_0^t \dd t' \varepsilon_{\alpha\beta} \Phi_\beta \delta \mu_5 = \vartheta g\varepsilon_{\alpha\beta}\Phi_\beta \mu_5 t.
\end{align}
By applying Eq.~(\ref{eq:U(1)Amu5}) to $W[\tilde \Phi,\Phi,\tilde \mu_5,\mu_5]$, we obtain its variation as
\begin{align}
\delta W&= \vartheta \int \dd t \int \dd \rv \,  \mu_5  \left[ \frac{\delta W}{\delta \mu_5} + g  \varepsilon_{\alpha\beta} t \frac{\delta W}{\delta \Phi_\alpha}\Phi_\beta  \right]\\
&= \vartheta \int \dd t \int \dd \rv \, \frac{\delta \Gamma}{\delta N_5}  \left[  N_5 + g  \varepsilon_{\alpha\beta} t \frac{\delta \Gamma}{\delta \Phi_\alpha}\Phi_\beta   \right],
\end{align}
where we use Eq.~(\ref{eq:id1}). 
Therefore, invariance of $W[\tilde \Phi, \Phi,\tilde \mu_5,\mu_5]$
leads to 
\begin{align}
\frac{\delta \Gamma}{\delta N_5}  \left[  N_5 + g  \varepsilon_{\alpha\beta} t \frac{\delta \Gamma}{\delta \Phi_\alpha}\Phi_\beta  \right]=0.
\end{align}
By taking a variation of this equation with respect to $\tilde \Phi_\alpha$ and $\Phi_\beta$, we obtain
\begin{align}
\label{eq:WT}
\frac{\delta^3 \Gamma}{\delta N_5\delta \tilde \Phi_\alpha \delta \Phi_\beta}N_5=-g\varepsilon_{\gamma\beta} t \left( \mu_5 \frac{ \delta^2 \Gamma}{\delta \tilde \Phi_\alpha  \delta \Phi_\gamma}+ j_\gamma \frac{\delta^2 \Gamma}{\delta N_5 \delta\tilde \Phi_\alpha } \right),
\end{align}
where we use Eqs.~(\ref{eq:id1}) and (\ref{eq:id3}). We take the functional derivative of Eq.~(\ref{eq:WT}) with respect to $\mu_5$ and set all the external fields to zero. In frequency space, we obtain,
\begin{align}
\frac{\delta^3 \Gamma}{\delta N_5\delta \tilde \Phi_\alpha \delta \Phi_\beta}=-\i\frac{g\varepsilon_{\gamma\beta}}{\chi_5} \left.\frac{\partial }{\partial \omega}\frac{ \delta^2 \Gamma}{\delta \tilde \Phi_\alpha  \delta \Phi_\gamma}\right|_{\omega\rightarrow0}
=\frac{g\varepsilon_{\gamma\beta}}{\chi_5} \left( 1-\i \left.\frac{\partial \Sigma_{\alpha\gamma}(\zev,\omega)}{\partial \omega}\right|_{\omega\rightarrow0}\right).
\end{align}
Here, we use $\chi_5=\delta N_5/\delta \mu_5$ and Eqs.~(\ref{eq:id2}) and (\ref{eq:DysonG}). By using Eqs.~(\ref{eq:id4}), (\ref{eq:bareV}) and (\ref{eq:mathcalVdef}), we finally arrive at Eq.~(\ref{eq:V}) when $\alpha\neq\beta$. Note here that one can write $\varepsilon_{\gamma\beta}\Sigma_{\alpha\gamma}=\varepsilon_{\alpha\beta}\Sigma_{11}\ ({\rm for}\ \alpha\neq\beta)$ by using $\Sigma_{11}=\Sigma_{22}$, which can be shown from Eq.~(\ref{eq:Sigma_alal}).

\chapter{Linear-stability analysis on the fixed points (iii) and (iv)}

\label{sec:crossover}

We here show that the parameter region that leads to the fixed point (iv) under RG is much larger than (iii) within the linear-stability analysis around the fixed point (i). 
Let us look at the full propagator of the conserved fields, $D_{ij}=\bar D_{ij}(w,h)$ as a function of two relevant parameters $w$ and $h$, which is transformed under the rescaling as 
\begin{align}
\label{eq:D1}
\bar D_{ij}(w,h)\sim \bar D_{ij}(l^{y_w}w,l^{y_h}h).
\end{align}
Here, we assume that $w$ and $h$ are scaled by the eigenvalues $y_w\equiv 7\epsilon/10,\,y_h\equiv 1-\epsilon/4$ obtained in Eq.~(\ref{eq:eigenvalues}), and we omit the overall factor. When the RG flow approaches $l\sim h^{-1/y_h}$, $\bar D_{ij}$ becomes a function of $wh^{-\varphi}$ with $\varphi\equiv y_w/y_h$ being the crossover exponent. Then, $\bar D_{ij}$ behaves differently depending on whether $wh^{-\varphi}\gg 1$ or $wh^{-\varphi}\ll 1$: in the former region, we see the behavior of the fixed point (iii); in the latter region, we see the behavior of the fixed point (iv). Since $\varphi=O(\epsilon)$ and $w\ll1$ near the fixed point (i), the latter parameter region is much larger than the former. Note also that these two regions should be continuously connected and the crossover between the fixed point (iii) and (iv) takes place. The right-hand side of the original RG equations (\ref{eq:REf})--(\ref{eq:REh}) are smooth functions of all the parameters, unless $w$ and $h$ are simultaneously infinity.

%In addition, by considering the initial parameter region with $w \ll 1$, we also find the crossover between the fixed points (i) and (iv) depending on the magnitude of the magnetic field $B$ and the reduced temperature $\tau$ as follows. We recall that $r\approx \tau$ is also a relevant parameter. Similarly to the discussion in the previous paragraph, RG transformation from the fixed point (i) by $l\sim \tau^{-1/\nu}$ yields the propagator as a function of $\tau h ^{-\varphi'}$. Here, $\nu$ defined by Eq.~(\ref{eq:eps}) is the critical exponent of $\tau$, and $\varphi'\equiv(\nu y_h)^{-1}$ is the crossover exponent. Therefore, the dynamic universality class belongs to that of model A governed by the fixed point (iv) in the regime $\tau\ll h^{\varphi'}$ and to that of model E governed by the fixed point (i) in the regime $\tau\gg h^{\varphi'}$ for a fixed magnetic field.

\newpage \thispagestyle{empty}

\end{document}